\documentclass[aps,prc,twocolumn,superscriptaddress,showpacs,nofootinbib,floatfix]{revtex4}
\usepackage{graphicx}
\usepackage{dcolumn}
\usepackage{bm}

\newcommand{\AuAu} {\mbox{$\mbox{Au}+\mbox{Au}$}}
\newcommand{\dAu} {\mbox{$\mbox{d}+\mbox{Au}$}}

\newcommand{\avgcostwodpsi} {\mbox{$\langle \cos{2\left(\Psi_1 - \Psi_2 \right)} \rangle $}}

\newcommand{\dngdy}{\mbox{$dN_g/dy$}} 
\newcommand{\dphi}{\mbox{$\Delta \phi$}}

\newcommand{\Leps}{\mbox{$L_\varepsilon$}}

\newcommand{\Mgg}  {\mbox{$m_{\gamma \gamma}$}}

\newcommand{\Npart}{\mbox{$N_{part}$}}

\newcommand{\piz} {\mbox{$\pi^{0}$}}
\newcommand{\pizs} {\mbox{$\pi^{0}$'s}}
\newcommand{\pp} {\mbox{$\mbox{p}+\mbox{p}$}}
\newcommand{\pt} {\mbox{$p_T$}}

\newcommand{\rhoL} {\mbox{$\rho {L}$}}
\newcommand{\rhoLsq} {\mbox{$\rho {L}^2$}}
\newcommand{\rhoLxy} {\mbox{$\rho {L_{xy}}$}}
\newcommand{\rhoLsqxy} {\mbox{$\rho {L^2_{xy}}$}}

\newcommand{\RAAphi} {\mbox{$R_{AA}(\Delta \phi)$}}
\newcommand{\RAAphipt} {\mbox{$R_{AA}(\Delta \phi, \pt)$}}
\newcommand{\RAApt} {\mbox{$R_{AA}(p_T)$}}
\newcommand{\raa} {\mbox{$R_{AA}$}}
\newcommand{\RAA} {\mbox{$R_{AA}$}}

\newcommand{\Rphipt} {\mbox{$R(\Delta \phi, \pt)$}}

\newcommand{\Sloss} {\mbox{$S_{\rm loss}$}}

\newcommand{\snn} {\mbox{$\sqrt{s_{NN}}$}}

\newcommand{\TAA}{\mbox{$T_{AA}(x,y)$}}

\newcommand{\vt} {\mbox{$v_{2}$}}

\newcommand{\wrt} {with respect to}

\begin{document}

\title{A Detailed Study of High-\pt\ Neutral Pion Suppression and \\
 Azimuthal Anisotropy in \AuAu\ Collisions at \snn\ = 200 GeV}

\newcommand{\abilene}{Abilene Christian University, Abilene, TX 79699, USA}
\newcommand{\acadsin}{Institute of Physics, Academia Sinica, Taipei 11529, Taiwan}
\newcommand{\banaras}{Department of Physics, Banaras Hindu University, Varanasi 221005, India}
\newcommand{\barc}{Bhabha Atomic Research Centre, Bombay 400 085, India}
\newcommand{\bnl}{Brookhaven National Laboratory, Upton, NY 11973-5000, USA}
\newcommand{\caucr}{University of California - Riverside, Riverside, CA 92521, USA}
\newcommand{\ciae}{China Institute of Atomic Energy (CIAE), Beijing, People's Republic of China}
\newcommand{\cns}{Center for Nuclear Study, Graduate School of Science, University of Tokyo, 7-3-1 Hongo, Bunkyo, Tokyo 113-0033, Japan}
\newcommand{\columbia}{Columbia University, New York, NY 10027 and Nevis Laboratories, Irvington, NY 10533, USA}
\newcommand{\dapnia}{Dapnia, CEA Saclay, F-91191, Gif-sur-Yvette, France}
\newcommand{\debrecen}{Debrecen University, H-4010 Debrecen, Egyetem t{\'e}r 1, Hungary}
\newcommand{\fsu}{Florida State University, Tallahassee, FL 32306, USA}
\newcommand{\gsu}{Georgia State University, Atlanta, GA 30303, USA}
\newcommand{\hiroshima}{Hiroshima University, Kagamiyama, Higashi-Hiroshima 739-8526, Japan}
\newcommand{\ihepprot}{Institute for High Energy Physics (IHEP), Protvino, Russia}
\newcommand{\isu}{Iowa State University, Ames, IA 50011, USA}
\newcommand{\jinrdubna}{Joint Institute for Nuclear Research, 141980 Dubna, Moscow Region, Russia}
\newcommand{\kaeri}{KAERI, Cyclotron Application Laboratory, Seoul, South Korea}
\newcommand{\kangnung}{Kangnung National University, Kangnung 210-702, South Korea}
\newcommand{\kek}{KEK, High Energy Accelerator Research Organization, Tsukuba-shi, Ibaraki-ken 305-0801, Japan}
\newcommand{\kfki}{KFKI Research Institute for Particle and Nuclear Physics (RMKI), H-1525 Budapest 114, POBox 49, Hungary}
\newcommand{\korea}{Korea University, Seoul, 136-701, Korea}
\newcommand{\kurchatov}{Russian Research Center ``Kurchatov Institute", Moscow, Russia}
\newcommand{\kyoto}{Kyoto University, Kyoto 606-8502, Japan}
\newcommand{\labllr}{Laboratoire Leprince-Ringuet, Ecole Polytechnique, CNRS-IN2P3, Route de Saclay, F-91128, Palaiseau, France}
\newcommand{\lawllnl}{Lawrence Livermore National Laboratory, Livermore, CA 94550, USA}
\newcommand{\losalamos}{Los Alamos National Laboratory, Los Alamos, NM 87545, USA}
\newcommand{\lpc}{LPC, Universit{\'e} Blaise Pascal, CNRS-IN2P3, Clermont-Fd, 63177 Aubiere Cedex, France}
\newcommand{\lund}{Department of Physics, Lund University, Box 118, SE-221 00 Lund, Sweden}
\newcommand{\muenster}{Institut f\"ur Kernphysik, University of Muenster, D-48149 Muenster, Germany}
\newcommand{\myongji}{Myongji University, Yongin, Kyonggido 449-728, Korea}
\newcommand{\nagasaki}{Nagasaki Institute of Applied Science, Nagasaki-shi, Nagasaki 851-0193, Japan}
\newcommand{\newmex}{University of New Mexico, Albuquerque, NM 87131, USA}
\newcommand{\nmsu}{New Mexico State University, Las Cruces, NM 88003, USA}
\newcommand{\ornl}{Oak Ridge National Laboratory, Oak Ridge, TN 37831, USA}
\newcommand{\orsay}{IPN-Orsay, Universite Paris Sud, CNRS-IN2P3, BP1, F-91406, Orsay, France}
\newcommand{\pnpi}{PNPI, Petersburg Nuclear Physics Institute, Gatchina, Russia}
\newcommand{\riken}{RIKEN (The Institute of Physical and Chemical Research), Wako, Saitama 351-0198, JAPAN}
\newcommand{\rikjrbrc}{RIKEN BNL Research Center, Brookhaven National Laboratory, Upton, NY 11973-5000, USA}
\newcommand{\saispbstu}{St. Petersburg State Technical University, St. Petersburg, Russia}
\newcommand{\saopaulo}{Universidade de S{\~a}o Paulo, Instituto de F\'{\i}sica, Caixa Postal 66318, S{\~a}o Paulo CEP05315-970, Brazil}
\newcommand{\seoulnat}{System Electronics Laboratory, Seoul National University, Seoul, South Korea}
\newcommand{\stonybrkc}{Chemistry Department, Stony Brook University, SUNY, Stony Brook, NY 11794-3400, USA}
\newcommand{\stonycrkp}{Department of Physics and Astronomy, Stony Brook University, SUNY, Stony Brook, NY 11794, USA}
\newcommand{\subatech}{SUBATECH (Ecole des Mines de Nantes, CNRS-IN2P3, Universit{\'e} de Nantes) BP 20722 - 44307, Nantes, France}
\newcommand{\tenn}{University of Tennessee, Knoxville, TN 37996, USA}
\newcommand{\titech}{Department of Physics, Tokyo Institute of Technology, Tokyo, 152-8551, Japan}
\newcommand{\tsukuba}{Institute of Physics, University of Tsukuba, Tsukuba, Ibaraki 305, Japan}
\newcommand{\vandy}{Vanderbilt University, Nashville, TN 37235, USA}
\newcommand{\waseda}{Waseda University, Advanced Research Institute for Science and Engineering, 17 Kikui-cho, Shinjuku-ku, Tokyo 162-0044, Japan}
\newcommand{\weizmann}{Weizmann Institute, Rehovot 76100, Israel}
\newcommand{\yonsei}{Yonsei University, IPAP, Seoul 120-749, Korea}
\affiliation{\abilene}
\affiliation{\acadsin}
\affiliation{\banaras}
\affiliation{\barc}
\affiliation{\bnl}
\affiliation{\caucr}
\affiliation{\ciae}
\affiliation{\cns}
\affiliation{\columbia}
\affiliation{\dapnia}
\affiliation{\debrecen}
\affiliation{\fsu}
\affiliation{\gsu}
\affiliation{\hiroshima}
\affiliation{\ihepprot}
\affiliation{\isu}
\affiliation{\jinrdubna}
\affiliation{\kaeri}
\affiliation{\kangnung}
\affiliation{\kek}
\affiliation{\kfki}
\affiliation{\korea}
\affiliation{\kurchatov}
\affiliation{\kyoto}
\affiliation{\labllr}
\affiliation{\lawllnl}
\affiliation{\losalamos}
\affiliation{\lpc}
\affiliation{\lund}
\affiliation{\muenster}
\affiliation{\myongji}
\affiliation{\nagasaki}
\affiliation{\newmex}
\affiliation{\nmsu}
\affiliation{\ornl}
\affiliation{\orsay}
\affiliation{\pnpi}
\affiliation{\riken}
\affiliation{\rikjrbrc}
\affiliation{\saispbstu}
\affiliation{\saopaulo}
\affiliation{\seoulnat}
\affiliation{\stonybrkc}
\affiliation{\stonycrkp}
\affiliation{\subatech}
\affiliation{\tenn}
\affiliation{\titech}
\affiliation{\tsukuba}
\affiliation{\vandy}
\affiliation{\waseda}
\affiliation{\weizmann}
\affiliation{\yonsei}
\author{S.S.~Adler}	\affiliation{\bnl}
\author{S.~Afanasiev}	\affiliation{\jinrdubna}
\author{C.~Aidala}	\affiliation{\bnl}
\author{N.N.~Ajitanand}	\affiliation{\stonybrkc}
\author{Y.~Akiba}	\affiliation{\kek} \affiliation{\riken}
\author{J.~Alexander}	\affiliation{\stonybrkc}
\author{R.~Amirikas}	\affiliation{\fsu}
\author{L.~Aphecetche}	\affiliation{\subatech}
\author{S.H.~Aronson}	\affiliation{\bnl}
\author{R.~Averbeck}	\affiliation{\stonycrkp}
% ...
\collaboration{PHENIX Collaboration} \noaffiliation

\date{\today}

\begin{abstract}
Measurements of neutral pion (\piz) production at mid-rapidity in
$\snn = 200$~GeV \AuAu\ collisions as a function of transverse
momentum, \pt, collision centrality, and angle \wrt\ 
reaction plane are presented. The data represent the final
\piz\ results from the PHENIX experiment for the first RHIC
\AuAu\ run at design center-of-mass-energy. They include
additional data obtained using the PHENIX Level-2 trigger
with more than a factor of three increase in statistics over
previously published results for \mbox{$\pt > 6$~GeV/$c$}.
We evaluate the suppression in the yield of high-\pt\
\piz's relative to point-like scaling expectations using
the nuclear modification factor \raa. We present the \pt\
dependence of \raa\ for nine bins in collision centrality.
We separately integrate \raa\ over larger \pt\ bins to show
more precisely the centrality dependence of the high-\pt\
suppression. We then evaluate the dependence of the
high-\pt\ suppression on the emission angle  \dphi\ of the
pions \wrt\ event reaction plane for 7 bins in collision
centrality. We show that the yields of high-\pt\ \piz's
vary strongly with \dphi, consistent with prior
measurements \cite{Adcox:2002ms,Ackermann:2000tr}. We show
that this variation persists in the most peripheral bin
accessible in this analysis. For the peripheral bins we
observe no suppression for neutral pions produced aligned
with the reaction plane while the yield of \piz's produced
perpendicular to the reaction plane is suppressed by a
factor of $\sim2$. We analyze the combined centrality and
\dphi\ dependence of the \piz\ suppression in different
\pt\ bins using different possible descriptions of parton
energy loss dependence on jet path-length averages to
determine whether a single geometric picture can explain the observed
suppression pattern.
\end{abstract}


\maketitle

\tableofcontents


\section{Introduction}
\label{sec:intro}

High transverse momentum particles resulting from hard
scatterings between incident partons have become one of the
most effective tools for probing the properties of the
medium created in ultra-relativistic heavy ion collisions
at RHIC. Data from the four RHIC experiments have
unequivocally established the phenomenon of high transverse
momentum hadron suppression in \AuAu\ compared to
(appropriately scaled) \pp\
collisions~\cite{Adcox:2001jp,Adcox:2002pe,Adams:2003kv,Adler:2003qi,Adler:2003au,Arsene:2003yk,Back:2003qr},
while the lack of similar suppression in \dAu\ collisions
\cite{Adler:2003ii,Adams:2003im,Back:2003ns,Arsene:2003yk}
provides strong evidence that the suppression is not due to
modification of parton distributions in the incident
nuclei.  This suppression has been observed for a large
variety of hadron species, at highest $p_T$ for \piz\ and
most recently $\eta$ \cite{Adler:2006hu} supporting further
the notion that energy loss occurs at the parton level.
Conversely, direct photon measurements by the PHENIX
collaboration show that the yield of hard photons in \AuAu\
collisions is consistent with \pp\ expectations scaled by
the number of incoherent nucleon-nucleon collisions
\cite{Adler:2005ig} and, thus, provide final confirmation
that hard scattering processes occur at rates expected from
point-like processes. This observation makes definitive the
conclusion that the suppression of high-\pt\ hadron
production in \AuAu\ collisions is a final-state effect.
Measurements of azimuthal angle correlations between hadron
pairs resulting from fragmentation of hard-scattered
partons into jets have provided additional confirmation of
final-state medium effects on these partons
\cite{Adler:2002ct}.

Predictions of high-\pt\ suppression were made before the start of
RHIC operation \cite{Bass:1999zq} and confirmation of these
predictions may be considered one of the key successes of the RHIC
program so far. The suppression of high-\pt\ single hadrons was
predicted to result from the energy loss of hard-scattered quarks and
gluons in the hot and dense QCD medium created in ultra-relativistic
heavy ion collisions (see \cite{Gyulassy:2003mc,Kovner:2003zj} and
references therein). In the canonical models, medium-induced gluon
bremsstrahlung is expected to dominate the energy loss process
\cite{Wang:1992xy}, and calculations of the high-\pt\ suppression
factor incorporating this effect have been able to successfully
describe the experimental measurements
\cite{Vitev:2002pf,Wang:2002ri,Eskola:2004cr}; however, recent
measurements of heavy quark suppression pose some questions to this
canonical view. Nonetheless, from comparisons of the energy loss
calculations with the experimental data, estimates of the initial net
color charge density which is usually expressed in terms of a gluon
rapidity density, \dngdy, have been obtained yielding $\dngdy \approx
1000$ and, assuming thermalization, estimates of the initial energy
density have produced values in excess of ${10~\rm GeV/fm^3}$
\cite{Gyulassy:2004zy,Wang:2004dn}.

However, in spite of this success, there are still a number of
outstanding issues with the interpretation of the \AuAu\ high-\pt\
single-hadron suppression. Since the properties of the medium created in
heavy ion collisions are not \emph{a priori} known, the energy-loss
calculations necessarily use the observed suppression to infer initial
parton densities, usually through an intermediate parameter that
appears in the energy loss calculations.
Although the initial parton density obtained by such ``tomographic'' studies
has to be consistent with the final (measured) total particle multiplicity,
it is fair to acknowledge that the \pt\ dependence of the suppression
(rather than its absolute magnitude) is a more discriminating observable
to test the various energy loss models. For \piz\ spectra, the
suppression in central \AuAu\ collisions at $\snn = 200$~GeV is
found to be approximately constant with \pt\ over the range, $3 <
\pt < 10$~GeV/$c$. While the different energy loss calculations can
reproduce this \pt-independent suppression, the detailed
explanation of the constancy is different in each model.
The effects invoked to explain the \pt\ dependence of the observed
\AuAu\ high-\pt\ suppression include: finite-energy
effects, absorption of energy from the medium, evolution from
incoherent (Bethe-Heitler) to coherent (Landau-Pomeranchuk-Migdal or LPM)
radiation with increasing
parton energy \cite{Aurenche:2000gf}, the \pt-dependent mixture of quark and gluon
contributions to the hard-scattered parton spectrum, the increasingly larger
exponent of the underlying (power-law) parton \pt\ spectra~\cite{Eskola:2004cr},
and shadowing/EMC effect~\cite{Arneodo:1992wf}. While most calculations of the high-\pt\ suppression in
\AuAu\ collisions account for shadowing/EMC modifications
of the nuclear parton distributions and for the relative mixture of
quarks and gluons in the hard-scattered parton spectra, finite-energy
corrections, absorption of energy from the medium, and the description
of the energy loss process itself differs from calculation to
calculation. Clearly the central \AuAu\ single-particle spectra are not
sufficient, by themselves, to validate or exclude any of the different
energy loss models; we must use more ``differential'' probes of
medium-induced energy loss to better understand the phenomenon.

A robust prediction of non-Abelian parton energy loss calculations is
that the average energy loss as a function of the in-medium path
length $L$ shows a quadratic dependence $\propto L^2$
~\cite{Baier:2000mf}. Such a behavior predicted for a
{\it static} QCD medium, turns into an effective $\propto
L$-dependence in an expanding QGP~\cite{Gyulassy:2001nm}.  In
principle, the centrality dependence of the high-\pt\ suppression
\cite{Adler:2003qi,Back:2003qr,Adams:2003kv} provides an effective
test of energy-loss calculations because the length of the path of the
partons in the medium will change between peripheral and central
collisions. However, the energy loss calculations also have to account
for changes in the initial properties of the medium with centrality
and the extra flexibility in the description of the initial conditions
means that the measured centrality dependence of the high-\pt\
suppression also does not stringently constrain energy loss models
\cite{Drees:2003zh}. However, the path length of the parton in the
medium can also be controlled by selecting high-\pt\ hadrons in
different bins of azimuthal angle difference from the event-by-event
determined reaction plane. Indeed, shortly after experimental
observations of azimuthal anisotropy were reported
\cite{Adler:2002ct,Adcox:2002ms}, arguments were made that the
high-\pt\ anisotropy in non-central collisions was due to the spatial
asymmetry of the medium and the resulting \dphi\ dependence of parton
path lengths \cite{Gyulassy:2000gk,Gyulassy:2001kr}. However, recent
analyses have argued that the large azimuthal anisotropies at high
\pt\ cannot be accounted for by energy loss alone -- at least when
realistic nuclear geometry is used to describe the spatial asymmetry
of the initial state \cite{Drees:2003zh,Shuryak:2001me,Muller:2002fa}.
Some of these analyses were based on a picture of the energy loss
process in which quarks or gluons that have emitted radiation
effectively disappear from the steeply falling high-\pt\ spectrum
because they are overwhelmed by partons of lower energy that escape
from the medium losing little or no energy. In this picture, the
medium effectively attenuates the high-\pt\ quarks and gluons and the
high-\pt\ spectrum is dominated by partons originating near the
surface -- i.e. partons originating in the ``corona''
\cite{Drees:2003zh,Shuryak:2001me,Muller:2002fa}. Then, the azimuthal
anisotropy could be largely determined by the  shape of the surface
\cite{Shuryak:2001me}. However, it has been separately argued that
fluctuations in the number of emitted gluons may be large and such
fluctuations may weaken the corona effect \cite{Gyulassy:2001nm}.

In this paper we present measurements of \piz\ production in $\snn
= 200$~GeV \AuAu\ collisions from the PHENIX experiment at RHIC. These
data, obtained during Run-2 operation of RHIC in 2002, include
additional data obtained with the PHENIX Level-2 trigger which
improved the total statistics by a factor of $\sim$ 3 compared to the
prior analysis in \cite{Adler:2003qi}. The analyses presented here
have also benefitted from advanced electromagnetic calorimeter
calibrations and from improved understanding of the systematic errors
in the \piz\ measurement in course of the direct photon analysis
presented in \cite{Adler:2005ig}, where the \piz\ decay photons
provide the main source of background. With the improved statistics,
the \pt\ reach of the data is extended to higher \pt, allowing us to
test whether the suppression starts to diminish above 10~GeV/$c$ in
\pt. In addition, we extend the measurement of the centrality
dependence of the suppression up to 8~GeV/$c$.

We present measurements of the dependence of the \piz\ yield as a
function of the angle \dphi\ of the \piz\ \wrt\ the event reaction
plane. By measuring the high-\pt\ hadron suppression as a function of
\dphi, for a given centrality bin, we can keep the properties of the
medium fixed and vary only the average geometry of the jet
propagation in the medium. By comparing different centrality bins we
can, in principle, test how the initial properties of the medium
affect the induced energy loss. Traditionally, measurements of the
\dphi\ dependence of hadron yields have been analyzed in terms of the
elliptic flow parameter, \vt, and we note that the data presented
here were used to obtain measurements of \piz\ \vt\ for comparison to
inclusive photon \vt\  \cite{Adler:2005rg}. However, in this
publication we focus not on \vt, but explicitly on the suppression as
a function of \dphi, expressed in terms of the \dphi-dependent
nuclear modification factor \RAAphi. While the data presented this
way contain, in principle, the same information as the combination of
\dphi-averaged \raa\ and \vt, \RAAphi\ provides a useful alternative
way to evaluate the dependence of high-\pt\ suppression on geometry
because it effectively combines \RAApt\ and \vt\ into a single set of
data.  We analyze the combined \dphi\ and centrality dependence of the
high-\pt\ suppression in the context of different path-length and
density  dependencies of the parton energy loss process to evaluate
whether any geometric picture can simultaneously describe the
centrality and \dphi\ dependence of the observed high \pt\ deficit.

\section{Experimental Details}
\label{sec:experiment}

The data presented in this paper were obtained during Run-2 operation
of the PHENIX experiment \cite{Adcox:2003zm} at the Relativistic Heavy
Ion Collider facility at Brookhaven National Laboratory
\cite{Hahn:2003sc}. The primary detectors used to obtain the presented
results were the PHENIX central arm spectrometers, particularly the
electromagnetic calorimeters
\cite{Aphecetche:2003zr}, and the two beam-beam counters (BBC's)
\cite{Allen:2003zt}. In addition, the PHENIX zero-degree calorimeters
\cite{Adler:2000bd} were used for triggering and centrality determination.

Two-photon decays of neutral pions were measured in the PHENIX
electromagnetic calorimeter, located at a radial distance of
$\sim$5.1~m from the beam-line, which has a pseudo-rapidity acceptance
of $-0.35 < \eta < 0.35$ and covers $\pi$ radians in azimuth. The
electromagnetic calorimeter is divided into eight sectors, with each
sector covering the full pseudo-rapidity range and $\pi/8$ in
azimuth. The calorimeter consists of two distinct parts using
different technologies. A lead-scintillator sandwich calorimeter
(PbSc) with $5~{\rm cm} \times 5~{\rm cm}$ towers covers 3/4 (6
sectors) of the central arm acceptance. A lead-glass \v{C}erenkov
calorimeter (PbGl) with $4~{\rm cm} \times 4~{\rm cm}$ towers covers
the remaining 1/4 (2 sectors) of the central arm acceptance.  The
corresponding $\Delta \eta \times \Delta \phi$ acceptance of a single
tower at $\eta = 0$ is $0.011^2$ and $0.0075^2$ for the PbSc and PbGl
calorimeters, respectively.

The event reaction plane in \AuAu\ collisions was measured in the two
BBC's. Each BBC consists of 64 hexagonal, quartz \v{C}erenkov
radiators closely packed around the beam pipe, in an approximately
azimuthally symmetric configuration. The beam-beam counters, located
144~cm in each direction from the nominal center of the interaction
diamond, are used to count charged particles produced in the
pseudo-rapidity range $3.0 < |\eta| < 3.9$. The distribution of
particles over the individual channels of the BBC's allows measurement
of the azimuthal distribution, $dN_{\rm ch}/d\phi$, of charged
particles within this pseudo-rapidity acceptance. The BBC's also
provide measurement of the collision vertex position along the
interaction diamond with a resolution of 0.6~cm \cite{Allen:2003zt}.

The data presented here were obtained using the PHENIX minimum-bias
Level-1 trigger, based on the BBC's and the PHENIX zero-degree
calorimeters, that selects $92.2^{+2.5}_{-3.0}\%$ of the total \AuAu\
hadronic interaction cross-section of 6.9~b \cite{Adler:2003qi}. For
a subset of the data, events selected by the Level-1 trigger were
subjected to software Level-2 trigger filtering after full assembly
of events in the PHENIX event builder \cite{Adler:2003zu}. A software
algorithm performed a crude reconstruction of electromagnetic
clusters by summing the pedestal-subtracted and gain-calibrated
energies of ``tiles'' made of adjacent $4 \times 4$ calorimeter
towers groups. The tiles are allowed to overlap such that every
possible such tile that can be constructed in each calorimeter is
tested. One of the Level-2 triggers ("LVL2A") selected events in
which at least one cluster (tile) had energy $> 3.5$~GeV. Another
Level-2 trigger ("LVL2B") selected events in the 50-92\% centrality
range (50\% most peripheral events) with at least one cluster having
energy $> 1.5$~GeV.

The measurements presented in this paper were obtained from 31.4~M
minimum bias triggers and approximately 1.7~M Level-2 trigger
selected events. Of the Level-2 triggered events, 743 K events were
selected by the higher energy LVL2A trigger and the remainder were
selected by the peripheral, lower-energy LVL2B
trigger. Taking into account their rejection factors, the two
triggers sampled the equivalent of 44.4~M (LVL2A) and 28.7~M (LVL2B)
minimum-bias triggers, respectively.  The difference is due to different
online trigger pre-scale factors.  Thus, the combined event sample
contains approximately a factor of 2.5-3 (considering both triggers
over all centralities) more \piz's above 6~GeV/$c$ than previously
published Run-2 \piz\ measurements \cite{Adler:2003qi}.

\section{Data Analysis}
\label{sec:ana}

\subsection{Event Selection and Centrality}
\label{sec:eventana}

In the offline analysis, the timing difference measured between the
two PHENIX BBC's is used to determine the position of the collision
vertex along the beam axis and to select events with vertex position
within 30~cm of the nominal center of the detector for subsequent
analysis. The energies measured in the zero-degree calorimeters and
the charged particle multiplicity measured in the BBC's are used to
determine the collision centrality~\cite{Adcox:2000sp}. For the \piz\
spectrum measurements presented here the total measured centrality
range (0-92.2\%) is subdivided into 9 bins: 0-10, 10-20, 20-30,
30-40, 40-50, 60-70, 70-80, 80-92.2\%. For the reaction plane
dependent analysis, the most central and two most peripheral bins are
excluded, the peripheral due to their large uncertainty in the
reaction plane resolution, and the 0-10\% bin simply because of its
smaller intrinsic eccentricity.  Additionally, we present also
combined 0-20\%, 20-60\% and 60-92\% data sets for comparison with
other PHENIX analyses of high \pt\ hadron production that use such
centralities.

\subsection{Reaction Plane Measurement}
\label{sec:reacpl}


\begin{figure}[tbh]
\includegraphics[width=0.7\linewidth]{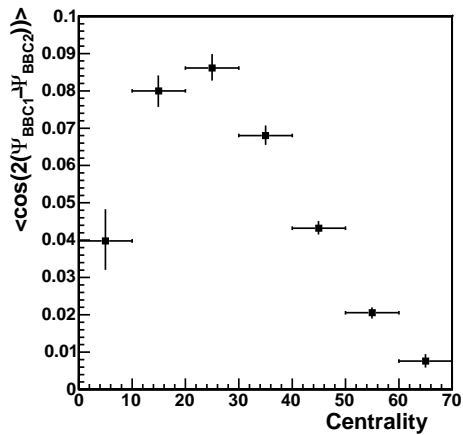}
   \caption{Resolution of the reaction plane determined in the BBC
    versus centrality.  The actual resolution is two times the
   square root of the quantity shown in the y-axis. 
   \label{fig:rp_reso} }
\end{figure}


\begin{table}[h]
   \caption{\label{tab:rp_reso}
Relative systematic uncertainty of the reaction plane resolution for the
   centrality bins used for the reaction plane analysis.  
(see Fig.~\protect\ref{fig:rp_reso}) }
\begin{ruledtabular}\begin{tabular}{crrc}
 & Centrality & Syst. error & \\ \hline
 & 0-10\% & 20.3\% & \\
 & 10-20\% & 5.1\% & \\
 & 20-30\% & 3.9\% & \\
 & 30-40\% & 3.8\% & \\
 & 40-50\% & 4.1\% & \\
 & 50-60\% & 4.6\% & \\
 &  60-70\% & 22.5\% & \\
\end{tabular}\end{ruledtabular}
\end{table}

PHENIX has previously published measurements of elliptic flow using
an event-by-event measured reaction plane
\cite{Adler:2005rg,Adler:2003kt,Adler:2005ab}, and the same technique
is used for the analysis presented here.  Each BBC detector consists
of 128 quartz radiators placed in hexagonal, roughly concentric rings
whose light is individually collected by Photomultiplier tubes (PMT's).
The calibrated charge from each radiator is converted into
an estimate for the number of charged particles within the acceptance
of each detector, $N_i$, using the measured single particle peak
centroid.

For the reaction plane measurement the measured $N_i$ values are
corrected such that the weight of the inner rings which have the
fewest radiators covering the full azimuthal angle range is reduced.
Then, in terms of the corrected $N_i$ values, $N_i^{\rm adj}$, the
angle of the reaction plane $\Psi$ is obtained from the formula
\begin{equation}
\tan{(2\Psi)} = \frac{\sum_{i} N_i^{\rm adj} \sin{(2\phi_i)} -
                     \langle \sum_{i} N_i^{\rm adj} \sin{(2\phi_i)}\rangle }
                    {\sum_{i} N_i^{\rm adj} \cos{(2\phi_i)} -
                     \langle \sum_{i} N_i^{\rm adj} \cos{(2\phi_i)}\rangle },
\label{eq:psirp}
\end{equation}
where $\phi_i$ represents the azimuthal angle of the center of a given
radiator $i$.  The subtraction of the average centroid position in
Eq.~(\ref{eq:psirp}) removes the bias in the reaction plane
measurement resulting from non-zero angle of the colliding beams,
non-uniformities in detector acceptance, and other similar
effects. The average is taken over many events localized in time with
the event in question.  A final correction is applied to remove
non-uniformities at the 20\% level in the $\Psi$ distribution.

Because the above-described procedure can also be applied
individually to each BBC, we have a redundant
measurement of the reaction plane in the north and in the south, and
we exploit this to determine the resolution of the full reaction
plane measurement using standard procedures \cite{Poskanzer:1998yz}.
The resolution of the reaction plane is directly reflected in the
quantity \avgcostwodpsi\ where $\Psi_1$ and $\Psi_2$ are the reaction
plane angles measured in each of the two beam-beam counters
individually and the average is taken over events.
Figure~\ref{fig:rp_reso} and Table.~\ref{tab:rp_reso} show
the variation of this quantity. The needed correction
factors can be derived from this using Eq.~(\ref{eq:rphicorr}) in
Sec.~\ref{sec:raaphi}, \RAAphi \emph{~Measurement}, where the reaction
plane corrections are described in more detail.

The systematic errors associated with the measurement of the reaction
plane come dominantly from how well the resolution is known.  The
uncertainty on this quantity is also shown with Figure
\ref{fig:rp_reso} for all but the most peripheral centralities.  This
error is determined by observing comparison of the similarly
calculated quantity $\langle \sin{2(\Psi_1-\Psi_2)} \rangle$ which
should by definition be equal to zero.  The value of $\langle
\sin{2(\Psi_1-\Psi_2)} \rangle$ is found to be consistent with 0 for
all centralities.  The mean size of its fluctuations around 0 are
compared to the size of the \avgcostwodpsi\ to derive the systematic
errors in the table.  Since the value of \avgcostwodpsi\ decreases
dramatically in the lower multiplicity peripheral events, the
relative size of the error increases.  The size of this relative
error is also cross checked by comparing it to the relative error on
elliptic flow ($v_2$) measurements which is directly comparable
since, as discussed in Sec.~\ref{sec:raaphi}, \RAAphi
\emph{~Measurement}, the resolution correction for $v_2$ is a plain
multiplicative factor.  For the cross check, the $v_2$ error is
derived by taking the difference of $v_2$ made with reaction planes
from the BBC North and BBC South separately.

Because of the large rapidity gap between the PHENIX BBC's and the
PHENIX Central Arm ($\Delta\eta > 2.7-4.0$), the measurements made in
the BBC's are assumed to have no correlations (except collision
geometry) with processes detected in the central arm that would
affect the results presented in Sec.~\ref{sec:raaphi}, \RAAphi
\emph{~Measurement}.  Specifically, PYTHIA studies \cite{Jia:2006sb}
indicate that any large rapidity-gap production correlated with jets
(and thus the hard \piz\'s we study) detected in the central arm have a
negligible effect on reaction plane determination even for the most
peripheral events considered in this paper.  Further, we average both
the North and South BBC, which are separated by $\Delta\eta > 6.0$,
making potential effects of this nature especially unlikely.

\subsection{Neutral Pion Detection}
\label{sec:pi0}

The detection of neutral pions is one of the major sources of
information on identified particle production at high $p_T$ at RHIC,
and PHENIX has already published the results of a number of \piz\
measurements in different colliding systems
\cite{Adcox:2001jp,Adler:2003qi,Adler:2003pb,Adler:2003ii,Adler:2004ps}.
Here we will describe the technique for obtaining \piz\ yields as a
function of \pt\ and centrality, which is now well established within
PHENIX.

Neutral pions are detected via their $\pi^0 \rightarrow \gamma+\gamma$
decay channel. Due to the relatively short mean lifetime of neutral
pions of about $10^{-16}$\,s, typical of electromagnetic decays, the
pions decay close to the interaction point ($c\tau \approx
25$~nm). This makes the decay vertex well known and the pions can be
reconstructed via an invariant mass analysis of photon pairs measured
by the EMCal.

In the EMCal, hits or clusters are reconstructed by finding contiguous
calorimeter towers with pulse heights above the ADC pedestal value.
In order to obtain a cleaner sample of electromagnetic hits shower shape cuts
are applied to select candidate photons and time-of-flight cuts are
applied to reject slow hadrons. For the PbSc we require measured
cluster times to be $t_{\rm clust} < L/c \pm 1.2$~ns where $L$ is the
straight-line path from the collision vertex to the reconstructed
cluster centroid. For the PbGl we require reconstructed clusters to
have times, $t_{\rm clust} < L/c \pm 2$~ns; the difference is due to
the intrinsic timing resolutions of the two calorimeter
technologies.

The energy of each EMCal cluster is corrected for angular dependence
and non-linearity based on test beam results and simulation.  The
linearity corrections for both detector types are different with the
PbGl showing a stronger dependence on the energy. The correction
factors for a photon with a detected energy of 1 GeV (10 GeV) are 1
(0.95) for the PbSc and 1.05 (0.975) for the PbGl, respectively. The
PbGl calorimeter also shows a stronger variation of the measured
photon energy with the angle of incidence on the detector surface, at
$20^{\circ}$ the measured energy is reduced by 5\% compared to
perpendicular incidence ($0^{\circ}$), while in the PbSc the effect is
only of the order of 2\%.

In a typical \AuAu\ central event the EMCal detects $>$ nearly 300
clusters corresponding to an occupancy of $\sim$ 10\% and therefore a
non-negligible probability of cluster overlaps. To minimize the
effects of cluster overlaps in high multiplicity events, the energy
of each cluster in the PbSc calorimeter is determined not only from
the sum of all contiguous towers with deposited energy above a given
threshold (15 MeV was our default value) but also, alternatively,
``extrapolating'' the measured ``core energy" of the 4--5 central
towers assuming a standard electromagnetic shower profile in an event
with zero background. For this latter case, the {\it ecore} energy
was computed from the experimentally measured center of gravity,
central shower energy, and impact angle in the calorimeter using a
parameterized shower profile function obtained from electromagnetic
showers measured in the beam tests. Such an {\it ecore} energy
represented an estimate of the true energy of a photon impinging on
the PbSc unbiased by background contributions from other particles
produced in the same event and depositing energy in the neighborhood
of a given cluster. The use of {\it ecore} instead of the total
cluster energy for photon reconstruction, helped to reduce
considerably the effects of cluster overlaps in central \AuAu\
collisions.

For a photon pair originating from a $\pi^0$ decay the invariant mass
\begin{eqnarray}
\label{eq:inv_mass}
 m_{\gamma\gamma} & = \sqrt{ \left( P_{\gamma_1} + P_{\gamma_2} \right)^2 }  & = \sqrt{2 E_{1} \cdot E_{2} \cdot \left( 1 -\cos{\theta_{12}} \right)}.
\end{eqnarray}
is identical to the $\pi^0$ rest mass. However, due to the finite
energy and position resolution in the detection of the photon pair,
the actual reconstructed value is smeared around a mean value, which
can deviate from the nominal value. The reconstructed peak position is
also influenced by the high multiplicity in a heavy ion collision,
where overlapping clusters can shift the measured energy of each
photon.

With the invariant mass analysis, $\pi^0$'s cannot be identified
uniquely since all possible photon-photon combinations have to be
considered. This leads to a large combinatorial background, which
increases quadratically with the multiplicity.  The $\pi^0$ yield is
instead determined on a statistical basis, with the background
contribution established via a \emph{mixed event} technique as
described below.

\begin{figure}[ht]
 \includegraphics[width=0.8\linewidth]{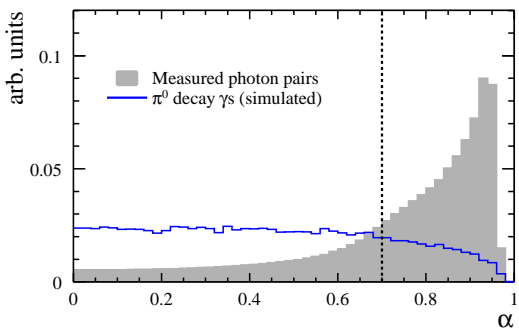}
   \caption{Asymmetry of photon pairs with $3\,\mbox{GeV}/c \le p_{T}
  < 5\,\mbox{GeV}/c$ within the acceptance of one PbGl sector,
  for simulated single $\pi^0$s and measured within minimum bias
  events. An asymmetry cut used during the analysis is also shown.
  (Due to the limited acceptance of the detector, the distribution of
  the energy asymmetry shows a slight decrease towards $\alpha = 1$).
   \label{fig:asym}
   }
\end{figure}

One possibility to reduce the combinatorial background is to make use
of the phase-space distribution of the photons in a $\pi^0$ decay.
 For the  $\pi^0 \rightarrow \gamma+\gamma$ decay, the two
photons have minimum opening angle
\begin{equation}
\tan \theta_{12}/2={m \over p},
\end{equation}
where $m$ is the \piz\ mass and $p$ its momentum, with $p\simeq p_T$
in the PHENIX central spectrometer. The angular distribution of the
$\gamma$ pair in the $\pi^0$ rest frame, $d\sigma/d\cos\theta^*$, is
constant, which leads to a flat distribution in the measured energy
asymmetry of the two photons from $\pi^0$ decay:
\begin{equation}
\alpha={{|E_1 - E_2|} \over {E_1 + E_2}}=\beta\, |\cos \theta^{*}| \qquad ,
\label{eq:asymmetry}
\end{equation}
where $\beta=p/E\sim 1$ is the velocity of the $\pi^0$. On the other
hand, high \pt\ combinatorial pairs are strongly peaked near $\alpha =
1$ because of the steeply falling spectrum of single photon
candidates. This is illustrated in Fig.~\ref{fig:asym}, where the
asymmetry distribution for photons from $\pi^0$s in a simulation is
compared to the measured asymmetry for photon candidate pairs in real
\AuAu\ collisions. In two independent analyses, asymmetry cuts of
$\alpha < 0.7$ and $\alpha < 0.8$ were employed, other values were
used as a cross-check and to verify the energy scale (see below).

\begin{figure}[hb]
\includegraphics[width=1.0\linewidth]{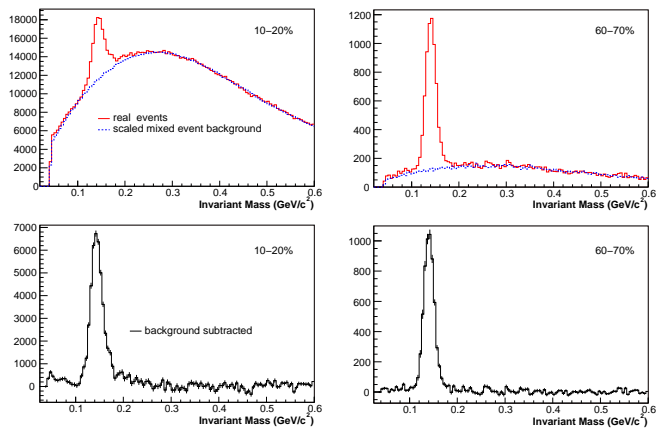}
 \caption{Invariant mass distributions of pairs of electromagnetic
   clusters passing photon selection cuts for pair transverse momenta
   satisfying $3.0<p_T<3.5$~GeV/$c$. Top panels: \Mgg\
   distributions in \AuAu\ events compared to a normalized mixed-event sample
   representing the combinatoric background. Bottom panels: The
   \Mgg\ distributions after subtraction of the combinatoric
   background. Left: 10-20\% centrality bin, Right: 60-70\% centrality
   bin.
   \label{fig:pizmass}
   \label{fig:mass_10_20}
   \label{fig:mass_60_70}
}
\end{figure}

Pairs of candidate photon clusters within the PbGl and the PbSc
calorimeter which satisfy the asymmetry cut are considered candidate
\piz's.  Fig.~\ref{fig:pizmass} shows example invariant mass
distributions for \piz\ candidates with $3.0 < p_T < 3.5$~GeV/$c$ in
\AuAu\ collisions for two different bins of collision centrality.
The background under the clear \piz\ mass peak in these figures is
due to combinatorial mixing of photons from two different decaying
\piz's or from pairs containing one or two non-photon clusters that
nonetheless pass the above-described cuts.

Such a combinatorial background can be determined by a so called
\emph{mixed event} technique. It is a widely used method to determine
the combinatorial background of combined particle properties, e.g. the
invariant mass of a photon pair. The basic idea is to compare the
result obtained by combining particles within one event to the result
for particle combinations from different events, which are \emph{a
priori} not correlated.

In the case of the $\pi^0$ invariant mass, the mixed event
distribution is determined by combining one photon candidate from the
current event with all photon candidates from previous events. The
number of previous events used for the pair combinations determines
the statistical error of the background and is limited basically by
computing resources. In this analysis a buffer of events is varied
from $\sim$ 3-10 previous events is used for the event mixing with
the current event.

In order to describe the combinatorial background correctly it is
essential that the events used for mixing have similar properties as
the real event and that they are not biased toward a certain reaction,
as e.g. events that are chosen because of a high-$p_{\rm T}$ photon
trigger. For this reason only minimum bias events are considered for
the determination of the background in both data sets and different
event classes for collision vertex, centrality and reaction plane are
employed.

It is self-evident that for the photons used in the event mixing the
same criteria are applied as for the pair combinations from one event,
such as PID cuts, cuts on bad modules, and the asymmetry cut. Other
properties valid \emph{a priori} for the real photon pairs, e.g. a minimum
distance that allows to distinguish them, have to be considered in
addition. In the analysis a minimum distance cut of a least 8\,cm is
required for each photon pair combination, within one event and for
mixed events, respectively.

For a given $\pt$ bin the mixed-event background is normalized to the
same-event invariant mass distribution outside the range of the \piz\
peak
by scaling the mixed-event background with a function $f(m_{\rm
inv})$.  This scaling function is determined by fitting the ratio of
the same-event and mixed-event invariant mass distribution for $\pt$
bins up to $3\,\mbox{GeV}/c$ with a linear function. This is needed
because at low $\pt$ correlations in the real-event background due to
overlapping clusters cannot be reproduced by the mixed-event
technique. For the $\pt$ bins above $3\,\mbox{GeV}/c$, a constant is
used. To cross-check the result, the linear and the constant scaling
function are also determined over the complete invariant mass region,
including the \piz\ peak, which is taken into account by an
additional Gaussian in the fit function (e.g. a linear plus a
Gaussian function).

The determination of the scaling function for large pair-$\pt$ is
limited by statistics in the real event sample and does not lead to
stable results. Instead a constant scaling factor is used if the ratio
of the invariant mass distribution shows bins with zero entries in the
fit region. The scaling factor is determined by integrating the real
and the mixed invariant mass distributions in the range with the peak
region excluded.

The scaled mixed-event background is subtracted from the same-event
distribution to produce a statistical measure of the true \piz\
yield. The result of such a subtraction procedure is shown in the
bottom plots of Fig.~\ref{fig:pizmass}.  The raw \piz\ yield is
obtained in each $p_T$ bin by integrating the subtracted invariant
mass distribution in a range around the peak mean ($m_{\pi^0}$) of
$\pm$ 3 times the Gaussian width ($\sigma_{\pi^0}$) of the $\pi^0$
peak. Values of the mean and $\sigma_{\pi^0}$, can be seen in Figure
\ref{fig:cmp_embed_real}.  Varying the size of the integration window
results in slightly different results, which contributes to the
overall systematic uncertainty of the measurement, discussed in the
\emph{Systematic Errors} section \ref{sec:sub_syserrors} of the next
chapter.

Residual differences between the mixed background and the foreground
are still apparent in some \pt\ bins especially below $\sim$ 2 GeV/c.
Cluster merging, cluster splitting (fluctuations in the 2-D $\phi-z$
energy profile cause multiple local maxima which are incorrectly
separated into distinct clusters), anti-neutron annihilation and even
second order residual physics correlations such as 3 and multi-body
decays, flow, HBT, etc., can all cause such differences. These
remaining differences are compensated by the shape of the scaling
function.  In addition, as a systematic check, the shape of the
remaining background after subtraction is also fit with various low
order polynomial functions and potential contributions to the peak
yield are considered in the determination of the total systematic
error from the peak extraction procedure.

The values of the peak width and mean are extracted in one initial
analysis of the invariant mass distribution in which a
$\pt$-dependent parametrization is determined for different
centralities. The use of predefined values for the position and
spread of the $\piz$ peak has the advantage that even in $\pt$
regions where no fit to the subtracted invariant mass distribution is
possible, the integration region is well defined just by
extrapolation from low $\pt$.

\subsection{\piz\ Spectrum Measurement}
\label{sec:pi0spectrum}

For the reaction-plane independent \piz\ spectrum measurement in a
given centrality class $cent$, the aforementioned analysis is applied
in $\Delta \pt = 0.5$~GeV/$c$ bins for $\pt > 1$~GeV/$c$.  We cease
attempting to extract \piz\ yields at high \pt\ when the number of
pairs within the selected (background-subtracted) \piz\ mass window
falls below 4 counts. We then correct the resulting raw \piz\ spectrum
for the geometric acceptance $a_{\Delta y}(p_T)$, the overall
detection efficiency $\varepsilon_{cent}(\pt)$, which accounts
for the cluster cut efficiency, the \piz\ mass cut efficiency, for
losses due to cluster overlaps in high multiplicity events, for cuts
on bad modules and for the calorimeter energy and position
resolution. In addition a correction for conversion losses ($c_{\rm
conv}$) in the material of the PHENIX central arms and for the
branching ratio of the two photon decay ($c_{\gamma\gamma}$) is applied:

\begin{widetext}
\begin{equation}
 \frac{1}{2\pi \pt} \frac{d^2N^{\pi^{0}}_{cent}}{d\pt dy} \equiv
\frac{1}{2\pi \pt N^{event}_{cent}} \times
\frac{1}{a_{\Delta y}(p_T) \varepsilon_{cent}(\pt) c_{\rm conv} c_{\gamma\gamma}} \times
\frac{N^{\pi^{0}}_{cent}(\Delta \pt)}{\Delta \pt \, \Delta y}.
\label{eq:invyld}
\end{equation}
\end{widetext}

\subsubsection{Acceptance and Detector Efficiency}

The geometric acceptance of the EMCal for the $\pi^0
\rightarrow \gamma\gamma$ decay is evaluated using a
Monte-Carlo program that generates $\pi^0$s in a rapidity
interval $\Delta y$ with the same vertex distribution and
rapidity distribution as observed in real events and
contains the complete geometry information of the EMCal.
The $\pi^0$ decay is calculated via JETSET routines that
are part of the PYTHIA event generator
\cite{Sjostrand:2000wi}. For each $\pi^0$ it is verified
that both decay photons hit the detector. The resulting
\pt\ distribution of accepted $\pi^0$s is divided by the
transverse momentum distribution of the generated $\pi^0$s
and provides the geometrical acceptance of the PbSc and
PbGl, respectively.


The detection efficiency is determined using GEANT to simulate the
complete response of the calorimeter to single \piz\ decays. The data
from each simulated \piz\ is, then, embedded into real \AuAu\ events
by adding the EMCal tower information of the simulated $\pi^0$ to the
tower information of the real event and recalculating the EMCal
clusters. The efficiency for detecting the embedded \piz\ is then
again determined by comparing the input \pt\ spectrum to the
reconstructed \pt\ spectrum of the embedded \pizs.  Using this
technique we determine ``efficiency'' corrections that account for
the energy resolution and position resolution of the calorimeter, as
well as for the losses due to overlapping clusters in a real event
environment. In addition, the embedding allows for a precise
determination of the effect of edge cuts and bad modules. Though
these effects can be in principle considered as acceptance
corrections, they depend not only on the geometry but also on the
energy deposition of an electromagnetic shower in the different
calorimeter towers.

In the embedding procedure the effects of photon conversions are also
included, as the GEANT simulation considers the material budget in
front of the EMCal and the information for decay-photon conversions
is retained.  The final conversion correction, which is factorized
from the rest of the efficiency for book-keeping purposes, is
evaluated by comparing the \piz\ yield with and without including
conversions in the simulation.  The final conversion correction,
constant with \pt\,  depends on the photon PID cuts and material in
front of each individual sector and ranges from 6-8\% in PbGl and
9-10\% in PbSc. Comparing this to the sheer probability of a \piz\
having at least one photon which converts, 21\% PbGl and 14\% PbSc,
we see that a large portion of these \piz\ are still reconstructable.

For the embedding, the input \piz\ spectrum is weighted to match a
functional form fit to the measured \piz\ spectrum so that the
correct folding of the \piz\ spectrum with the resolution is
obtained.  This procedure is iterated, with the fit of the \pt\
dependence of the input weights adjusted as the estimate of the
efficiency correction improves, until the procedure converges within
the nearly \pt-independent statistical error of the embedded sample,
approximately 3\%.

\begin{figure}[h]
\includegraphics[width=1.0\linewidth]{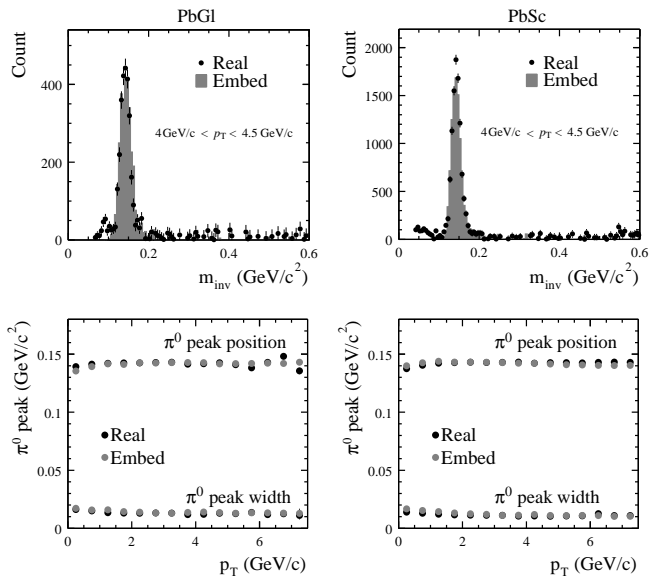}
 \caption{Comparison of the $\pi^0$ peak position in real events
  measured with the PbGl and the PbSc with the results obtained for
  embedded $\pi^0$s.  The $p_T$-dependence of the peak
  position and width is due to calorimeter energy resolution as
  discussed in the text.
 \label{fig:cmp_embed_real}
 }
\end{figure}

Fig.~\ref{fig:cmp_embed_real} compares the invariant-mass peak after
background subtraction in the real data and the invariant mass peak
of the embedded \piz\ for the two different detector types. The
measured \piz\ peak position is shifted from the nominal value of
approximately 134.98~MeV due to the finite energy resolution of the
detector in combination with the steeply falling spectrum and due to
the additional effect of overlapping clusters. As illustrated the
effects are well reproduced by the embedded \pizs.

\subsubsection{Trigger Efficiency}

\begin{figure}[h]
\includegraphics[width=0.9\linewidth]{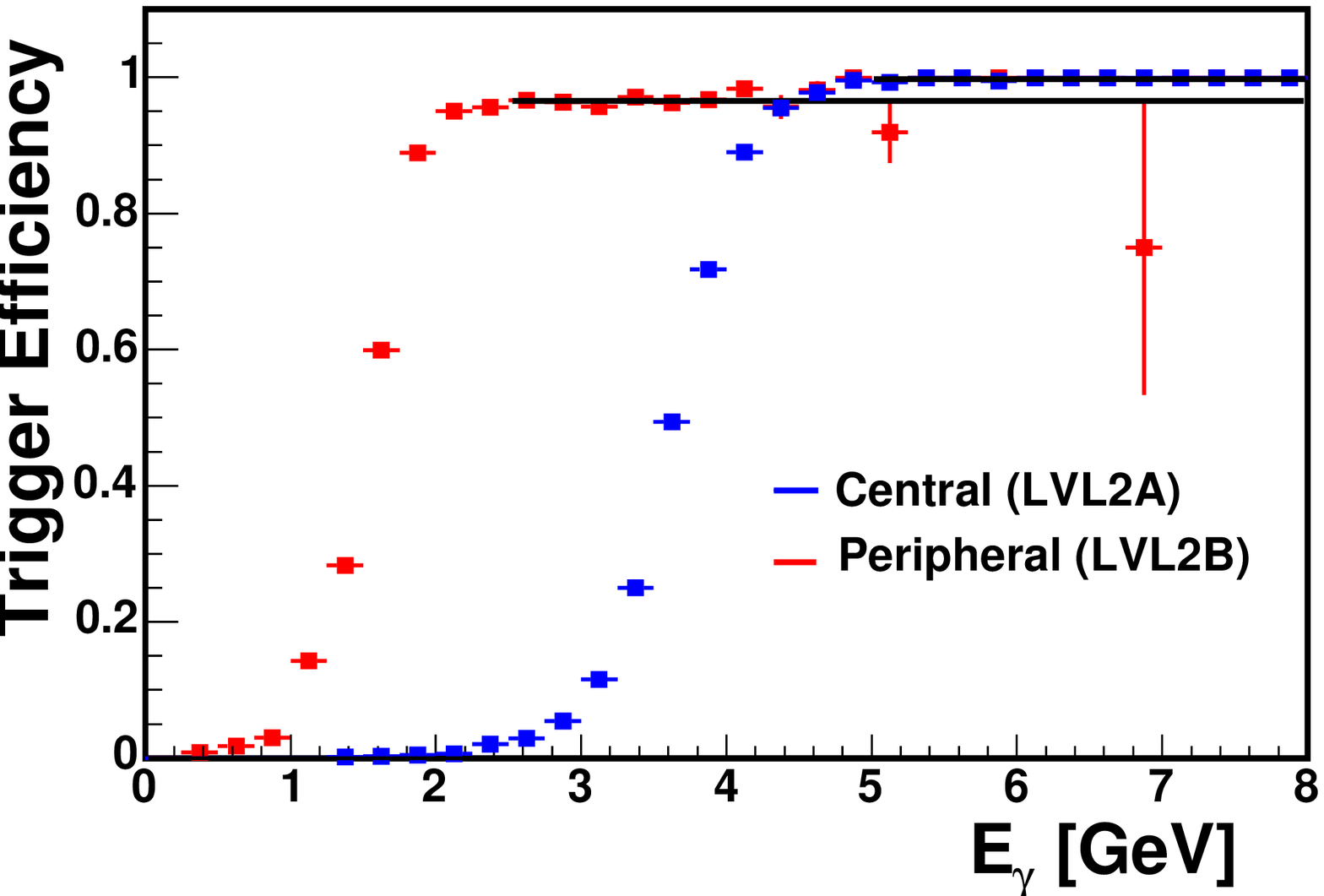}
\includegraphics[width=0.9\linewidth]{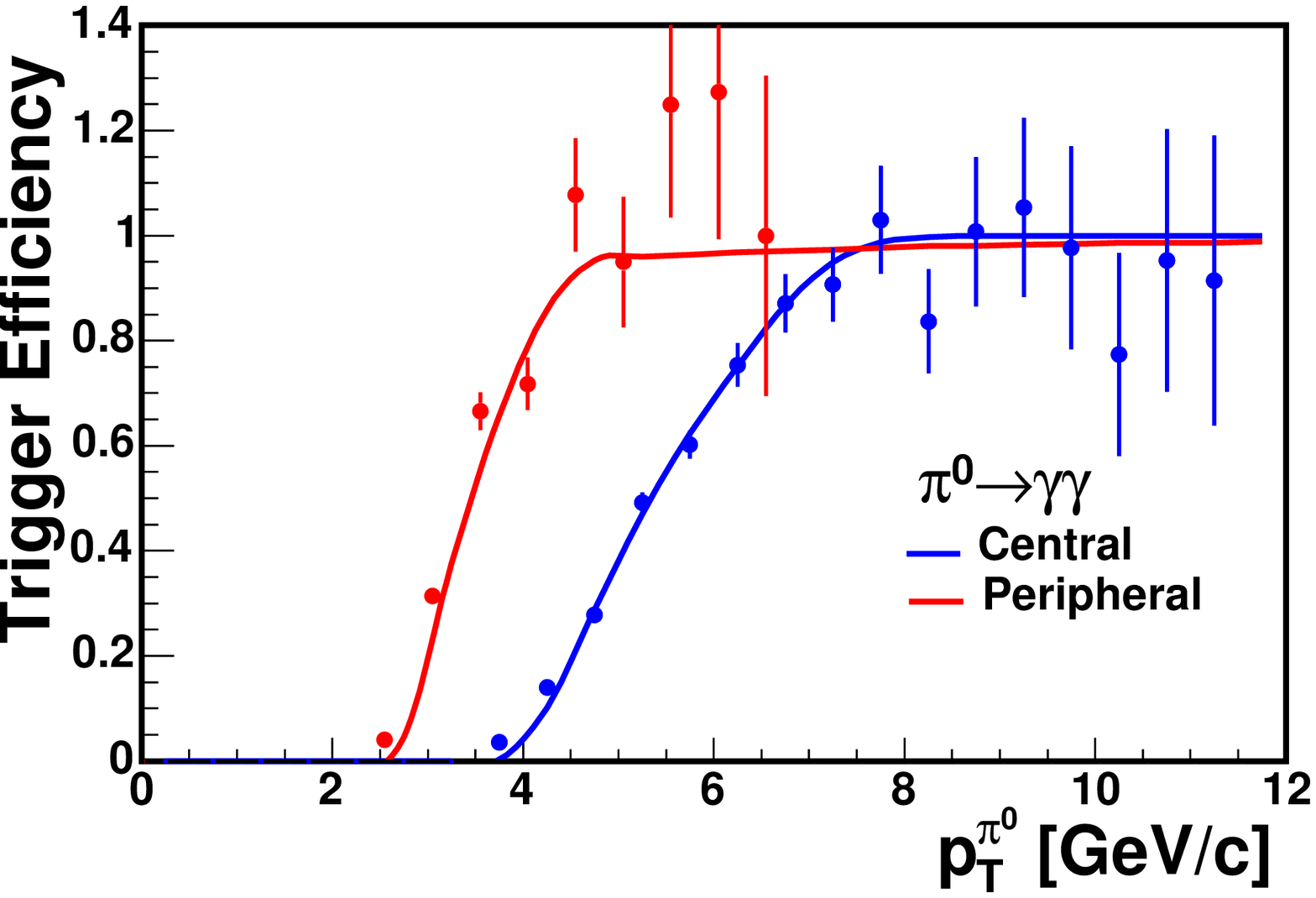}
 \caption{(color online) (a) Measured efficiency of single clusters
 of LVL2A (blue) and LVL2B (red) triggers as discussed in the text.  The black lines
 are constant value fits to the plateau efficiency, greater than 99.7
 (95\%) for LVL2A (LVL2B). (b) Efficiency for neutral pion detection of the
 triggers as a function of \piz\ \pt, calculated (solid curves) based upon
 the efficiencies in ~(a)~ and, as a
 cross check (data points), compared to ratio of per equivalent
minimum-bias event yields in the full trigger sample with
the same in the true minimum bias sample. Since the latter
is the ratio of two separate data samples, independent
statistical fluctuations, as well as $\sim$ 8\% systematic
effects in the yield extraction (discussed section
\ref{sec:sub_syserrors}) in either sample can cause this
measured ratio to be greater than 100\%.
 \label{fig:trig1eff}
}
\end{figure}

The efficiency of the Level-2 trigger is separately
evaluated by processing recorded minimum-bias events with
the Level-2 trigger and evaluating the efficiency for the
trigger to select events containing a high \pt\ cluster.
This analysis shows complete (100\%) efficiency for the
LVL2A trigger at momenta well-above the trigger threshold
of 3.5 GeV/$c$ (95\% above 1.5 GeV/$c$ for LVL2B) for
obtaining clusters that also pass all offline cluster cuts.
This is demonstrated in Figure \ref{fig:trig1eff}
\emph{(a)}. The ``plateau'' values are determined from
fitting the region above the turn-on also shown.

The related trigger efficiency of reconstructed \piz's is calculated
from fast MC based on these measured single cluster efficiencies. The
calculation is performed both by using a integrated Gaussian fit to
the single cluster efficiency and by directly using the finely binned
histogram and constant plateau fit.  Both methods give consistent
results.  The result for the latter method is shown in Figure
\ref{fig:trig1eff} \emph{(b)}, solid curves.  The calculation is
cross-checked, as demonstrated by the data points in Figure
~\ref{fig:trig1eff} \emph{(b)}  which show the ratio of the yield
from the two Level-2 trigger samples per \emph{equivalent} number of
minimum bias events to the same from the true minimum bias sample
itself.  We combine the yields obtained in the minimum bias event
sample and the LVL2A (LVL2B) trigger sample above a cut-off of
6.5~GeV/$c$ (3.5 ~GeV/c) where the trigger reaches efficiencies
greater than 0.4 such that the correction factor is not allowed to be
large.  A conservative error of 3\% is assigned to the efficiency
calculations, resulting in a total error of $\sim$ 3-5\%, based on
the three studies:  $1)$ comparisons of the data shown in Figure~
\ref{fig:trig1eff}, $2)$ comparisons of the two calculational
methods, and $3)$ a study of the yields in the subsample of minimum
bias events which also fired the triggers, similar to 1).

\subsubsection{Other Corrections}

The calculated corrections are applied to the raw $\pi^0$
yield as given by Eq.~(\ref{eq:invyld}).
Table~\ref{tab:yieldcorr_pi0} shows the correction in
central collisions for two different bins in transverse
momentum and for the PbGl and PbSc, respectively. As
discussed above the effect of the cut on bad modules is
included in the efficiency correction, due to its
dependence on the depth of the electromagnetic shower.

\begin{table}
\caption{Corrections in the PbGl and PbSc to the raw \piz\ yield in
central collisions (0-10\%) and with TOF and shower shape cut
applied. The main part of the efficiency loss in PbGl is due to the
effect of bad module and edge cuts which is approximately 40\% at
high \pt\ for the PbGl and 20\% for the PbSc, respectively}
\label{tab:yieldcorr_pi0}
\begin{ruledtabular}\begin{tabular}{lcccc}
\hline
   & \multicolumn{2}{c}{PbGl} &   \multicolumn{2}{c}{PbSc}\\
\multicolumn{1}{c}{\pt} &\,3.25~GeV/$c$ & \,8.5~GeV/$c$  &\,  3.25~GeV/$c$ & \,8.5~GeV/$c$  \\
\hline
$a_{\Delta y = 0.9}$ & 0.068      & 0.080  & 0.216     & 0.246    \\
$\varepsilon$        & 0.351     & 0.358   & 0.455     & 0.515      \\
$c_{\rm conv}$ &  0.93        &  0.93  & 0.90   & 0.90     \\
\cline{2-5}
$c_{\gamma\gamma}$ &          \multicolumn{4}{c}{0.98798}    \\
\end{tabular}\end{ruledtabular}
\end{table}

Following the usual PHENIX procedure of modifying the quoted yield
values for each finite sized \pt\ bin such that the measurement
corresponds to \pt\ value at the bin center instead of the average
\pt\ of the bin~\cite{Lafferty:1994cj} (thereby facilitating taking
ratios of spectra from different collision systems), a final
correction is applied to the {\it yield} of each data point. Using a
continuous function which is fit to the data points, values for the
invariant yields at the centers of the chosen \pt\ bins are scaled by
the ratio of the fit value at the fit's average \pt\ to the fit value
at the bin center.  This is an iterative procedure similar to the
final efficiency correction described above, with a smaller
convergence criteria of $<$ 0.1\% of the previous correction.

\subsubsection{Systematic Errors}
\label{sec:sub_syserrors}

 Each correction of the raw yield following
Eq.~(\ref{eq:invyld}) is afflicted with its own
uncertainty, but already the determination of the $\pi^0$
raw yield itself is sensitive to the method of extraction.
In particular it is sensitive to the choice of the fit
function for the background scaling and the extraction
window. In principle, both should be taken into account by
the detector efficiency, but in the efficiency calculation
no background subtraction is necessary. For this reason the
systematic error of the peak extraction method is
determined in two steps: first via the comparison of the
raw yield obtained with two different fits for the
background scaling, and second through the comparison of
the fully corrected spectra for different sizes of the
extraction window, for the real data as well as for the
efficiency calculation.

The systematic error introduced by the efficiency calculation is
estimated by comparing the fully corrected spectra for different PID
criteria as well as for different additional smearing. The smearing
(or energy resolution in the simulation) is changed in a way that a
clear disagreement between the measured $\pi^0$ peak width and the
peak width from the embedding is observed.

Apart from the uncertainty of the efficiency, the main contribution to
the systematic error is the determination of the absolute energy
scale. Based on the comparison of the $\pi^0$ peak positions in the
data to the expectation from simulation the energy scale can only be
determined or confirmed with limited accuracy, $\Delta(E)/E = 1.6\%$
in the PbSc and, because of the smaller acceptance, $\Delta(E)/E =
2\%$ in the PbGl.

The additional contributions to the systematic error that have not
been discussed in detail involve the uncertainty of the conversion
correction (2.9\%) and of the acceptance calculation (2.5\%) both due
to small uncertainties in detector material and alignment.
Table~\ref{tab:syserr_pi0} provides a final overview of the various
contributions to the total error of the $\pi^0$ measurement in the
PbSc and the PbGl, respectively.

\begin{figure}[h]
\includegraphics[width=0.7\linewidth]{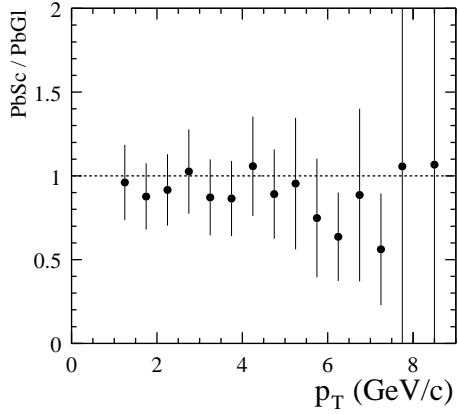}
 \caption{Comparison of fully corrected spectra for the PbGl and the
  PbSc for peripheral events.  Similar consistency is observed for
  all centralities.  The error bars represent the statistical and
  systematic uncertainties.} \label{fig:cmppi0_glsc}
\end{figure}
\begin{figure}[h]
\includegraphics[width=0.95\linewidth]{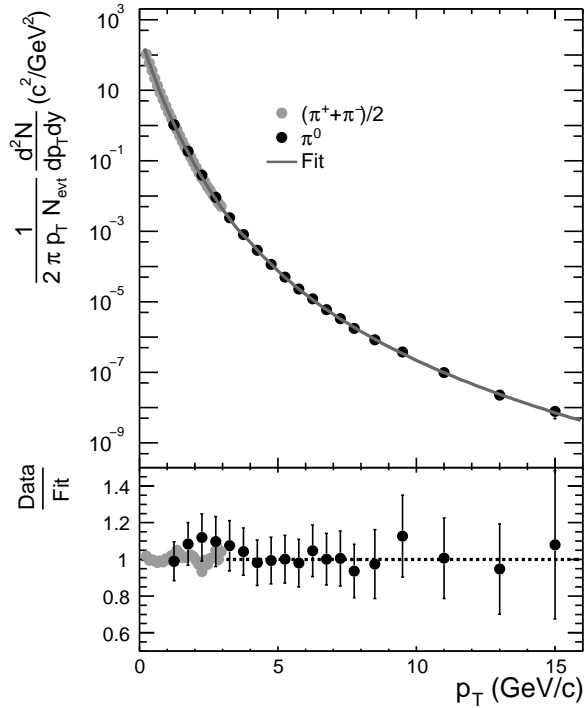}
 \caption{ Comparison of the combined $\pi^0$ result to the
  measurement of charged pions within the PHENIX experiment in minimum
  bias events. The fit considers the averaged result of the $\pi^+$
  and $\pi^-$ measurement \cite{Adler:2003au} below $\pt = 3\,\mbox{GeV}/c$
  and the $\pi^0$ data above.}
 \label{fig:cmp_charged}
\end{figure}

The most important cross check of the final result is the comparison
of the result for the two different detector types PbGl and PbSc,
which is shown for peripheral events in Fig.~\ref{fig:cmppi0_glsc}.
 A good agreement within the errors is seen and similar consistency
is found in all centralities.  Since they represent essentially
independent measurements, the two results are averaged and the total
error of the combined result is reduced using a standard weighted
least-squares method also described in \cite{Eidelman:2004wy}.  An
additional cross check of the final result based on iso-spin symmetry
is provided by the measurement of charged pions in the central arm
\cite{Adler:2003au}, this is shown for minimum-bias collisions in
Fig.~\ref{fig:cmp_charged}. The neutral pion measurement smoothly
extends the result for charged pions to larger transverse momenta.


\begin{table}
\caption{\label{tab:syserr_pi0}
Summary of the dominant sources of systematic errors on the
$\pi^0$ yields extracted independently with the PbGl and PbSc
electromagnetic calorimeters in central events for different \pt. For comparison the statistical uncertainty is also shown.}
\begin{ruledtabular}\begin{tabular}{lcccc}
    & \multicolumn{2}{c}{PbGl} &   \multicolumn{2}{c}{PbSc}\\
\multicolumn{1}{c}{\pt}(GeV/$c$) & 3.25 & 8.5  &  3.25 & 8.5 \\ \hline
Yield extraction &  8.7\%  &  6.\%  &  9.8\%    &   7.3\% \\
Efficiency &  11.4\%   &  11.4\%    &  11.4\%   & 11.4\% \\
Acceptance &  2.5\%    &  2.5\%     &  2.5\%    &  2.5\%  \\
Conversions & 2.9\%   & 2.9\%      &  2.9\%    & 2.9\%   \\
Level-2 data & -- & -- & -- & 3\% \\
Energy scale     &  13.8\% &   14.1\% &   10.5\% &   11.2\% \\
\hline
Total syst. & 20.5\% &   19.3\% &   18.8\% &   18.7\% \\
\hline
Statistical &  10.6\% &   50\% &   8.1\% &   26.6\% \\
\end{tabular}\end{ruledtabular}
\end{table}

\subsection{\RAApt ~Measurement}
\label{sec:raapt}

Using the invariant yields obtained from the
above-described analysis and the separately measured invariant
cross-section for \piz\ production in \pp\ collisions \cite{Adler:2003pb},
we calculate the nuclear modification factor, $R_{AA}$, according to
\begin{equation}
R_{AA}(p_T)\,=\,\frac{(1/N^{evt}_{AA})\,d^2N^{\pi^0}_{AA}/dp_T dy}{\langle T_{AA}\rangle \,\times\,
d^2\sigma^{\pi^0}_{pp}/dp_T dy},
\label{eq:R_AA}
\end{equation}
where $\langle T_{AA} \rangle$ is the average Glauber nuclear overlap function
for the centrality bin under consideration
\begin{equation}
\langle T_{AA}\rangle\equiv
\frac {\int T_{AA}(\mathbf{b})\, d\mathbf{b} }{\int (1- e^{-\sigma_{pp}^{inel}\, T_{AA}(\mathbf{b})})\, d\mathbf{b}} \quad ,
\label{eq:meanTAB}
\end{equation}
from which the corresponding average number of nucleon-nucleon collisions, 
$\langle N_{coll}\rangle=\sigma_{pp}^{inel} \langle T_{AA}\rangle$,
can be easily obtained.

\subsection{\RAAphi ~Measurement}
\label{sec:raaphi} The measurement of the raw \piz\ yield with
respect to the event reaction plane, $\dphi = \phi({\piz}) - \Psi$,
proceeds as described in Sec.~\ref{sec:pi0spectrum} for the \pt\
spectrum except that we measure the yields as a simultaneous function
of both \pt\ and \dphi. Because the beam-beam counters have $2\pi$
acceptance, PHENIX can measure the \piz\ yields with uniform
acceptance over $0 < \dphi < 2\pi$ even though the electromagnetic
calorimeters have only $1\pi$ nominal azimuthal acceptance. Since the
measurement of $\Psi$ is ambiguous {\it wrt} a $180^\circ$ rotation
of the reaction plane, and since we expect the \piz\ yields to be
symmetric {\it wrt} reflection around $\dphi = 0$, we measure the
\piz\ yields in 6 bins of $|\dphi|$ over the range $0 < |\dphi| <
\pi/2$. For each \pt\ bin we evaluate the ratio,
\begin{equation}
R(\dphi_i, \pt) = \frac{\Delta N(\dphi_i, \pt) }{\sum_{i = 1}^6{\Delta N(\dphi_i, \pt)}}
\label{eq:rdphi}
\end{equation}
where $ N(\dphi_i, \pt)$ is the measured number of \piz's in a given
$(\dphi,\pt)$ bin, $\dphi_i$ representing one orientation of $\dphi$.
Because the PHENIX central arm acceptance is effectively constant as
a function of \dphi\ and we do not expect any azimuthal-dependence of
our \piz\ efficiency corrections, $R({\dphi}_i, \pt)$ can be written
as:
\begin{equation}
R(\dphi_i, \pt) = R_{AA}(\dphi_i, \pt) / R_{AA}(\pt).
\label{eq:raaratio}
\end{equation}
Using the measured \RAApt\ values we can directly convert the
$R({\dphi}, \pt)$ to \RAAphipt\ without having to apply acceptance
and efficiency corrections to the reaction-plane dependent yields.
These corrections are already included in the \RAApt\ values as
described above.

\begin{figure}[tbh]
\includegraphics[width=0.95\linewidth]{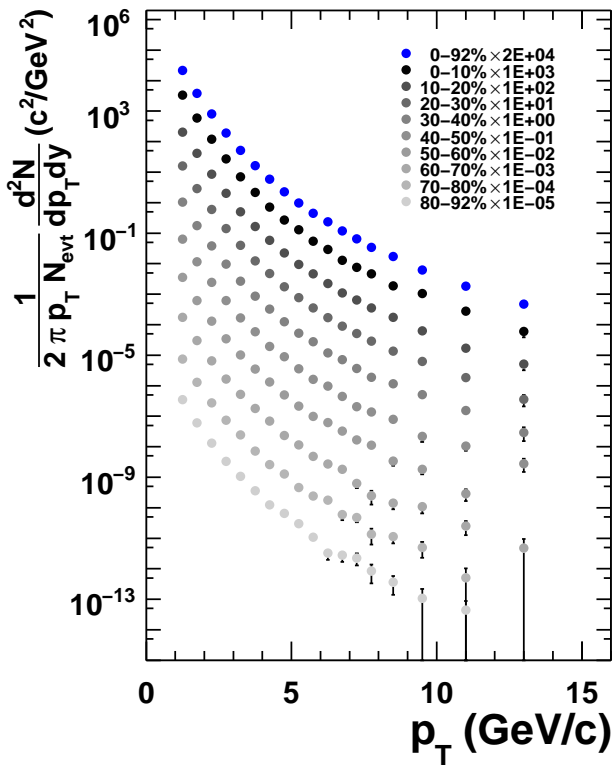}
 \caption{\label{fig:spectra}
(color online) Neutral pion invariant yields as a function of $p_{T}$
measured in minimum bias and 9 centrality classes in \AuAu\
collisions at \snn\ = 200~GeV. Spectra for different
centralities are scaled for clarity.  Errors are total
errors, full systematic and statistical added in
quadrature.}
\end{figure}

However, before applying this procedure we must first correct the
\Rphipt\ values for the finite resolution of the reaction plane
measurement. One goal of our measurement is to determine \RAAphipt\
without assuming any particular functional dependence on
\dphi. However, for purposes of correcting for reaction plane
resolution, we take advantage of the fact that the observed \piz\
yields and hence the nuclear modification vary with \dphi\ to first order as
\begin{equation}
R^{\rm raw}(\dphi, \pt) \approx R_0 \left (1 + 2\vt^{\rm raw}
\cos{(2\dphi)} \right). \label{eq:rdphidep}
\end{equation}
Where the superscript ${\rm raw}$ denotes the values not corrected
for the reaction plane resolution.  This resolution reduces \vt\ by
the factor $\sqrt{2\avgcostwodpsi}$ \cite{Poskanzer:1998yz}, which is
given by the independent measurement of $\Psi$ in the two BBC's shown
previously in Fig~\ref{fig:rp_reso}a).  For each \pt\ bin in a given
centrality class we fit the \Rphipt\ values to the functional form in
Eq.~(\ref{eq:rdphidep}) and then correct each measured \Rphipt\ value
according to
\begin{equation}
R^{\rm corr}(\dphi, \pt) = R^{\rm raw}(\dphi, \pt) \left( \frac{1 +
\vt^{\rm corr} \cos{(2 \dphi)}}{1 + \vt^{\rm raw} \cos{(2 \dphi)}}
\right) \label{eq:rphicorr}
\end{equation}
with $\vt^{\rm corr} = \vt^{\rm raw}/\sqrt{2\avgcostwodpsi}$. We
estimate the systematic error in the reaction plane resolution
correction by propagating the centrality dependent uncertainties in
\avgcostwodpsi from Fig~\ref{fig:rp_reso}b).  Of course, the
above-described correction only strictly applies if \RAAphi\ is
well-described by the functional form in Eq.~(\ref{eq:rdphidep}).
While we do observe some departure from this harmonic form in the
data, the differences are typically below 5\% so our correction will
not introduce a large error.

\begin{figure}[tbh]
\includegraphics[width=1.0\linewidth]{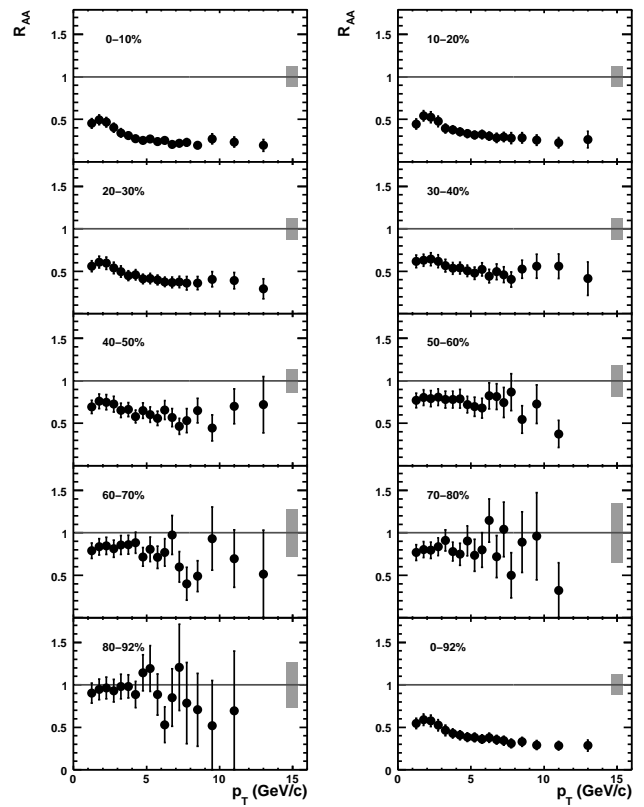}
\caption{Nuclear modification factor \raa\ for neutral pions as a
function of $p_{T}$ for different centralities. The shaded error band
around unity indicates the uncertainty in scaling factor $T_{AA}$ and
an overall scale uncertainty in the \pp\ reference.
 \label{fig:raa}
}
\end{figure}

\section{Results}
\label{sec:results}

\subsection{\piz\ Transverse Momentum Spectra and Nuclear Modification Factors} \label{sec:spectra}

The \piz\ invariant yields obtained using the procedure described in
Sec.~\ref{sec:pi0spectrum} are presented in Fig.~\ref{fig:spectra} as
a function of $p_T$ for the nine chosen centrality bins. With the
increased statistics included in this analysis, we have extended the
\pt\ range of the previous PHENIX measurement by at least 2~GeV/$c$
for all centrality bins. The \pt\ range of the central bin has been
extended from 10~GeV/$c$ to 14~GeV/$c$. Where the spectra overlap,
the results shown here are consistent with the previously published
results within systematic errors. The errors shown on the points in
Fig.~\ref{fig:spectra} include statistical errors and point-to-point
varying systematic errors. The appendix tabulates the $\pi^0$ spectra
plotted in Fig.~\ref{fig:spectra} (centralities: 0--10\%, 10--20\%,
..., 70--80\%, 80-92\% ) plus the combined spectra for centralities
0--20\%, 20--60\% and 60--92\%, which are used for comparison to
other neutral meson measurements~\cite{Adler:2006hu}. The spectra in
Fig.~\ref{fig:spectra} depart from the exponential-like shape above
3~GeV/$c$ which is consistent with the expectation that high-\pt\
hadron production is dominated by hard-scattering processes which
produce a 


\noindent power-law \pt\ spectrum for hadrons resulting from quark
and gluon fragmentation.

\begin{figure}[thb]
\includegraphics[width=1.0\linewidth]{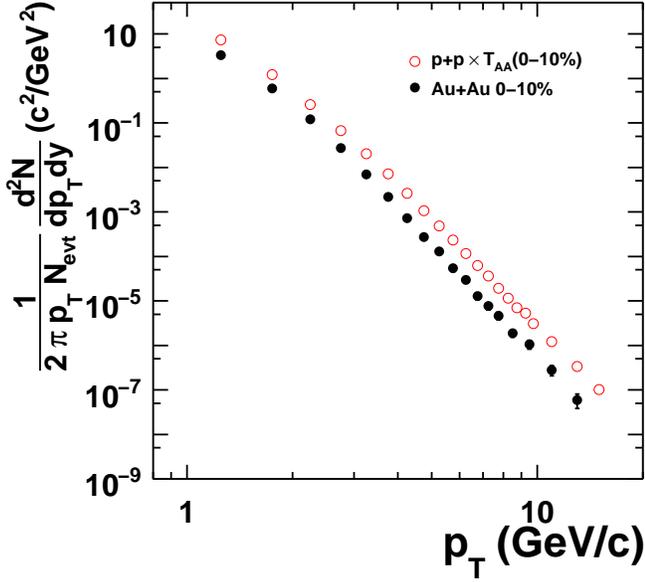}
    \caption[]{Log-log plot of central \AuAu\ and scaled \pp\
\piz\ $p_T$ distributions;
\label{fig:pizAuAuppll}
}
\end{figure}

In previous publications, we have established the suppression of
high-\pt\ \piz\ production in \AuAu\ collisions
\cite{Adcox:2001jp,Adler:2003qi}. This suppression cannot be easily
seen given the large range of invariant yield covered by
Fig.~\ref{fig:spectra}.

To evaluate the suppression of high-\pt\ \piz's, we show in
Fig.~\ref{fig:raa} the \pt\ dependence of the \piz\ nuclear
modification factor, \RAApt, for the nine individual bins of
collision centrality and for the full minimum-bias centrality range
0-92.2\%. We make use of the PHENIX Run3 \pp\ baseline \piz\ data.
\cite{Adler:2004ps}. The error bars on the data in Fig.~\ref{fig:raa}
include contributions from statistical errors in the \AuAu\ and \pp\
measurements and from the systematic errors that do not cancel
between the measurements. The separate band shown in each panel
indicates \pt-independent errors on the \raa\ measurement resulting
from uncertainties in estimating \TAA\ and systematic errors on the
normalization of the \AuAu\ and \pp\ measurements that do not cancel.
As in previously published papers, (\emph{e.g.}
\cite{Adcox:2002pe,Adler:2003au,Adams:2003kv}) a factor of $\sim 5$
high \pt\ \piz\ suppression in the most central \AuAu\ collisions,
$R_{AA}\approx$ 0.2, is observed, with the suppression approximately
\pt\ independent for \mbox{$\pt > 5$~GeV/$c$}.  The suppression at
high \pt\ decreases in more peripheral collisions such that the two
most peripheral bins have \raa\ values consistent with unity for $\pt
> 3$~GeV/$c$.

\subsection{Suppression via Spectrum Shift}

The suppression of high $p_T$ particles as shown above was determined
by comparison of the semi-inclusive measured yields as a function of
centrality in \AuAu\ collisions at $\sqrt{s_{NN}}=200$ GeV to the
$\langle T_{AA}\rangle$ scaled $p_T$ spectrum from \pp\ collisions
\cite{ppg044}. A direct comparison of the 0--10\% centrality \AuAu\
spectrum to the scaled \pp\ spectrum is shown in
Fig.~\ref{fig:pizAuAuppll} as a log-log plot to emphasize the pure
power law dependence of the data for $p_T > 3$ GeV/$c$. The
suppression is commonly expressed by taking \raa\ the ratio of the
point-like scaled semi-inclusive yield to the reference distribution
(Eq.~\ref{eq:R_AA}).

As illustrated in Fig.~\ref{fig:pizAuAuppll}, instead of viewing the
suppression in the nuclear modification factor as ``vertical''
reduction of the Au+Au yields, it can equally well be taken as a
``horizontal'' shift in the $\langle T_{AA}\rangle$ scaled \AuAu\
spectrum, such that
\begin{widetext}
\begin{equation}
{{(1/N^{evt}_{AA})d^2N_{AA}(p_T)/dp_T dy}\over {\langle
T_{AA}\rangle}}= \frac{d^2\sigma_{pp}(p_T^{'}=p_T+S(p_T))}{dp_T^{'}
dy} \times (1+dS(p_T)/dp_T) \label{eq:SpT} \qquad
\end{equation}
\end{widetext}
where the last term in parenthesis is the Jacobian, $d p_T^{'}/dp_T$.


Furthermore, owing to the pure power law of the \pp\ reference
spectrum, $Ed^3\sigma/dp^3\propto p_T^{-n}$ with $n = -8.10 \pm 0.05$
above $p_{T}\approx$ 4 GeV/$c$, the relative shift of the
spectra---assumed to be the result of energy loss for the \AuAu\
spectrum---is easily related to the equivalent ratio, $R_{AA}(p_T)$:
\begin{eqnarray}
\label{eq:RAA-SpT}
R_{AA}(p_T)&=&\frac{(p_T + S(p_T))^{-n+1}}{{p_T}^{-n+1}}(1+{dS(p_T)/dp_T}) \\ 
&=&(1+S(p_T)/p_T)^{-n+1} (1+{dS(p_T)/dp_T}) \qquad \nonumber
\end{eqnarray}
where the exponent is $n-1$ because the relevant shift is in the
$d\sigma/dp_T$ spectrum rather than in $d\sigma/p_T dp_T$. The fact
that the \AuAu\ and reference \pp\ $p_T$ spectra are parallel in
Fig.~\ref{fig:pizAuAuppll} provides a graphical illustration that the
fractional $p_T$ shift in the spectrum, $S(p_T)/p_T=S_0$, is a
constant for all $p_T > 3$ GeV/$c$, which also results in a constant
ratio of the spectra, $R_{AA}(p_T)$. For the constant fractional
shift, the Jacobean is simply $dS(p_T)/dp_T=S_0$ and
Eq.~(\ref{eq:RAA-SpT}) becomes:
\begin{equation}
R_{AA}(p_T)=(1+S_0)^{-n+2}, \qquad \label{eq:RAA-S0}
\end{equation}
\begin{equation} R_{AA}(p_T)^{1/(n-2)}={1 \over {1+S_0}} \qquad .
\label{eq:RAA-S0-2}
\end{equation}

\begin{figure}[h]
\includegraphics[width=1.0\linewidth]{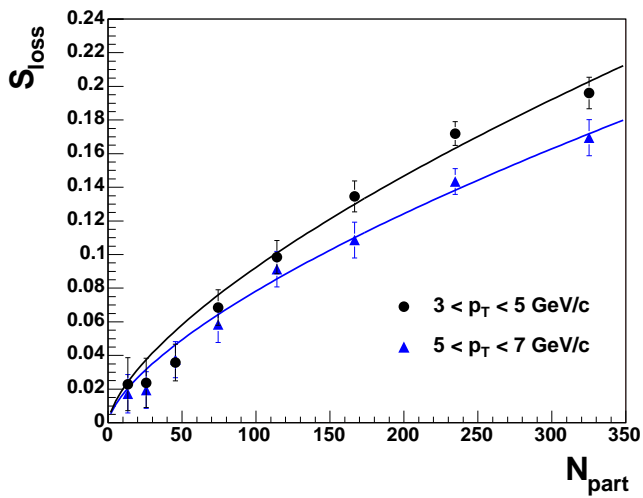}
  \caption[]{(color online) Fractional energy loss $S_{\rm loss}$ obtained from Eq.~(\ref{eq:Sloss})
versus centrality given by $N_{part}$.
The lines are fits of the form $\propto N_{\rm part}^{2/3}$ for each $p_T$ range.
\label{fig:sloss}}
\end{figure}

The effective fractional energy loss, $S_{\rm loss}$, is related to
the fractional shift in the measured spectrum, $S_0$. The hadrons
that would have been produced in the reference \pp\ spectrum at
transverse momentum $p_T+S(p_T)=(1+S_0)p_T$, were detected with
transverse momentum, $p_T$, implying a fractional energy loss:
\begin{equation}
 S_{\rm loss}=1-1/(1+S_0)= 1-R_{AA}(p_T)^{1/(n-2)}\qquad .
 \label{eq:Sloss}
 \end{equation}
The fractional energy loss $S_{\rm loss}$ as a function of centrality
expressed as $N_{\rm part}$ is shown in Fig.~\ref{fig:sloss} for two
different $p_T$ ranges, $3 <p_T < 5$ GeV/$c$ and $5 < p_T < 7$
GeV/$c$. There appears to be a small decrease of $S_{\rm loss}$ with
increasing $p_T$, but the main observation from Fig.~\ref{fig:sloss}
is that $S_{\rm loss}$ increases approximately like $N_{\rm part}^{2/3}$, as
suggested by GLV~\cite{Vitev:2006uc} and PQM \cite{Dainese:2005kb}.

 It is important to realize that the effective fractional energy loss,
 $S_{\rm loss}$ estimated from the shift in the $p_T$ spectrum is
 actually less than the real average energy loss at a given $p_T$,
 i.e. the observed particles have $p_T$ closer to the original value,
 than to the average. The effect is similar to that of ``trigger bias"
 \cite{Jacob:1978mj} where, due to the steeply falling spectrum, the
 $\langle z\rangle$ of detected single inclusive particles is much
 larger than the $\langle z\rangle$ of jet fragmentation, where
 $z=\vec{p}_{\pi^0}\cdot \vec{p}_{jet}/p_{jet}^2$. Similarly for a
 given observed $p_T$, the events at larger $p_T^{'}$ with larger
 energy loss are lost under the events with smaller $p_T^{'}$ with
 smaller energy loss.



It should be noted that fluctuations due to the variation of the
path length and densities traversed by different partons
also contribute to the difference between $\Sloss^{\rm true}$ and
$\Sloss^{\rm obs}$. However, as long as the dependences of the
induced energy loss on path length and parton energy approximately
factorize, these fluctuations will also produce a \pt-independent
reduction in $\Sloss^{\rm obs}$ compared to $\Sloss^{\rm true}$.


\subsection{Angle Dependence of High \pt\ Suppression} \label{sec:phi_sloss}

     In order to try to separate the effects of the density of the medium and
path length traversed, we study the dependence of the $\pi^0$ yield
\wrt\ the reaction plane. For a given centrality, variation of
\dphi\ gives a variation of the path-length traversed for fixed
initial conditions, while varying the centrality allows to determine
the effect of varying the initial conditions.

\begin{figure}[thb]
\includegraphics[width=1.0\linewidth]{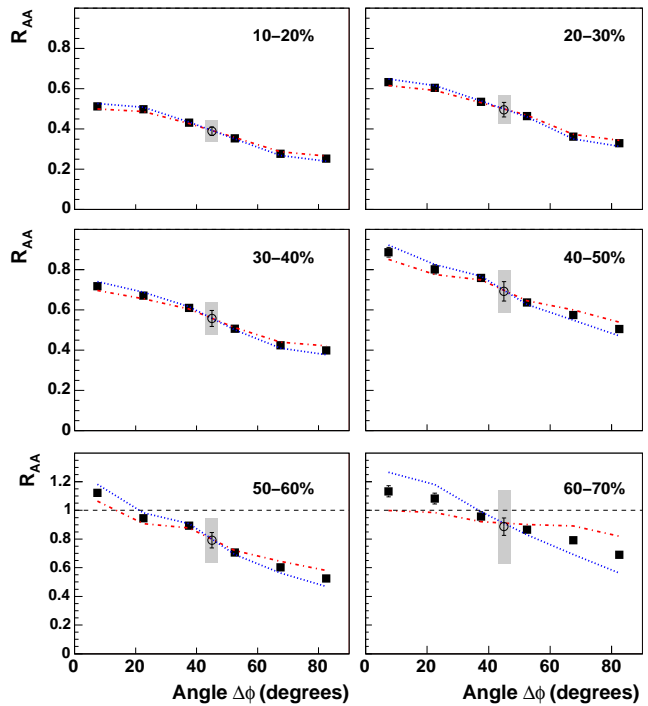}
  \caption{(color online) $R_{AA}$ vs. \dphi\ for $\pi^0$ yields integrated over $3 <
 p_T < 5$~GeV/$c$.  Most statistical errors are smaller than the size
 of the points.  The lines following the data points show the
 bin-to-bin errors resulting from the uncertainty in the reaction
 plane resolution correction (Fig.~\ref{fig:rp_reso}) and from
 bin-to-bin uncertainties in the $R_{AA}$ values.  The shaded band
 indicates the overall $R_{AA}$ uncertainty.
    \label{fig:raa_phi_int3}
    }
\end{figure}

\begin{figure}[thb]
\includegraphics[width=1.0\linewidth]{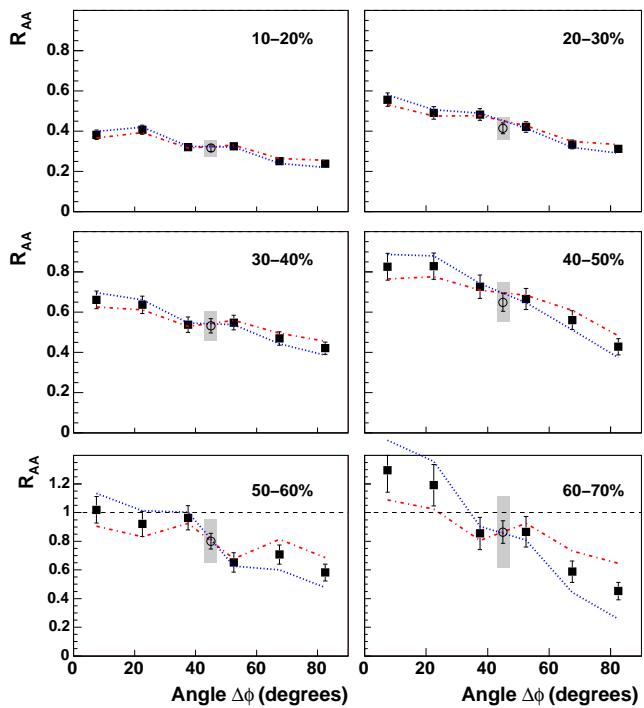}
  \caption{(color online) $R_{AA}$ vs. \dphi\ for $\pi^0$ yields integrated over $5 < p_T < 8$~GeV/$c$.
  The error lines and band are the same as in Fig.~\ref{fig:raa_phi_int3}.
    \label{fig:raa_phi_int5}
  }
\end{figure}

Figs.~\ref{fig:raa_phi_int3} and \ref{fig:raa_phi_int5} show the
nuclear modification factor $R_{AA}$ as a function of \dphi\
integrated over $ p_T \in \left] 3 \mbox{GeV}/c,5 \mbox{GeV}/c
\right]$ and $ p_T \in \left] 5 \mbox{GeV}/c,8 \mbox{GeV}/c \right]$,
respectively.  For all centralities (eccentricities) considered,
there is almost a factor of two more suppression out-of-plane (\dphi\
= $\pi$/2) than in-plane (\dphi\ = 0), something that is immediately
apparent in viewing the data in this fashion, explicitly displaying
information only implicit in \raa\, $v_2$, or the combination
thereof. Strikingly, in contradiction to the data the variation in
\raa\ \wrt\ the reaction plane expected by parton energy loss
models~\cite{Drees:2003zh,Dainese:2004te} should be much smaller for
the more peripheral bins than for the central bins.  As a result, the
suppression vanishes (and perhaps an enhancement is observed) for
smaller \dphi\ in the peripheral bins, corresponding to small
path-length traversed in the medium.
Although collective elliptic flow effects, usually not included in
those models, are known to boost in-plane (compared to out-of-plane)
particle production~\cite{Ackermann:2000tr,Adler:2003kt}, it is
unclear how such collective effects can still play such an important
role at the high $p_T$ bins considered.  This may point to the
possible need for a formation time before suppression can
occur~\cite{Pantuev:2005jt} and which could also explain why attempts
to describe the azimuthal asymmetry $v_2$ solely in terms of purely
geometrical energy loss have failed. 

\begin{figure}[h]
\includegraphics[width=1.0\linewidth]{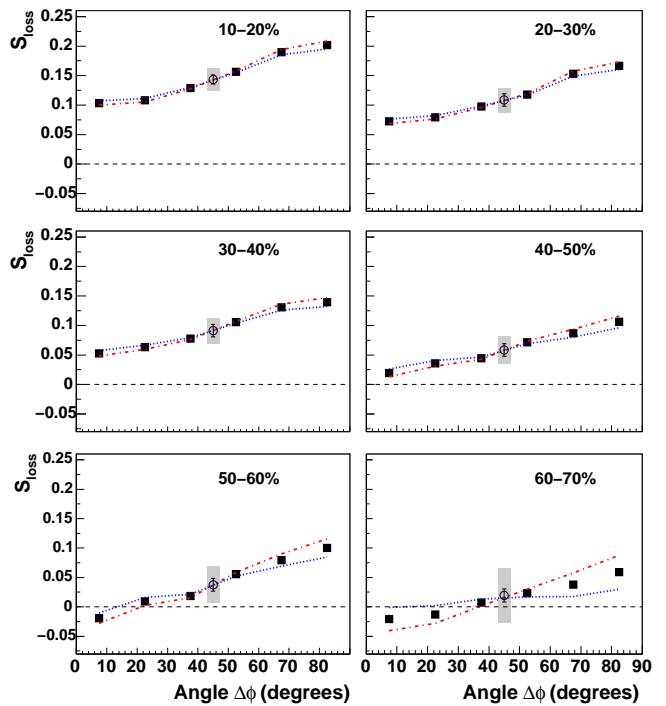}
 \caption{(color online) $S_{\rm loss}$ vs. \dphi\ for $\pi^0$ yields integrated over $3 < p_T < 5$~GeV/$c$.
 The statistical errors are smaller than the size of the points.
The lines following the data points show the
 bin-to-bin errors resulting from the uncertainty in the reaction
 plane resolution correction (Fig.~\ref{fig:rp_reso}) and from
 bin-to-bin uncertainties in the $S_{\rm loss}$ values.  The shaded band
 indicates the overall $S_{\rm loss}$ uncertainty.
   \label{fig:sloss_phi_int3}}

\end{figure}

\begin{figure}[h]
\includegraphics[width=1.0\linewidth]{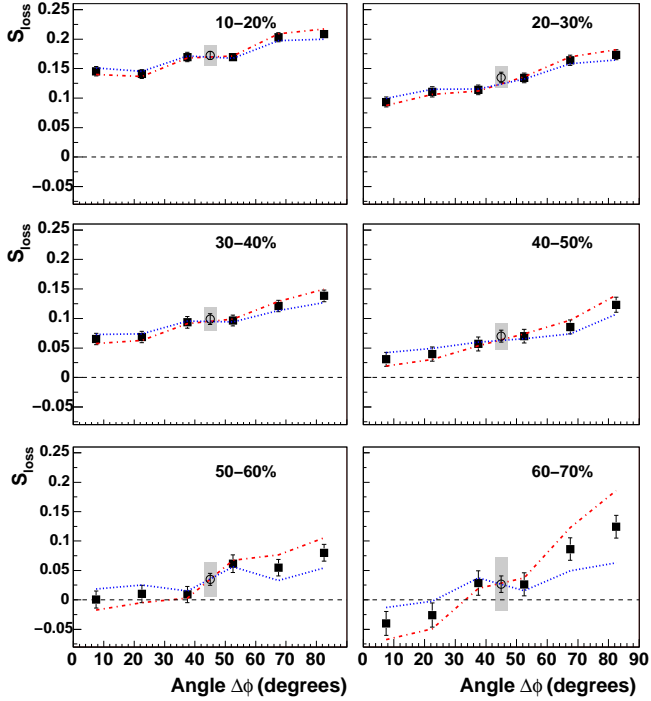}
  \caption{(color online) $S_{\rm loss}$ vs. \dphi\ for $\pi^0$ yields integrated over $5 < p_T < 8$~GeV/$c$.
  The error lines and band are the same as in Fig.~\ref{fig:sloss_phi_int3}.
\label{fig:sloss_phi_int5}
  }
\end{figure}

Figs.~\ref{fig:sloss_phi_int3} and \ref{fig:sloss_phi_int5} give the
angular dependence  in terms of the fractional energy loss
$S_{\rm loss}$, and provide essentially the same information as shown in
the plots of \RAAphi\ in Figs.~\ref{fig:raa_phi_int3} and
\ref{fig:raa_phi_int5}. Once again we see a large variation in energy
loss as a function of angle. 
All the measurements of $R_{AA}$ or equivalently $S_{\rm loss}$
vs. reaction plane and centrality, provide new constraints to models
of jet quenching.  To better understand the implications of the
results shown in these figures, we will attempt in the next Section to
find a common geometric description of the angle and centrality
dependences in terms of an estimated path length of the parton in the
medium.

\subsection{Path-length Dependence of Energy Loss}

To analyze the path length dependence of parton energy loss using the
data presented here we will use different methods for estimating the
path lengths of partons in the medium as a function of centrality and
\dphi. The ``standard'' approach would be to evaluate a
length-weighted integral of the color-charge density in the medium
along the parton path. We will adopt such an approach, described by
the parameter $L_{xy}$ defined below, but we will also consider two
other simplified approaches that may help indicate which physics is
most relevant in determining the observed suppression. We first
consider, simply, $L_{\varepsilon}$, the distance from the edge to
the center of the elliptical overlap zone of the \AuAu\ collision to
represent the average path length of a parton in the medium. Then we
try to weight the path length (or length-squared) traversed by a
parton from the center of the ellipse by the participant density in
the transverse plane: $\rho L(\Delta \phi)$ ($\rho L^2 (\Delta
\phi)$). Finally, we do the same path length weighting for partons
produced across the overlap ellipse, with hard-scattering production
points weighted by $T_{AA}(x,y)$: $\rho L_{xy}$ ($\rho L^2_{xy}$). It
is obvious that such a $\Delta \phi$ dependent analysis is not
possible from just a simple combination of \raa\ and $v_2$.

In detail, the three approaches considered here are as follows:
\begin{description}
\item (1)  The simplest picture for the angular dependence of the energy loss in
non-central collisions is that it is due to the asymmetric shape of
the overlap region of the colliding nuclei. Taking this idea to its extreme,
only the simplest length scale, the length of the overlap region
in a particular direction, matters.  To evaluate this length,
 we first estimated the RMS radius and eccentricity of an
ellipse approximating the shape of the overlap region from the
transverse distribution of the participant density calculated using
standard Glauber techniques. We then estimated the path length, \Leps,
of partons emitted at a given angle \dphi\ by evaluating the distance
from the center of the ellipse to the edge.  We thus ignore such
considerations as event by event eccentricity fluctuations and only consider
the average.
\item (2) Although the participant density is used to evaluate the dimensions of
the ellipse, the above analysis ignores the dependence of participant
density on position in the transverse plane. Thus as a natural extension
of the simple length scale in 1), for another analysis of
the dependence of energy loss on \dphi, we assume that the
color-charge density in the medium is proportional to participant
density and evaluate \rhoL, the integral of this density along the
path length of the particle. This quantity is proportional to the
opacity of the medium ($n=L/\lambda$) divided by some undetermined
cross-section. While the integral in principle extends to infinity the
participant density naturally cuts off the integral outside the
collision zone.
\begin{equation}
\rhoL = \int_0^{\infty} dr \; \rho_{\rm part} (r, \dphi).
\label{eq:rhoL}
\end{equation}
To account for the possible role of LPM coherence in the energy loss
process, we evaluate a similar quantity including an extra factor of
$r$ in the integrand.
\begin{equation}
\rhoLsq = \int_0^{\infty} dr \; r \rho_{\rm part} (r, \dphi).
\label{eq:rhoL2}
\end{equation}
We note that a Bjorken $1/\tau$ expansion of the medium would
approximately cancel one power of $r$ in the above expressions. Then,
\rhoL, might represent LPM energy loss in the presence of 1-D
expansion. In the above integrals we assume all jets originate at the
center of the collision region similar to our assumption for \Leps.

\item (3) A final refinement on our geometrical calculation evaluates
integrals like those in Eq.~(\ref{eq:rhoL}),(\ref{eq:rhoL2}) for jet
production points distributed over the collision region to better
account for geometric fluctuations.  We are using a Monte-Carlo
algorithm to sample jet production points $(x_0, y_0)$ according to
\TAA\ weighting and \dphi\ angles from a uniform distribution. For
each jet, we evaluate the integral of the color-charge density
($\propto$ participant density) along the path of the parton out of
the medium,
\begin{equation}
\rhoLxy = \int_0^{\infty} dl \; \rho_{\rm part} (x_0 + l\cos{\dphi},
y_0 + l\sin{\dphi} ).
\label{eq:rhoLxy}
\end{equation}
or
\begin{equation}
\rhoLsqxy = \int_0^{\infty} dl \; l \rho_{\rm part} (x_0 + l\cos{\dphi},
y_0 + l\sin{\dphi} ).
\label{eq:rhoLsqxy}
\end{equation}
The above Monte-Carlo sampling yields a distribution of \rhoLxy
(\rhoLsqxy) values for each centrality. The larger values of \rhoLxy
(\rhoLsqxy) correspond to larger energy loss which means these jets
will have smaller contribution to the observed yield. To take this
into account, a weighting factor is applied when evaluating $\langle
\rhoLxy \rangle$. We assume that the energy loss can be represented by
our empirical energy loss, \Sloss\ which we take to be proportional to
\rhoLxy (\rhoLsqxy) but with an undetermined multiplicative constant,
$\kappa$. We determine this constant in each centrality bin by
relating \Sloss\ to \RAA\ using Eq.~(\ref{eq:Sloss}) and then evaluating
the survival probability of each jet through
\begin{equation}
P_{\rm surv}(\rhoLxy) = 1 - \left(\kappa \, \rhoLxy\right)^{(n-2)}
\end{equation}
and requiring that the resulting suppression summed over all sampled
jets agrees with the measured \dphi-integrated \RAA\ for that
centrality bin. This determines the constant $\kappa(\Npart)$ and allows us to
evaluate a survival probability weighted average for $\rhoLxy$.
\end{description}

We now evaluate how well the three above-described treatments of the
geometry of the parton propagation in the medium perform in providing
a consistent description of the \dphi\ {\em and} centrality dependence
of \piz\ suppression.


\begin{figure}[tbh]
\includegraphics[width=0.48\linewidth]{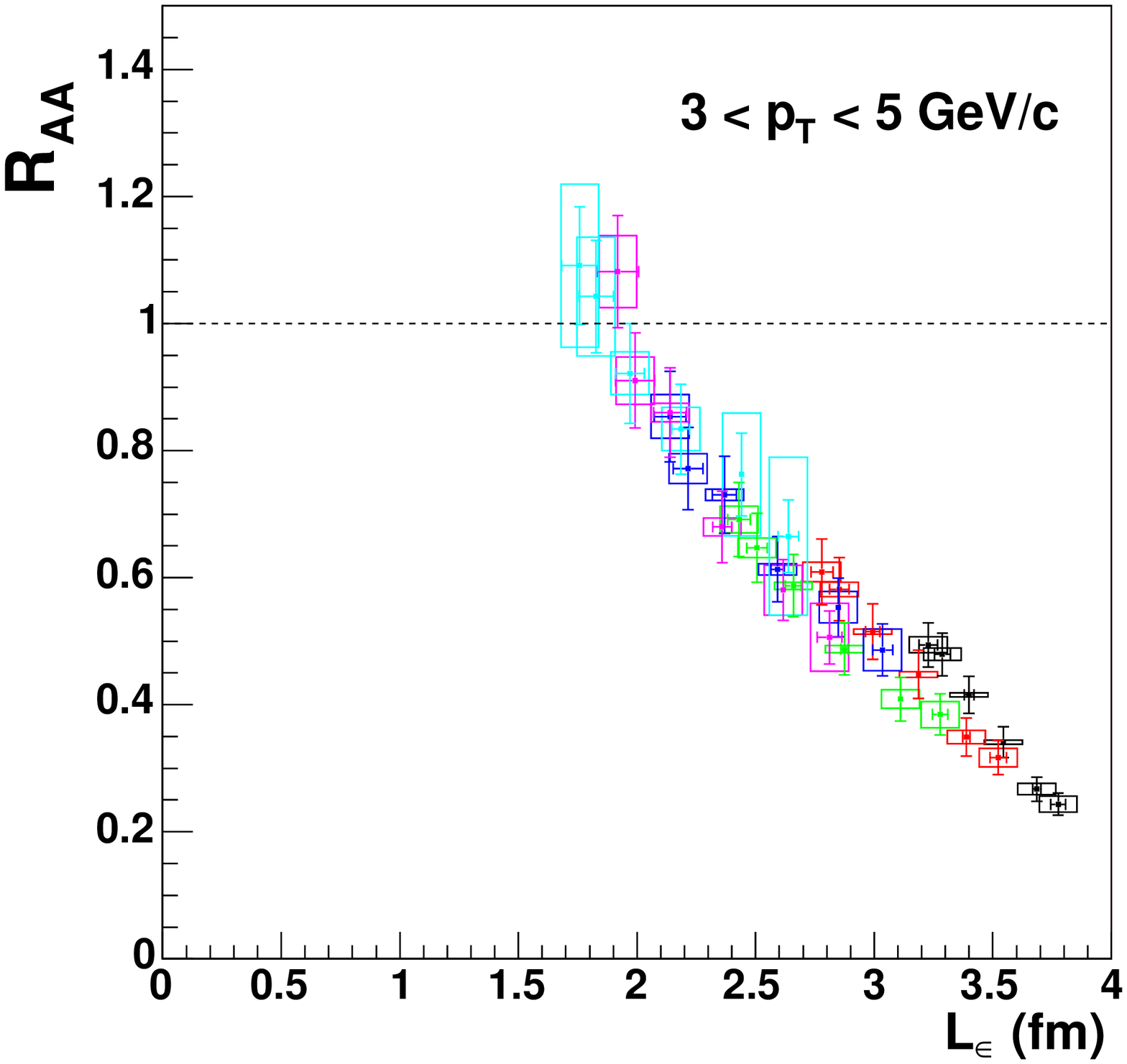}
\includegraphics[width=0.49\linewidth]{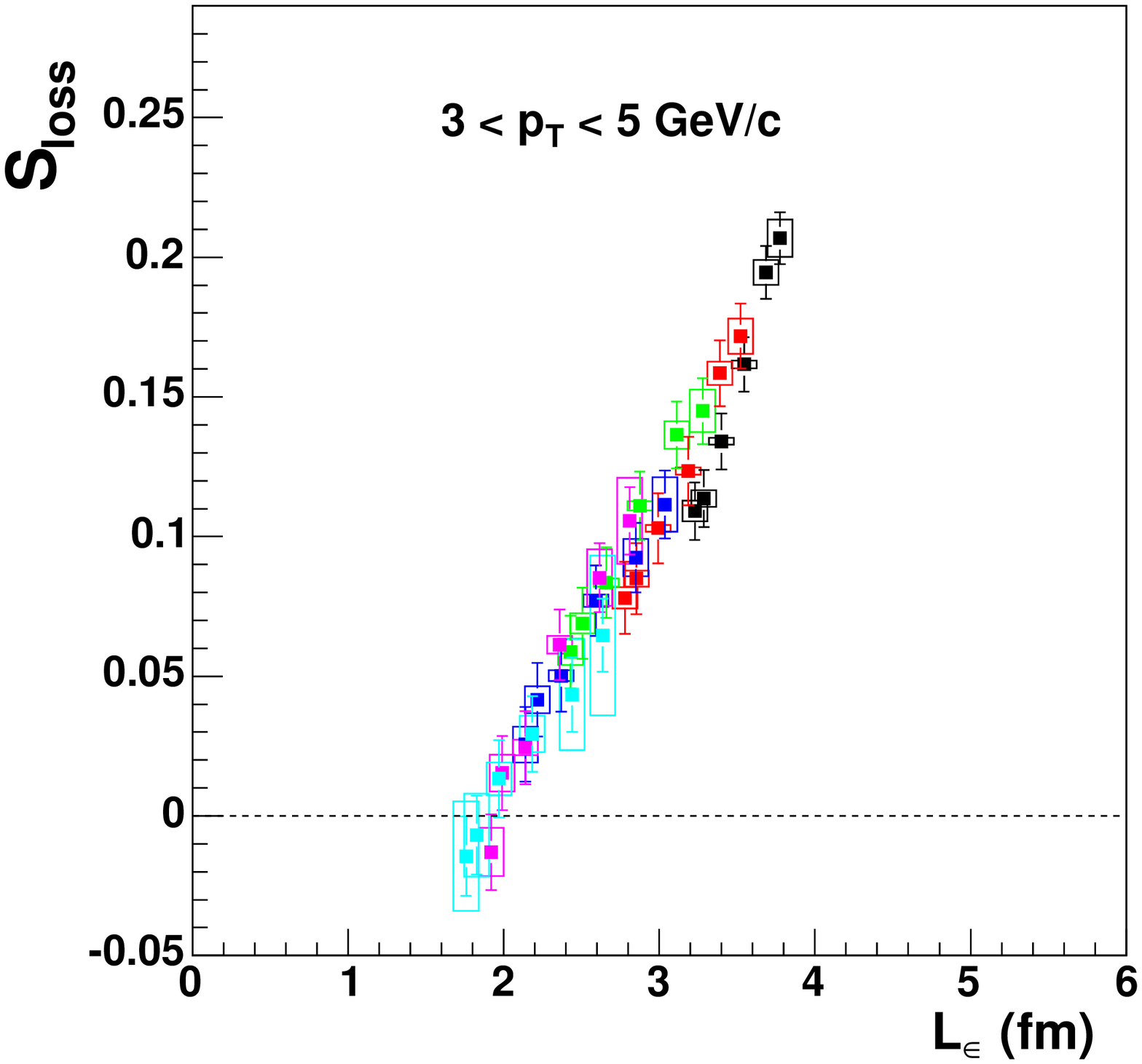}
\includegraphics[width=0.48\linewidth]{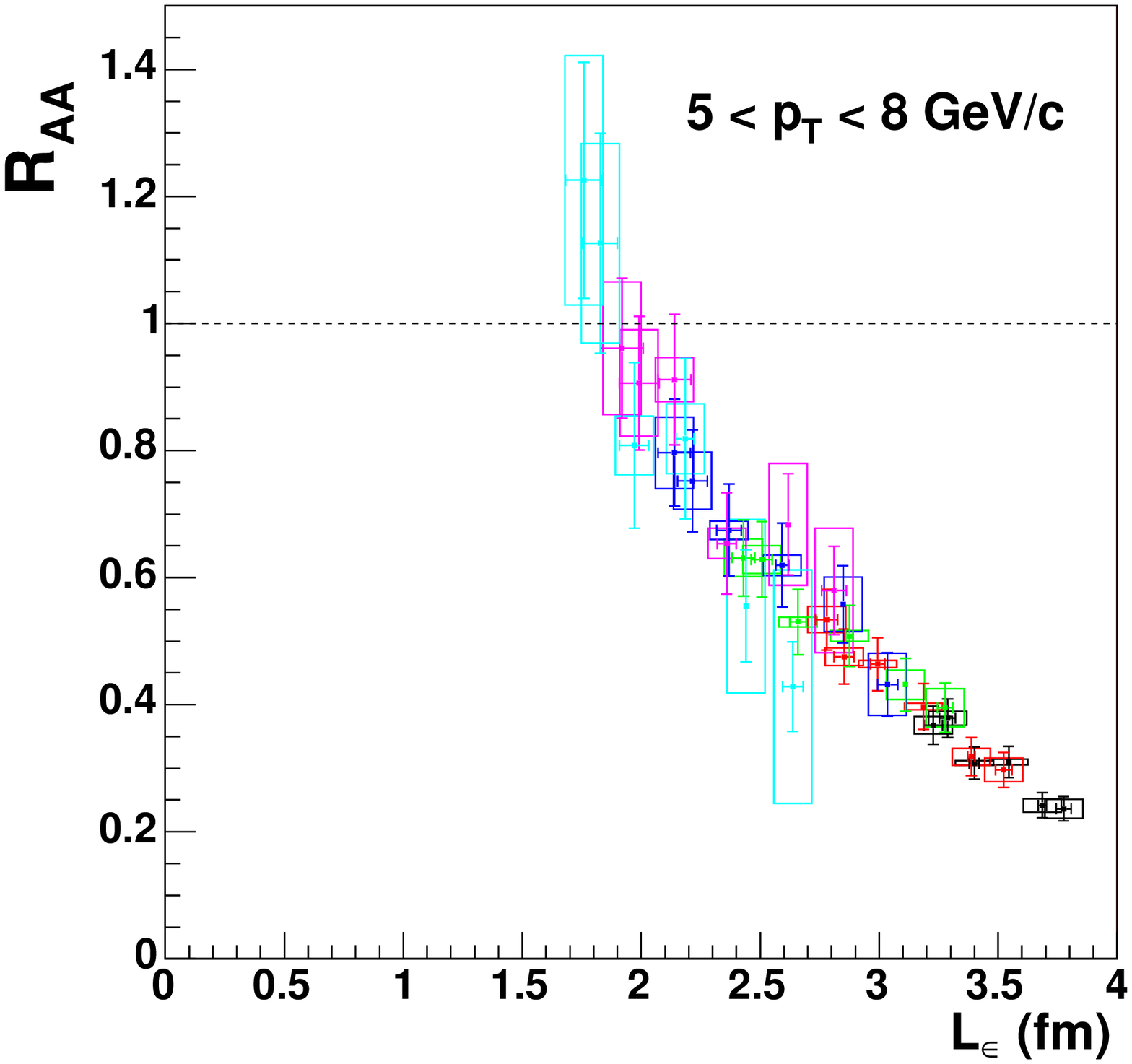}
\includegraphics[width=0.49\linewidth]{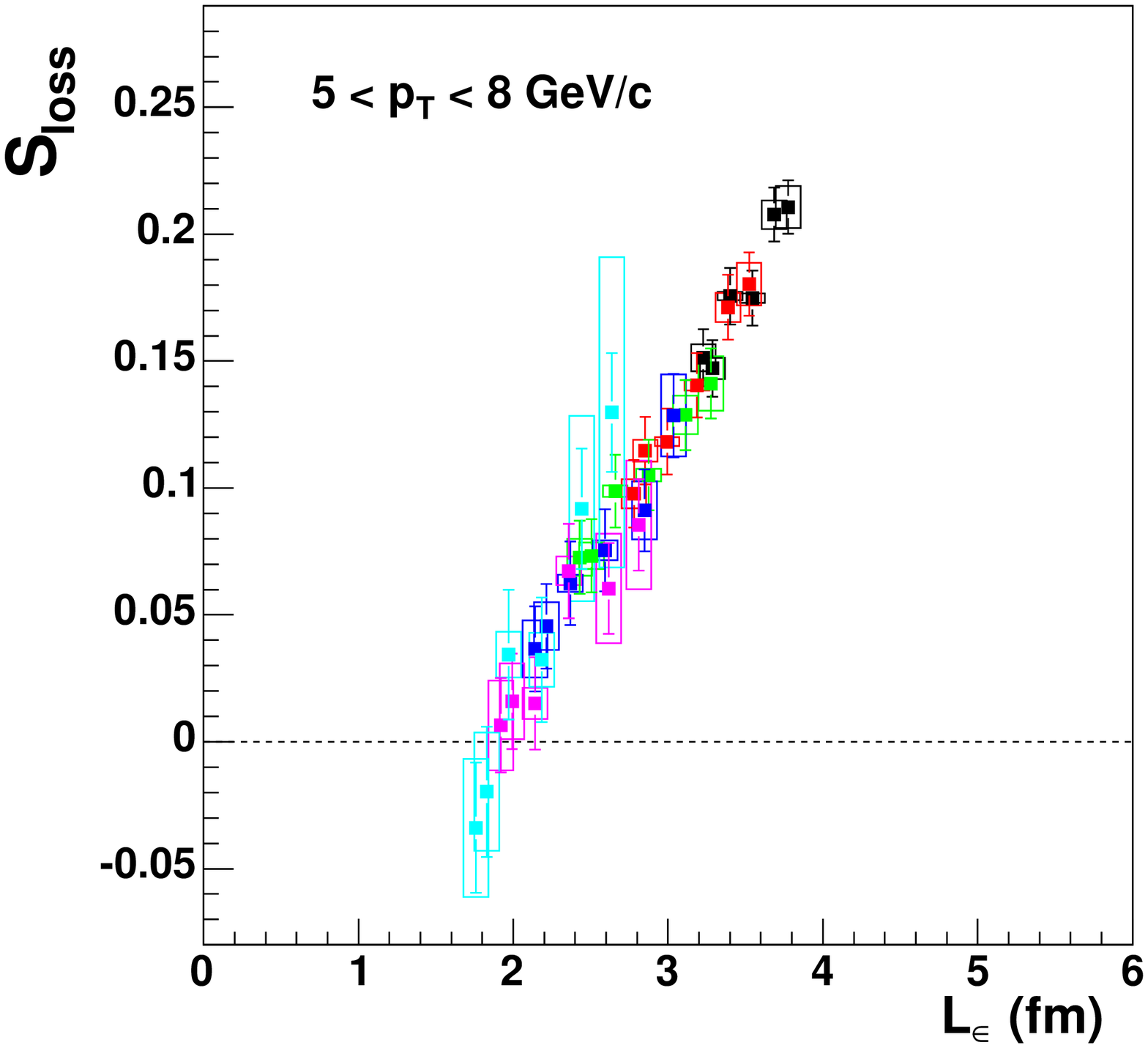}
\caption[]{(color online) \RAA\ and \Sloss\ vs. \Leps\, whose definition is
explained in the text.  Each data point represents a centrality bin
and \dphi\ (azimuth defined \emph{w.r.t.} the reaction plane) bin
combination.  The six centrality bins are denoted by different colors
as follows: cyan 60-70\%, mauve 50-60\%, blue 40-50\%, green 30-40\%,
red 20-30\%, black 0-10\%.  Within each centrality  group, the six
different data points correspond to the same \dphi\ bins as in Figs.
\ref{fig:raa_phi_int3}$-$ \ref{fig:sloss_phi_int5}. The height of the
bars around each data point represent the systematic error in $R_{AA}
(\Delta\phi)$ ($S_{\rm loss}$ corresponding to $L_{\varepsilon}$). }
\label{fig:Leps}
\end{figure}


\begin{figure}[tbh]
\includegraphics[width=0.49\linewidth]{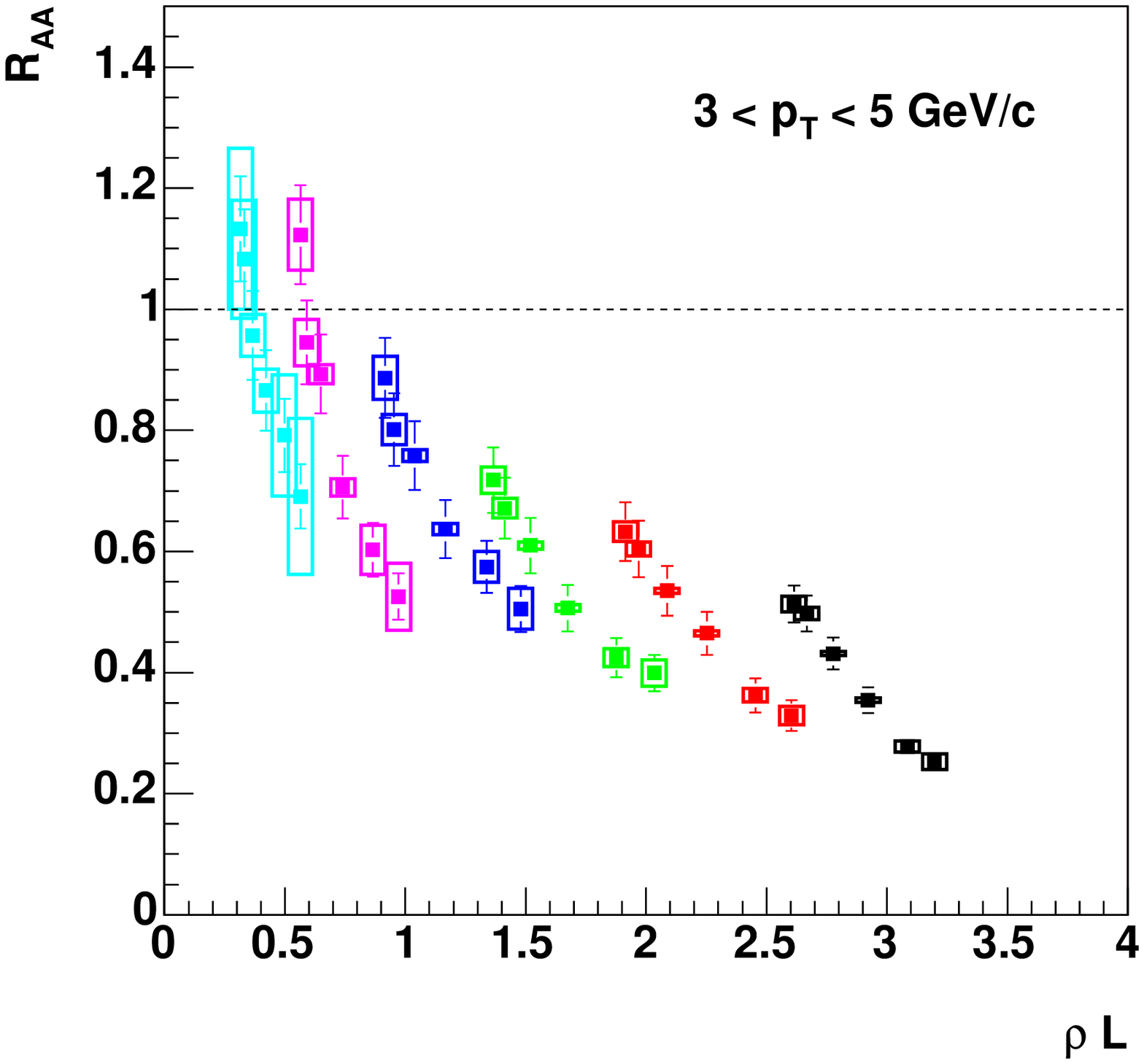}
\includegraphics[width=0.49\linewidth]{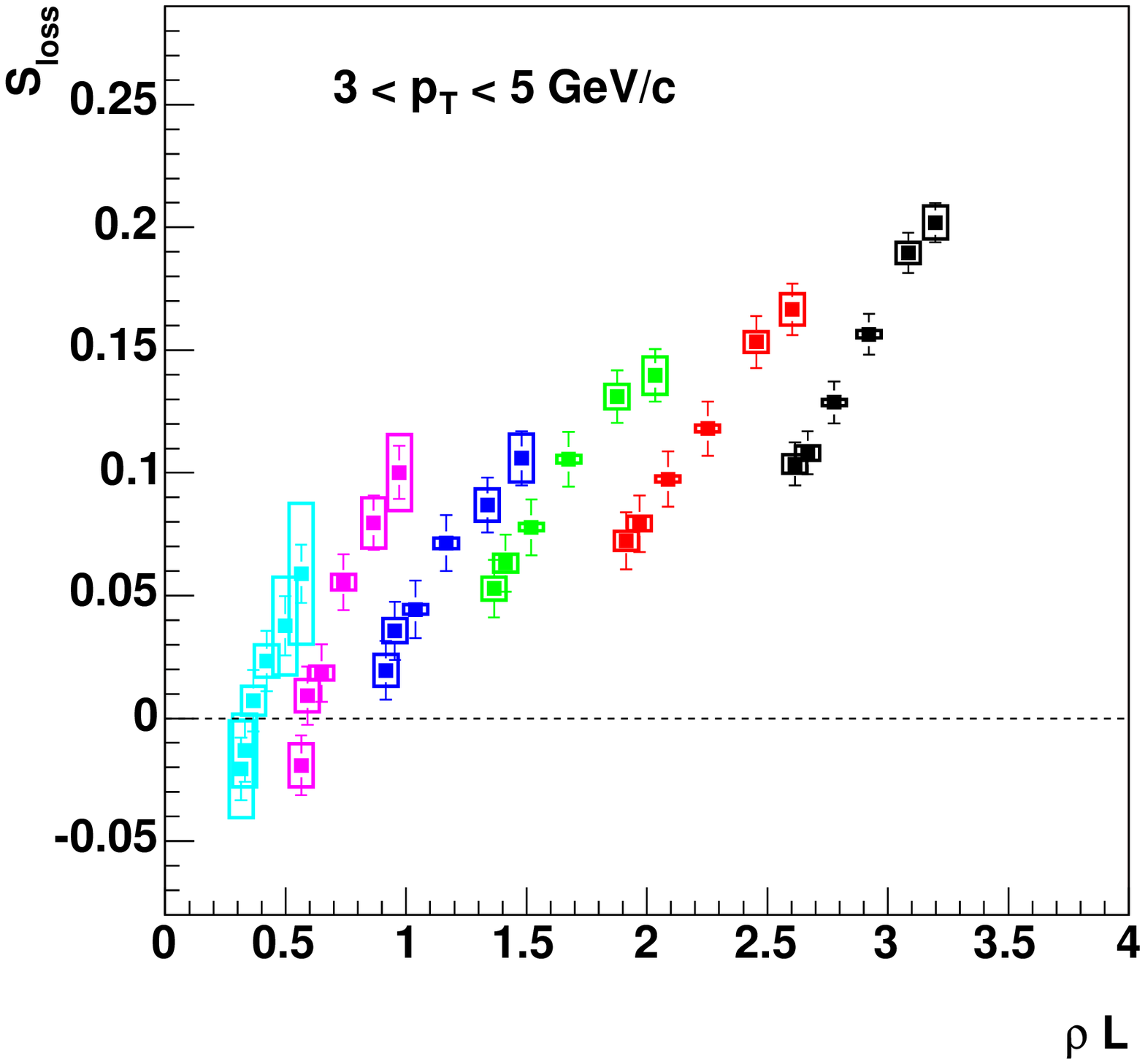}
\includegraphics[width=0.49\linewidth]{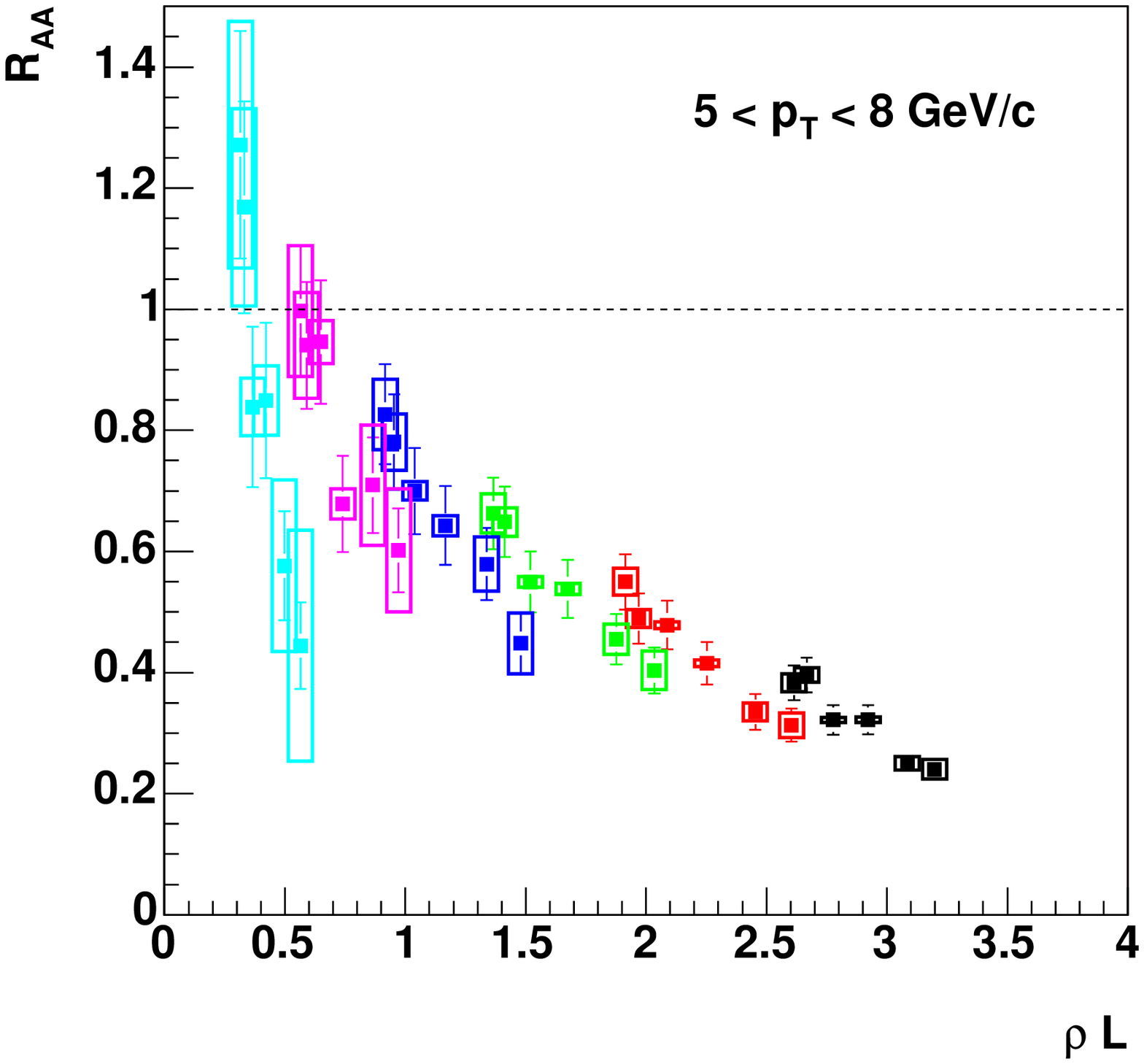}
\includegraphics[width=0.49\linewidth]{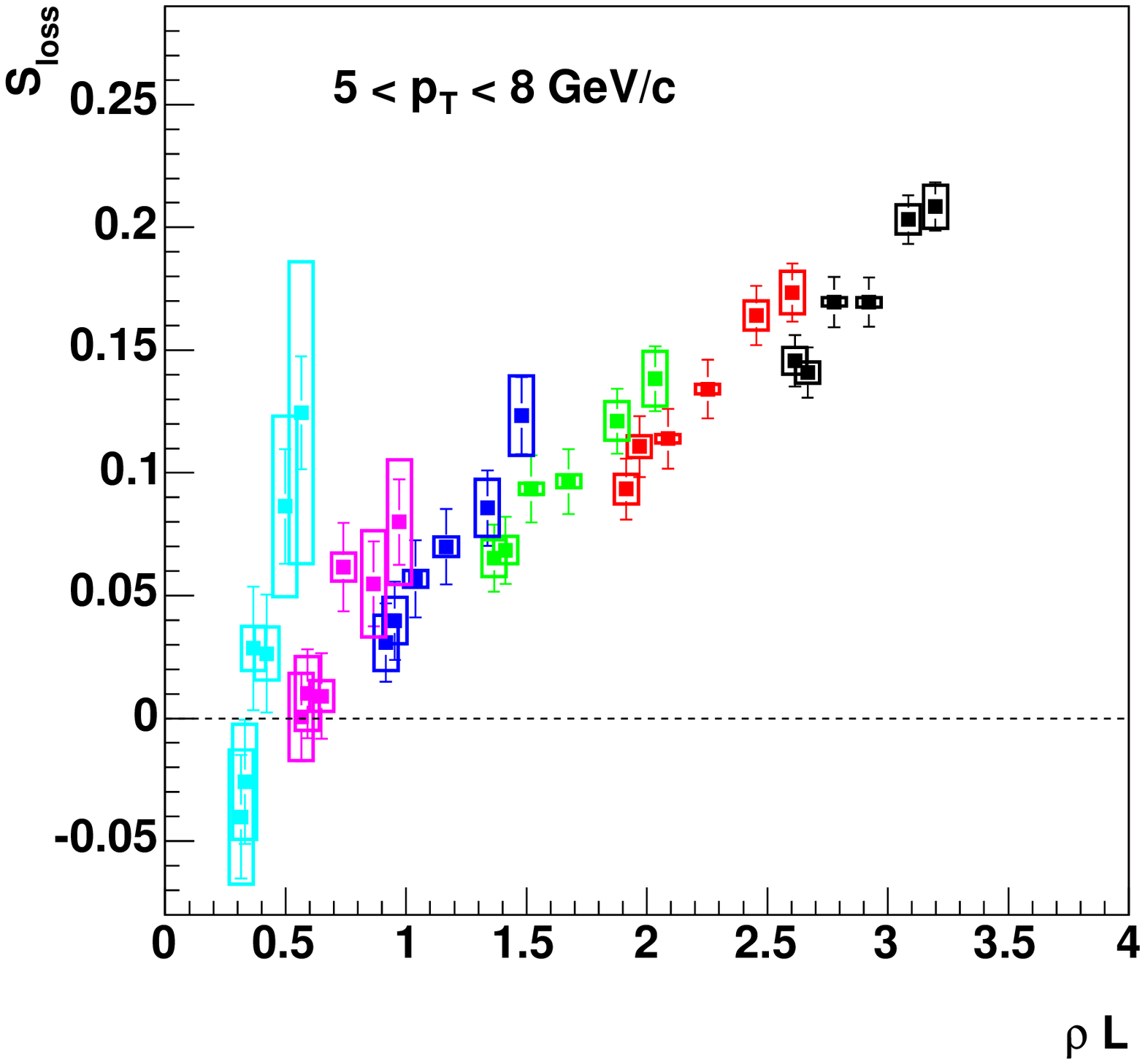}
\caption[]{(color online) \RAA\ and \Sloss\ vs. $\rhoL$, the density
weighted path-length. Colors/data points as in Fig. \ref{fig:Leps}.}
\label{fig:rhoL}
\end{figure}


\begin{figure}[tbh]
\includegraphics[width=0.49\linewidth]{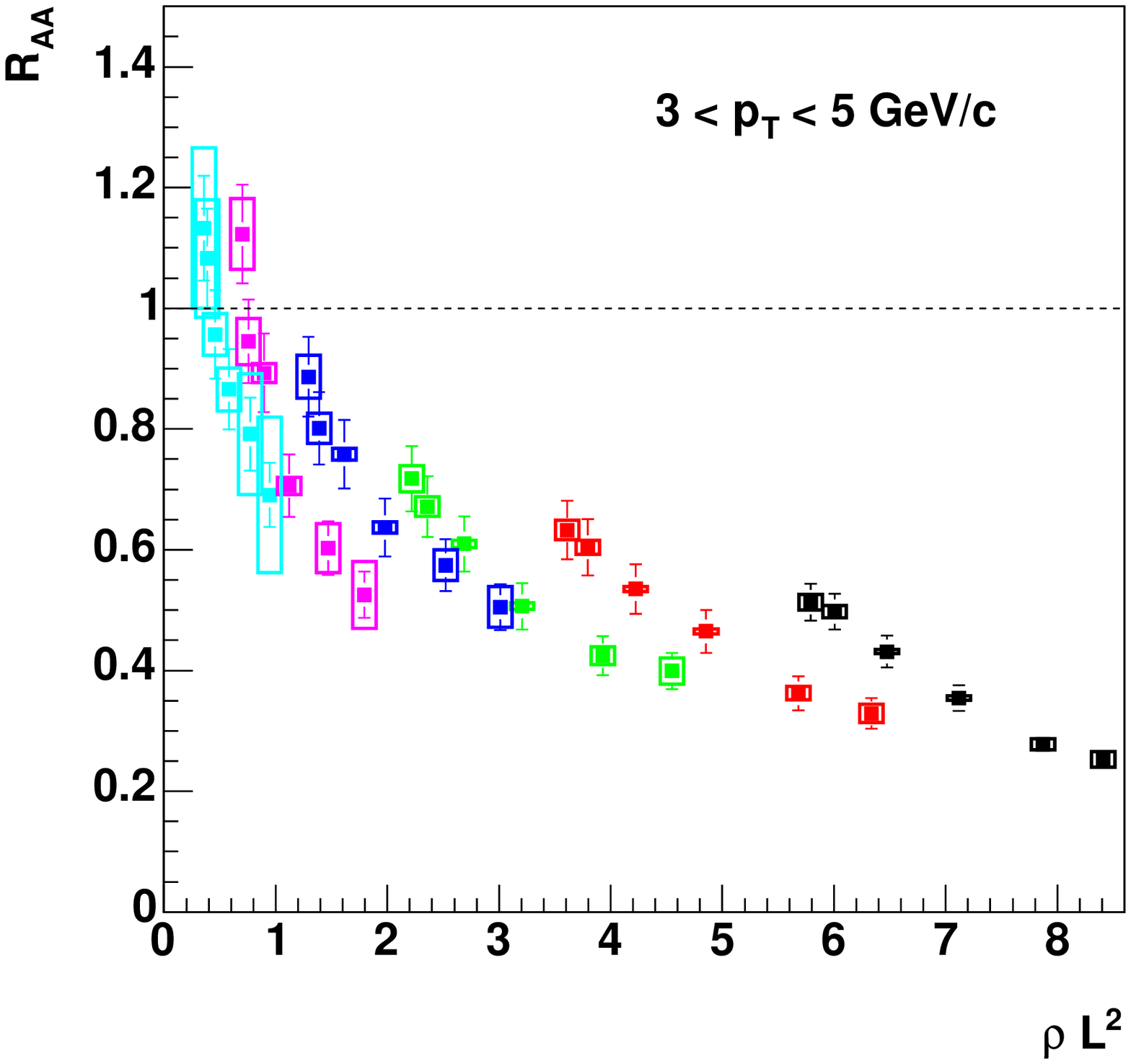}
\includegraphics[width=0.49\linewidth]{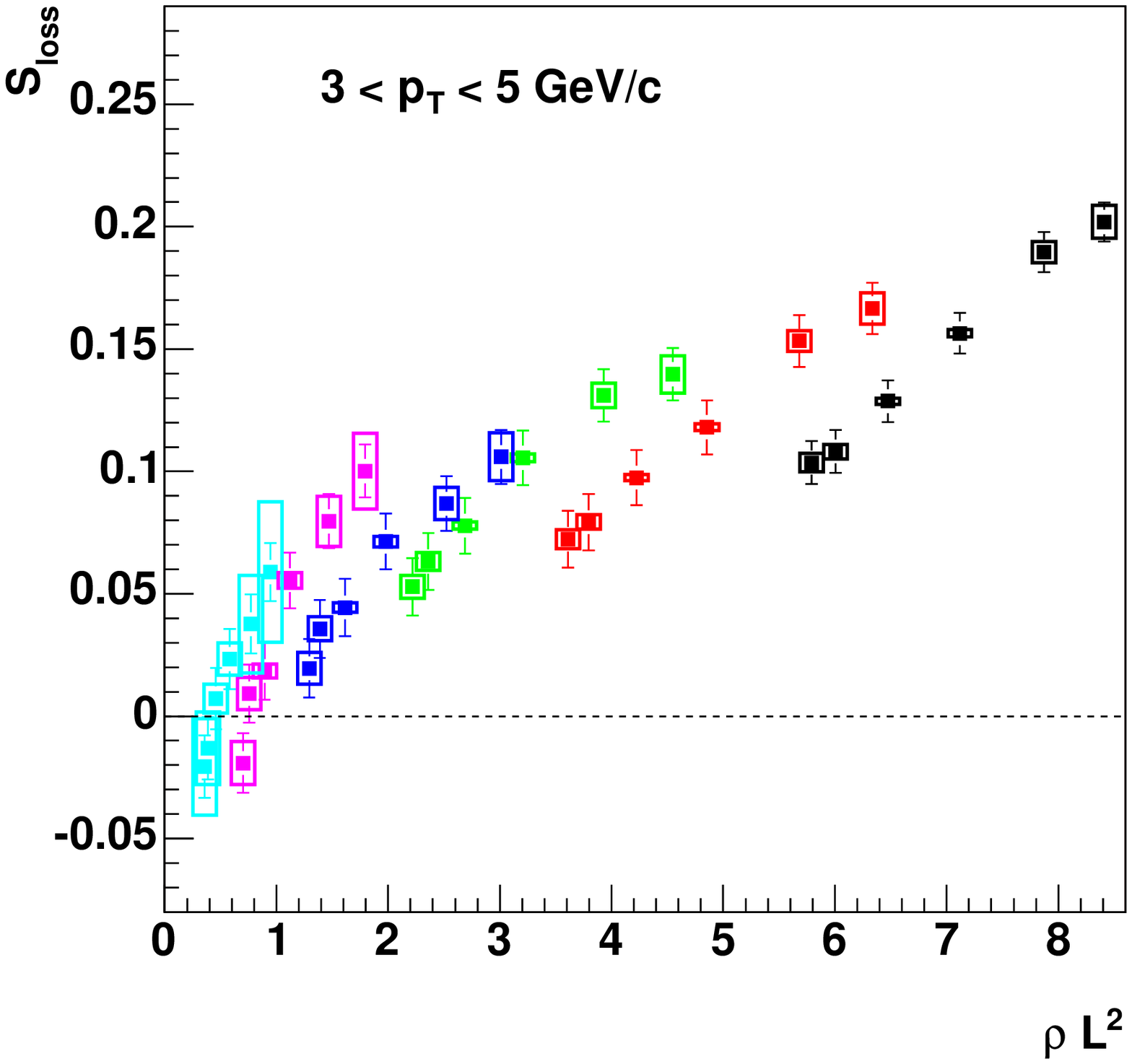}
\includegraphics[width=0.49\linewidth]{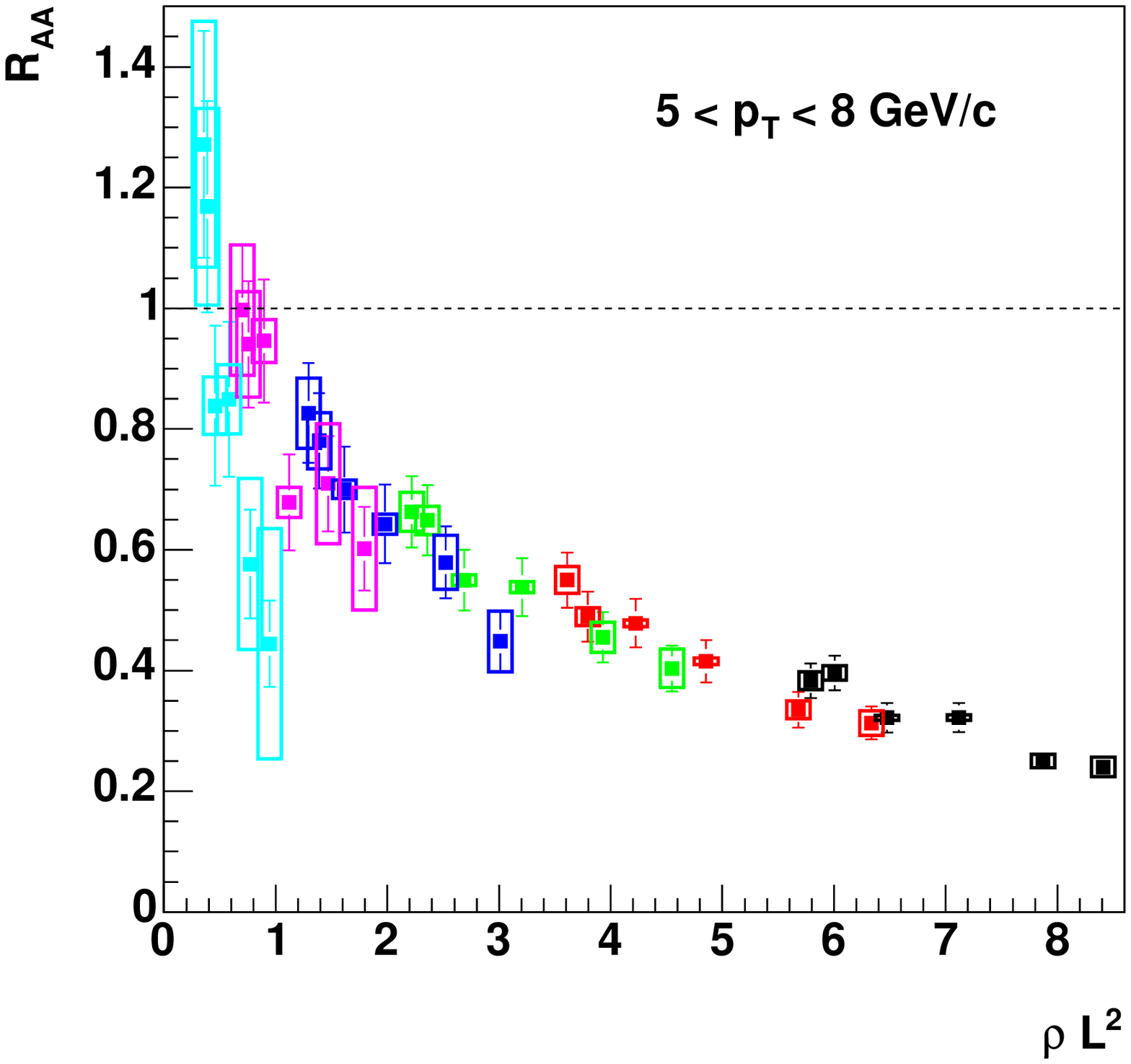}
\includegraphics[width=0.49\linewidth]{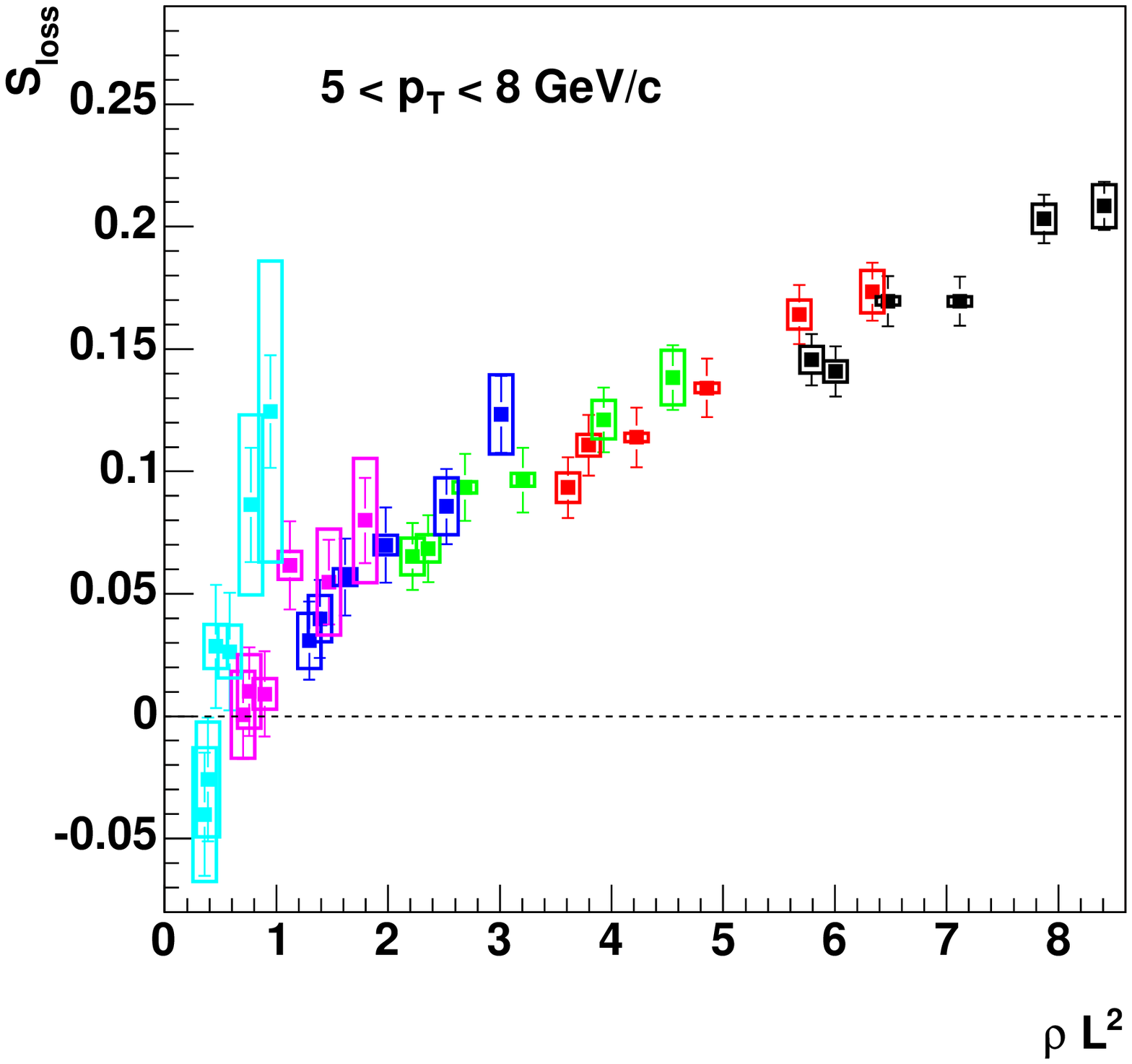}
\caption[]{(color online) \RAA\ and \Sloss\ vs. $\rhoL^2$, the density
weighted path-length squared.  Colors/data points as in Fig.
\ref{fig:Leps}.} \label{fig:rhoLsqr}
\end{figure}


\begin{figure}[tbh]
\includegraphics[width=0.49\linewidth,angle=0]{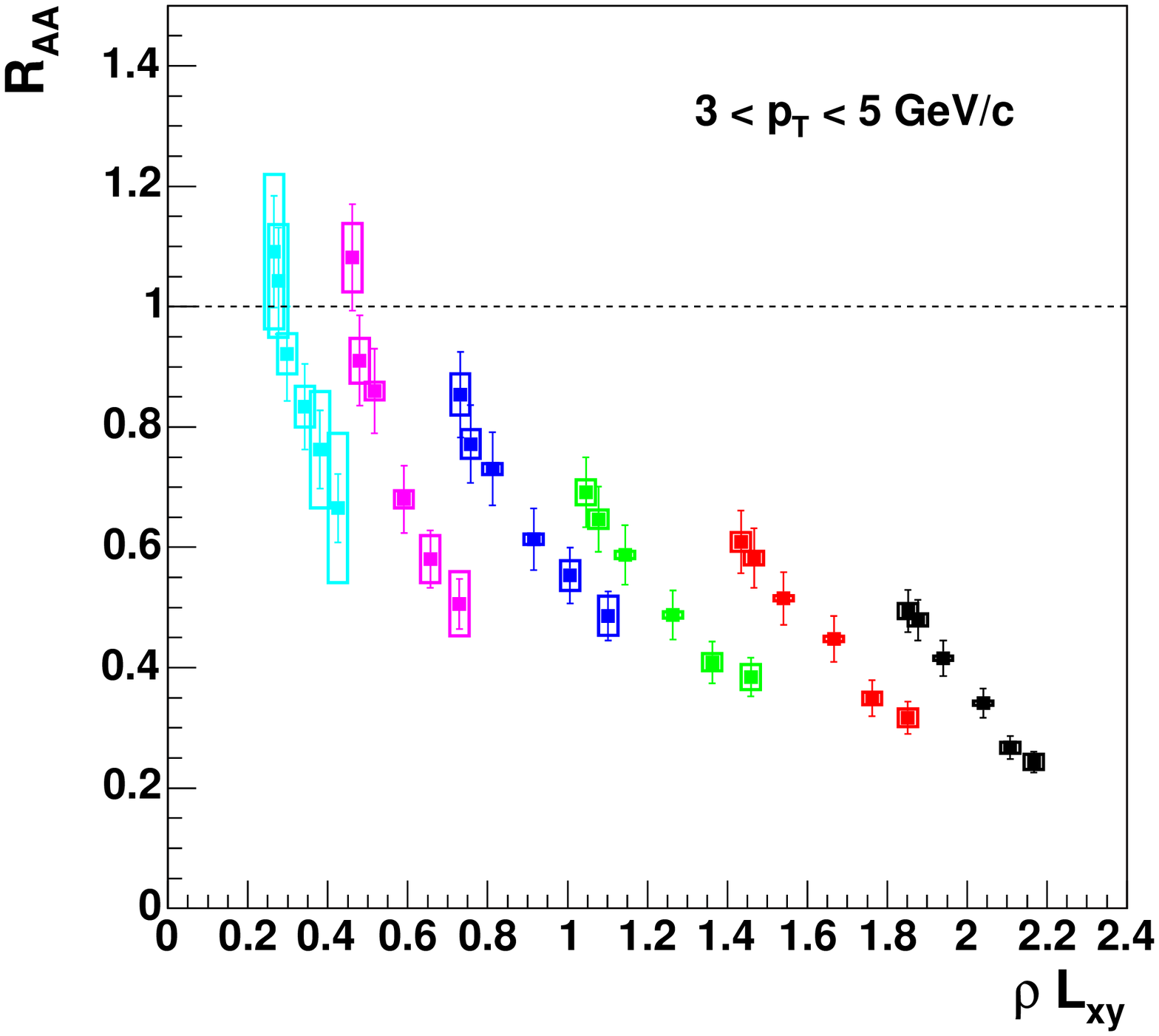}
\includegraphics[width=0.49\linewidth,angle=0]{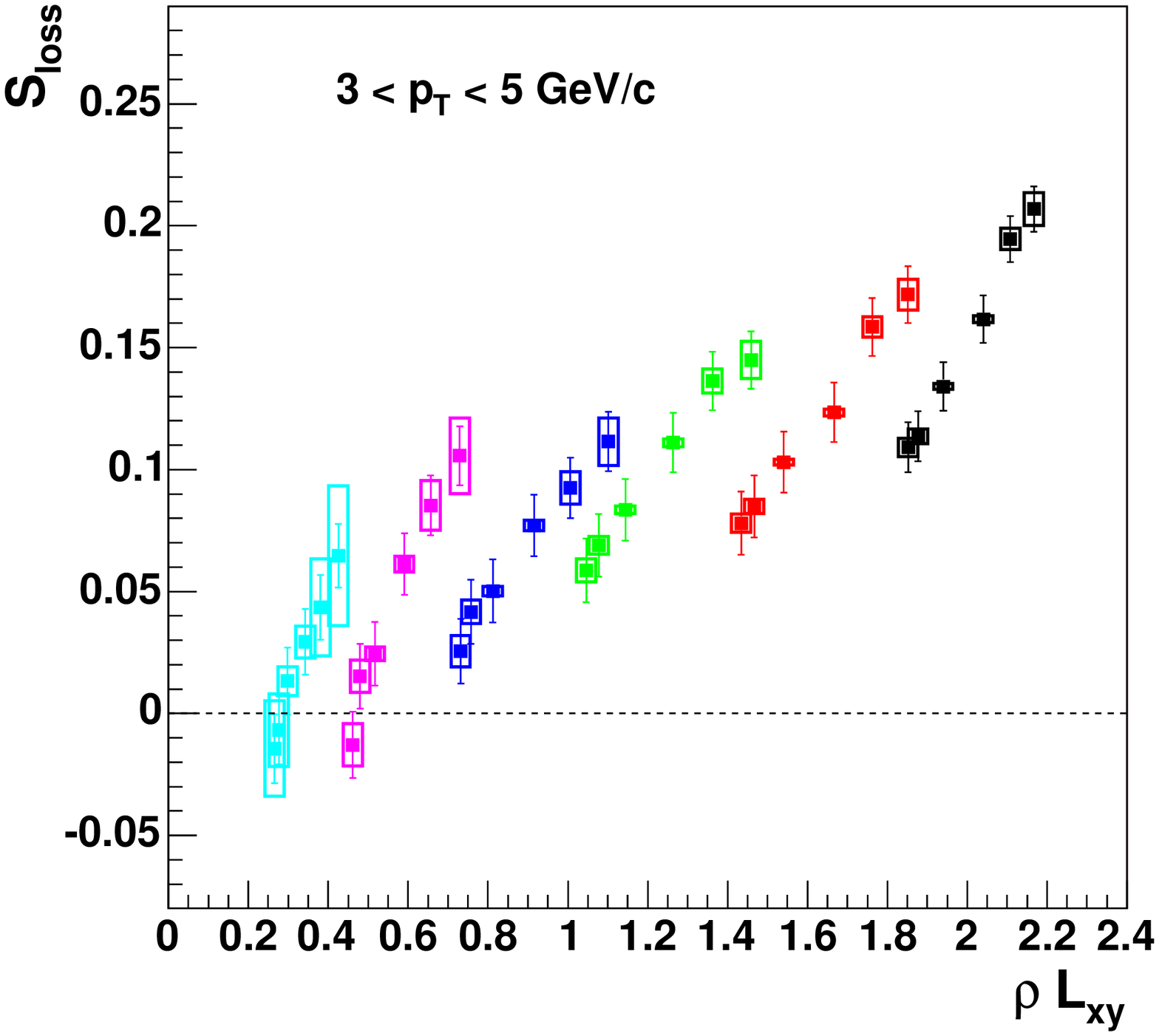}
\includegraphics[width=0.49\linewidth,angle=0]{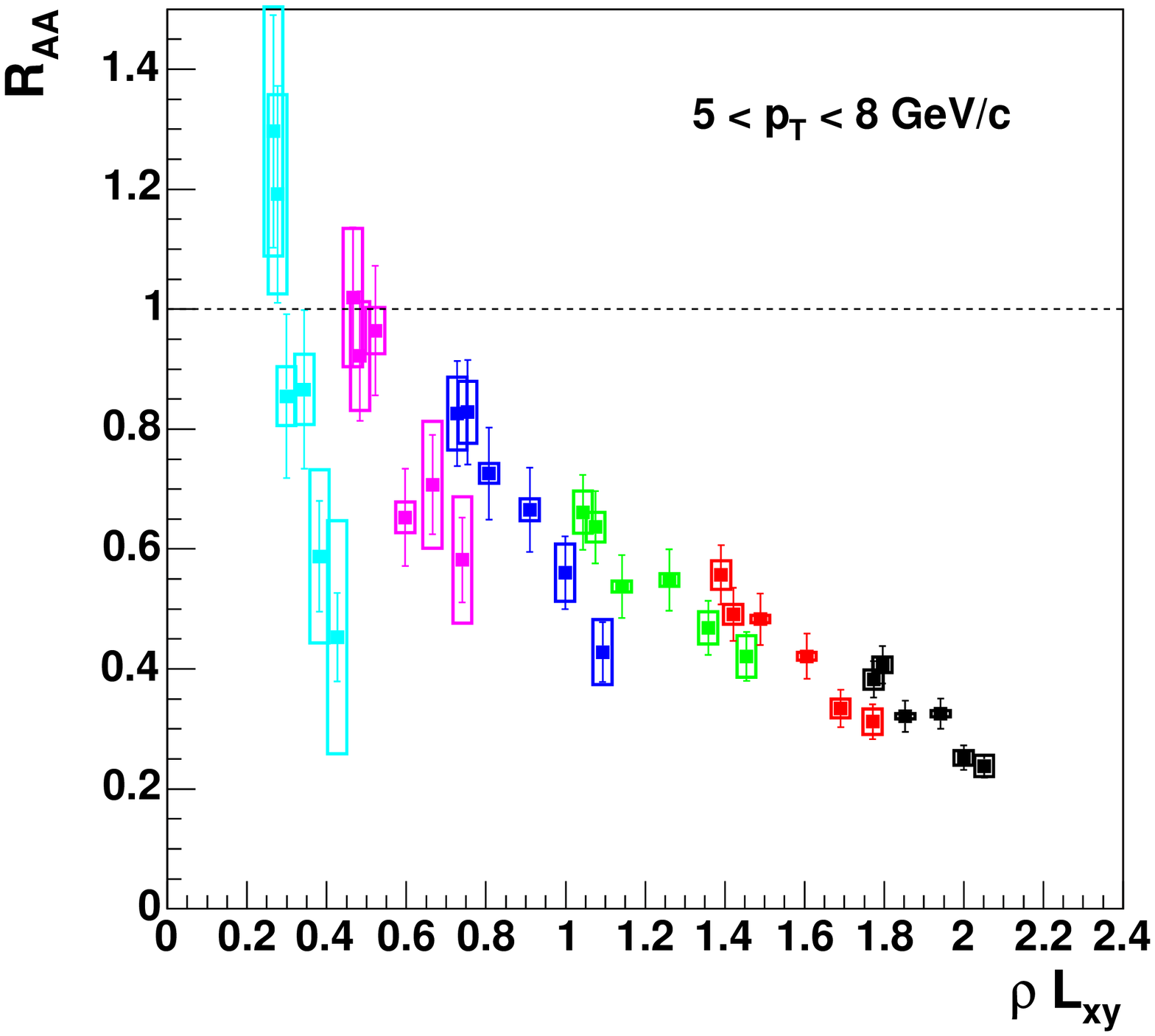}
\includegraphics[width=0.49\linewidth,angle=0]{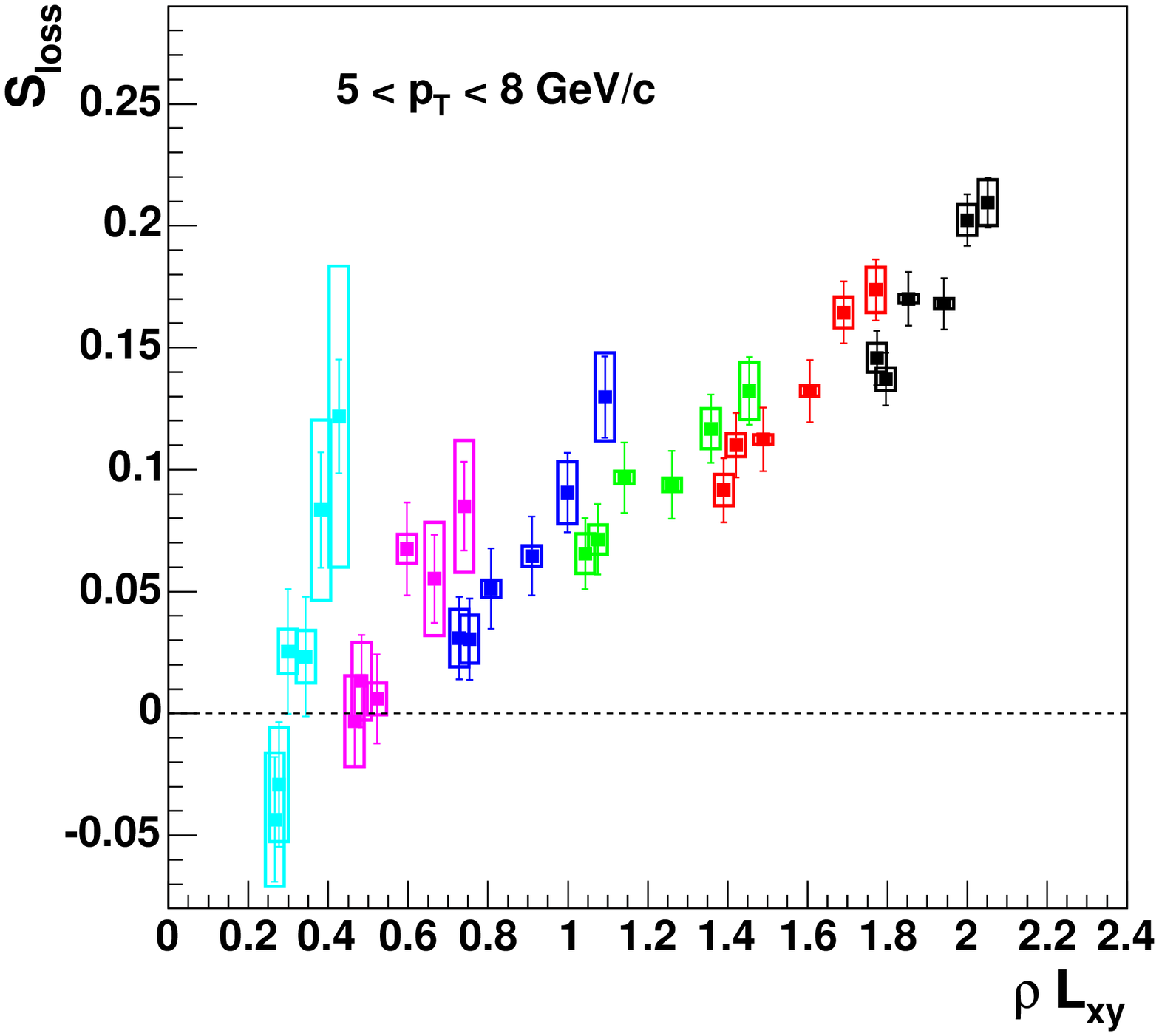}
\caption[]{(color online) \RAA\ and \Sloss\ vs. \rhoLxy\, whose definition
is explained in the text.  Colors/data points as in Fig.
\ref{fig:Leps}.} \label{fig:rhoLxy}
\end{figure}


\begin{figure}[tbh]
\includegraphics[width=0.49\linewidth,angle=0]{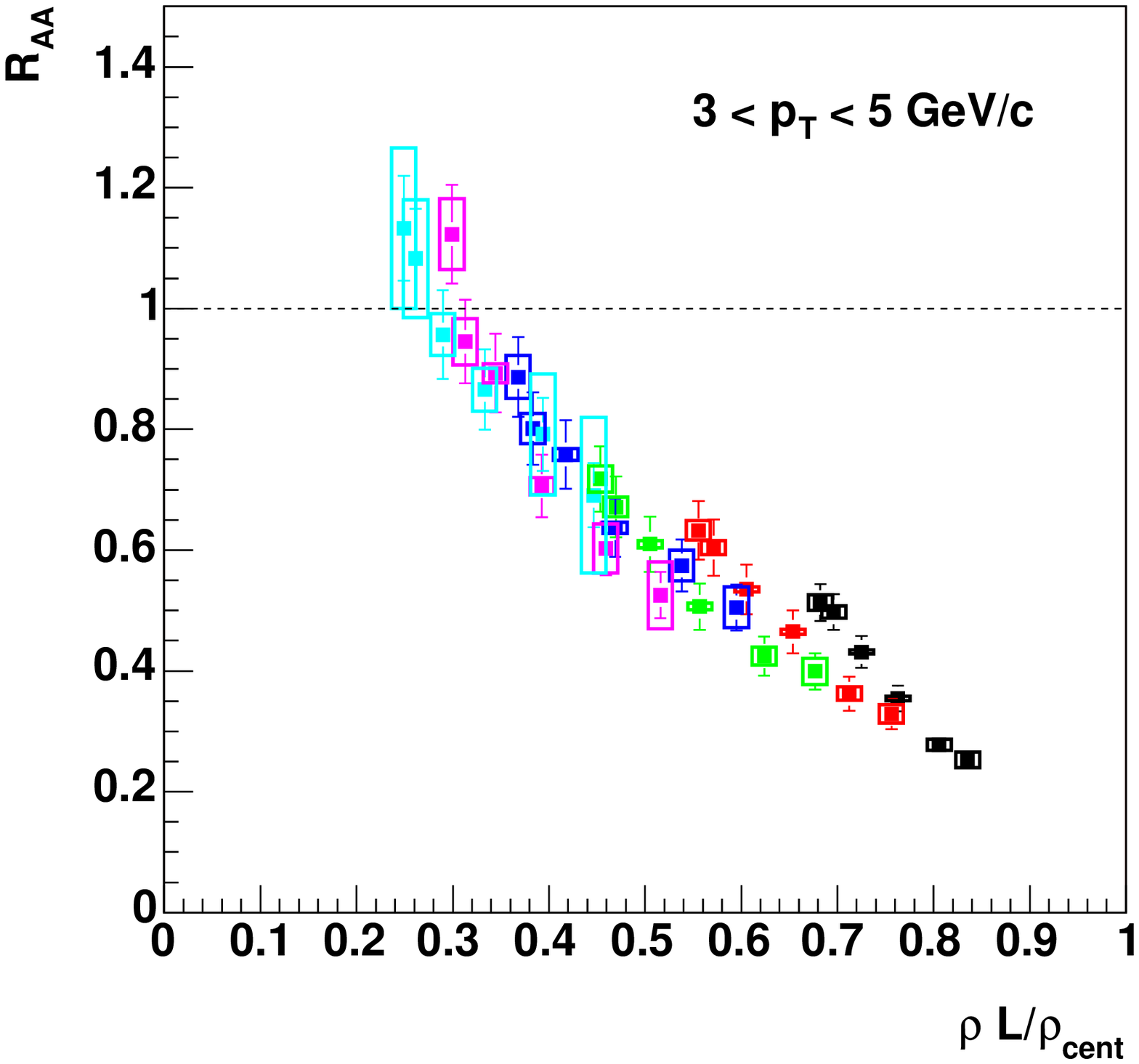}
\includegraphics[width=0.49\linewidth,angle=0]{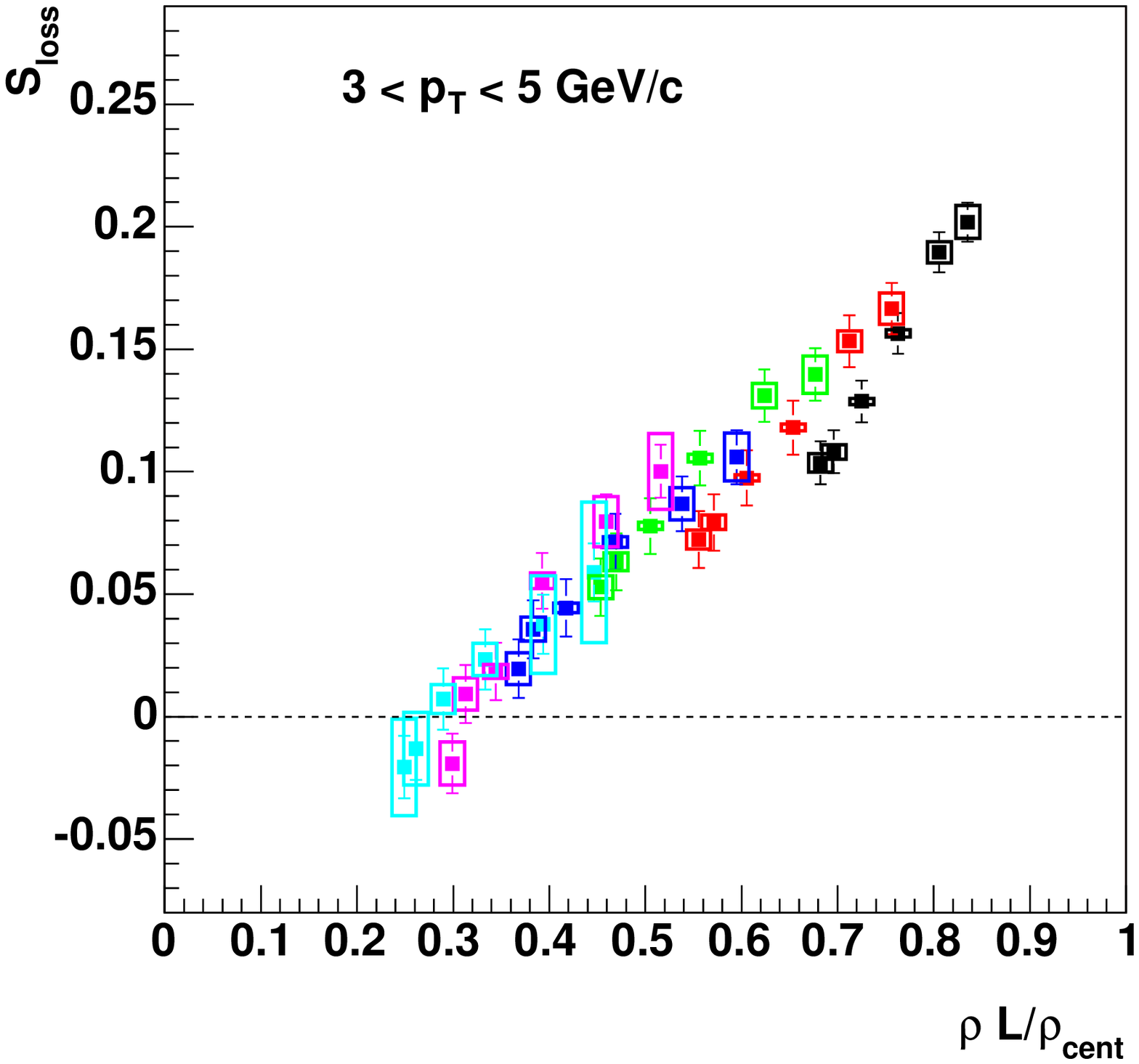}
\includegraphics[width=0.49\linewidth,angle=0]{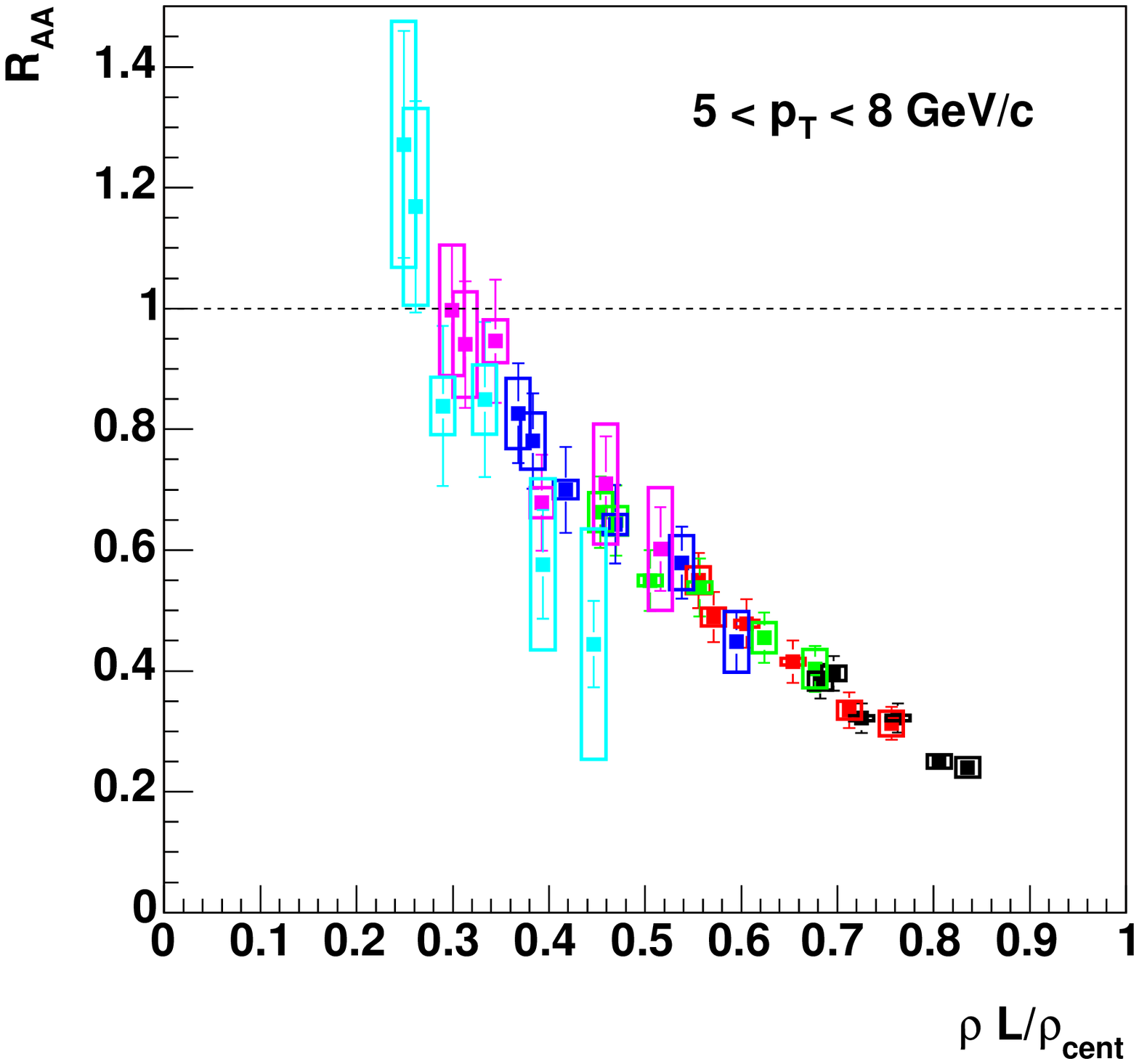}
\includegraphics[width=0.49\linewidth,angle=0]{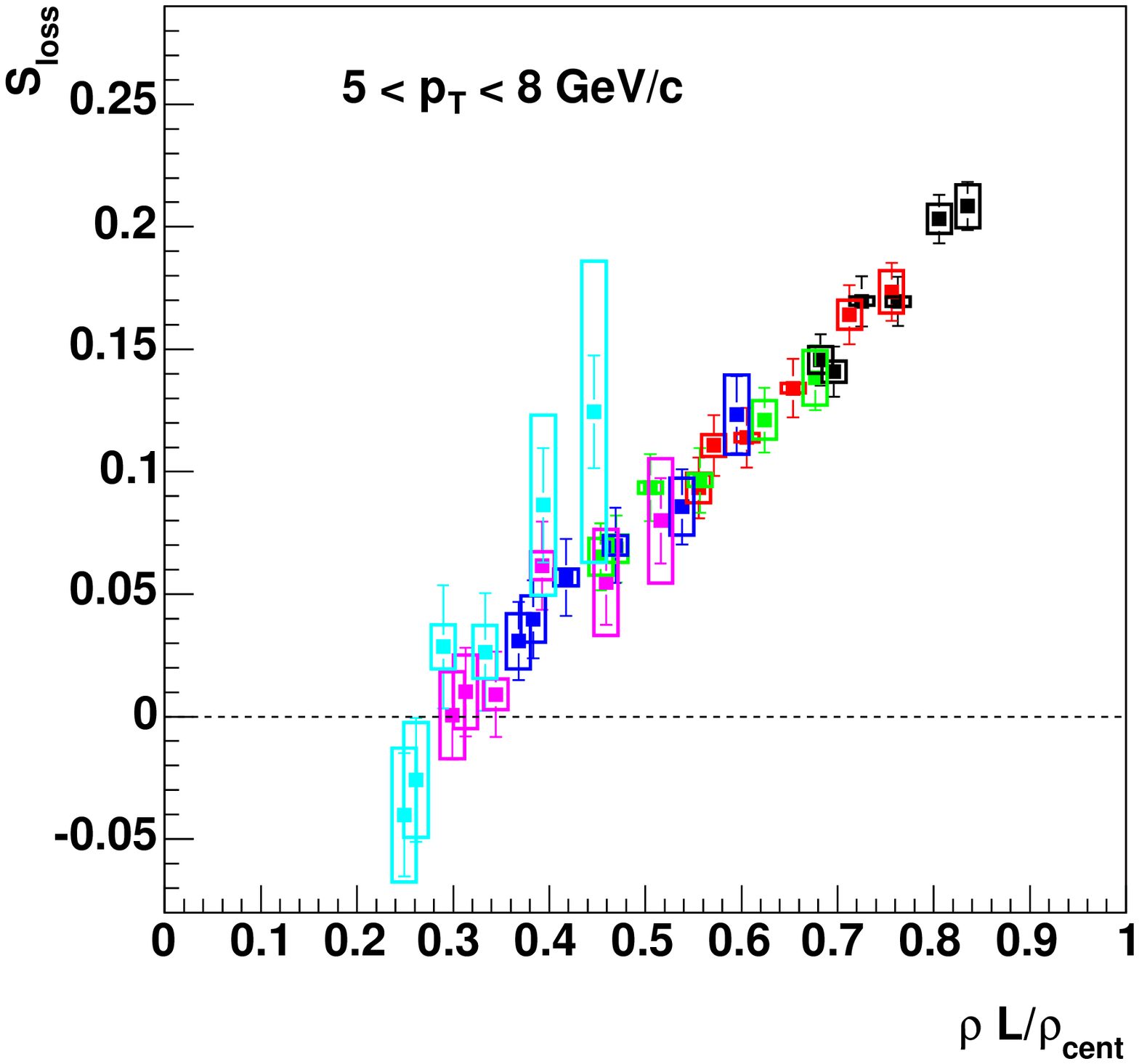}
\caption[]{(color online) \RAA\ and \Sloss\ vs. \rhoL\ normalized by the
most central ($b_x = b_y = 0$) density $\rho_{\rm cent}$.
Colors/data points as in Fig. \ref{fig:Leps}.}
\label{fig:rhoLdividebyrhocent}
\end{figure}

%
%
%



The plots shown in
Figs.~\ref{fig:sloss_phi_int5}-\ref{fig:rhoLdividebyrhocent}
illustrate the path length and color-charge dependence of suppression
using our empirical estimators. The systematic error in the
estimators due to the uncertainty of the overlap geometry parameter
in a centrality class are approximately 10-20\% and is not shown in
the figures.  This uncertainty is derived by propagating the impact
parameter and eccentricity uncertainties from the PHENIX Glauber MC
itself \cite{Adcox:2000sp, Adcox:2002ms}.

From Figs. ~\ref{fig:sloss_phi_int5}-\ref{fig:rhoLdividebyrhocent}
it is evident that the individual centrality bins exhibit roughly
parallel linear dependencies of the variables vs $\rho L_{xy}$ etc.
For the $3 <p_T<5$ GeV/c bin these slopes are such that the curves
are disjoint due to the steeper value of the slopes in each
centrality group (each color in the plots) compared to the bin-to-bin
trend.  For the higher $5<p_T<8$ GeV/c bin, the slopes in the
individual centralities flatten such that they follow the bin-to-bin
trend much better.  These are meant to be qualitative statements.  We
defer anymore quantitative tests, \emph{e.g.} statistical tests, to
subsequent datasets (\emph{e.g.} the larger PHENIX 2004-2005 Run4
dataset) with which we can improve statistical precision.

In this spirit, we note several other interesting qualitative
dependencies:

\begin{itemize}
\item $R_{AA}$ is universal as a function of $L_{\varepsilon}$ for all centrality
classes and both $p_T$ ranges.
\item $S_{\rm loss}$ is universal and is a linear function of
$L_{\varepsilon}$ for all centrality classes and both $p_T$ ranges.
\item Within our errors, we see no suppression $R_{AA} \approx 1$,
hence no apparent fractional energy loss $S_{\rm loss}$ for
$L_{\varepsilon} \leq 2$ fm. \item Neither $R_{AA}$ nor $S_{\rm loss}$
is universal as a function of $\rho L$, $\rho L^2$ or $\rho L_{xy}$
for $3 < p_T < 5$ GeV/$c$.
\item  For the higher $5 < p_T < 8$ GeV/$c$ \pt\ bin, $S_{\rm loss}$ ($R_{AA}$) approaches
universality as a function of $\rho L^2$, $\rho L$, and $\rho L_{xy}$
(possibly to a lesser extent for the latter two) but does not achieve
the level of universality found for $L_{\varepsilon}$.  The largest
deviations from universality in these quantities are towards the
longer axis (perpendicular to the event plane) in the more peripheral
events. The dependence of $S_{\rm loss}$ is reasonably linear as a
function of $\rho L$ but tends to level off at larger values of $\rho
L^2$.
\item When $\rho L$ is normalized by the central density
$\rho_{\rm part}(0,0)=\rho_{\rm cent}$, then $S_{\rm loss}$
($R_{AA}$) become universal in the quantities $\rho L/\rho_{\rm
cent}$ for both $p_T$ ranges with a linear dependence.  The
universality appears to become more exact in the higher $p_T$ range.
A similar improvement (not shown) of the qualitative universality for
$\rho L^2$ and $\rho L_{xy}$ is also observed when these quantities
are scaled in the same way by $\rho_{\rm cent}$.
\end{itemize}

The most important of these observations is the absence of
suppression for the same value of $L_{\varepsilon} \leq 2$ fm for
both $p_T$ ranges, $3\leq p_T\leq 5$ GeV/$c$ and $5\leq p_T\leq 8$
GeV/$c$.  This may suggest a ``formation time effect'' (see
\cite{Pantuev:2005jt}, \cite{Shuryak:2003ty}) or some other type of
emission zone which has generally not been taken into account in
parton energy-loss models.  The level of universal scaling with this
simple geometric quantity is surprising.

\section{Summary and Conclusions}
\label{sec:conclusion}

We have presented a detailed analysis of high-$p_T$ neutral pion
suppression as a function of transverse momentum, centrality and
angle with respect to reaction plane in \AuAu\ collisions at \snn\ =
200 GeV. The $\pi^0$ yields have been measured in the range
$p_T\approx$ 1 -- 14 GeV/$c$ in nine centrality bins and compared to
the $\pi^0$ differential cross-sections measured in \pp\ . The ratio
of \AuAu\ over \pp\ spectra (scaled by the number of equivalent
nucleon-nucleon scatterings) is reduced more and more for larger
centralities. The resulting suppression factor is, however,
independent of $p_T$ above $p_T\approx$ 4 GeV/$c$ for all
centralities. This observation can be interpreted as an indication of
a constant effective fractional energy loss, fixed $S_{\rm loss}$
``$p_T$ shift'', in the \AuAu\ compared to the \pp\ yields. The
dependence of $S_{\rm loss}$ in the centrality as given by the number
of participating nucleons $N_{part}$ follows an approximately $N_{\rm
part}^{2/3}$ law as predicted by parton energy loss models.

In order to constrain the ``jet quenching'' models with more
differential observables, we have experimentally tested the
path-length ($L$) dependence of the energy loss by exploiting the
spatial azimuthal asymmetry of the system produced in non-central
nuclear collisions. Due to the characteristic almond-like shape of the
overlapping matter produced in A+A reactions with finite impact
parameter, partons traversing the produced medium along the direction
perpendicular to the reaction plane (``out-of-plane'') will
comparatively go through more matter than those going parallel to it
(``in-plane''), and therefore are expected to lose more energy.

We have studied the suppression pattern along different \dphi\
trajectories with respect to the reaction plane determined with the
Beam-Beam-Counters at high rapidities. The measured
$R_{AA}(\Delta\phi)$ curves show clearly a factor of $\sim$2 more
suppression out-of-plane ($\Delta\phi$ = $\pi$/2) than in-plane
($\Delta\phi$ = 0) for all the centralities (eccentricities)
considered. Theoretical calculations of parton energy loss in an
azimuthally asymmetric medium predict a significantly smaller
difference between the suppression patterns for partons emitted at
$\Delta\phi$ = 0 and $\Delta\phi$ = $\pi$/2
\cite{Shuryak:2001me,Drees:2003zh,Muller:2002fa}.  The discrepancy is
stronger for more peripheral centralities (with correspondingly
larger eccentricities) and challenges the underlying in-medium
path-length dependence of non-Abelian parton energy loss. Although
elliptic flow effects are responsible for extra boost of in-plane
(compared to out-of-plane) pions, it is unclear how such collective
effects persist up to $p_T$ values as high as $\sim 8$ GeV/$c$.  We
have analyzed the observed reaction-plane and centrality dependence
of the nuclear modification factor with three different versions of a
Monte Carlo model with an increasing level of refinement in the
description of the azimuthal propagation of the parton in the medium.
For all three approaches we observe that the $\pi^0$ suppression
tends to vanish for values of the path-length $L\approx 2$ fm in the
two $p_T$ ranges considered, $3\leq p_T\leq 5$ GeV/$c$ and $5\leq
p_T\leq 8$ GeV/$c$. Such a result suggests either a formation time
effect or a surface emission zone that results in a $p_T$-independent
suppression and puts additional constraints to parton energy-loss
models.


We thank the staff of the Collider-Accelerator and Physics
Departments at Brookhaven National Laboratory and the staff
of the other PHENIX participating institutions for their
vital contributions.  We acknowledge support from the
Department of Energy, Office of Science, Nuclear Physics
Division, the National Science Foundation, Abilene Christian
University Research Council, Research Foundation of SUNY, and
Dean of the College of Arts and Sciences, Vanderbilt
University (U.S.A), Ministry of Education, Culture, Sports,
Science, and Technology and the Japan Society for the
Promotion of Science (Japan), Conselho Nacional de
Desenvolvimento Cient\'{\i}fico e Tecnol{\'o}gico and Funda\c
c{\~a}o de Amparo {\`a} Pesquisa do Estado de S{\~a}o Paulo
(Brazil), Natural Science Foundation of China (People's
Republic of China), Centre National de la Recherche
Scientifique, Commissariat {\`a} l'{\'E}nergie Atomique,
Institut National de Physique Nucl{\'e}aire et de Physique
des Particules, and Institut National de Physique
Nucl{\'e}aire et de Physique des Particules, (France),
Bundesministerium f\"ur Bildung und Forschung, Deutscher
Akademischer Austausch Dienst, and Alexander von Humboldt
Stiftung (Germany), Hungarian National Science Fund, OTKA
(Hungary), Department of Atomic Energy and Department of
Science and Technology (India), Israel Science Foundation
(Israel), Korea Research Foundation and 
Korea Science and Engineering Foundation (Korea), 
Russian Ministry of Industry, Science
and Tekhnologies, Russian Academy of Science, Russian
Ministry of Atomic Energy (Russia), VR and the Wallenberg
Foundation (Sweden), the U.S. Civilian Research and
Development Foundation for the Independent States of the
Former Soviet Union, the US-Hungarian NSF-OTKA-MTA, the
US-Israel Binational Science Foundation, and the 5th European
Union TMR Marie-Curie Programme.


\section*{Appendix:  Data tables of Au+Au $\rightarrow \pi^0 +X$ $p_T$ spectra}
\label{app:pi0_tables}

\begingroup \squeezetable

\begin{table}[bh]
\caption{Final combined PbSc+PbGl $\pi^0$ invariant yields $vs.$ $p_T$ for centrality 0-10\%.}
\begin{ruledtabular}\begin{tabular}{cccccc} \hline
$p_T$ & Yield & Stat. Error & \% & Sys. Error & \% \\ \hline
1.25 & 3.314E+00 & 2.518E-02 & 0.76 &  4.026E-01 & 12.15\\
1.75 & 5.981E-01 & 4.946E-03 & 0.83 &  6.784E-02 & 11.34\\
2.25 & 1.208E-01 & 1.253E-03 & 1.04 &  1.447E-02 & 11.98\\
2.75 & 2.718E-02 & 3.744E-04 & 1.38 &  3.521E-03 & 12.96\\
3.25 & 6.970E-03 & 1.270E-04 & 1.82 &  9.751E-04 & 13.99\\
3.75 & 2.158E-03 & 4.713E-05 & 2.18 &  2.686E-04 & 12.44\\
4.25 & 7.185E-04 & 2.133E-05 & 2.97 &  9.349E-05 & 13.01\\
4.75 & 2.715E-04 & 1.063E-05 & 3.92 &  3.575E-05 & 13.17\\
5.25 & 1.288E-04 & 5.931E-06 & 4.61 &  1.702E-05 & 13.21\\
5.75 & 5.417E-05 & 2.606E-06 & 4.81 &  7.731E-06 & 14.27\\
6.25 & 2.940E-05 & 1.560E-06 & 5.31 &  4.106E-06 & 13.97\\
6.75 & 1.280E-05 & 9.501E-07 & 7.43 &  1.922E-06 & 15.02\\
7.25 & 7.641E-06 & 6.459E-07 & 8.45 &  1.241E-06 & 16.24\\
7.75 & 4.630E-06 & 4.668E-07 & 10.08 &  7.508E-07 & 16.22\\
8.50 & 1.883E-06 & 1.809E-07 & 9.61 &  3.033E-07 & 16.11\\
9.50 & 1.057E-06 & 1.276E-07 & 12.07 &  1.952E-07 & 18.47\\
11.00 & 2.777E-07 & 4.274E-08 & 15.39 &  5.664E-08 & 20.39\\
13.00 & 5.941E-08 & 1.704E-08 & 28.87 &  1.222E-08 & 20.57\\
\hline
\end{tabular}\end{ruledtabular}
\end{table}

\clearpage

\begin{table}
\caption{Final combined PbSc+PbGl $\pi^0$ invariant yields $vs.$ $p_T$ for centrality 10-20\%.}
\begin{ruledtabular}\begin{tabular}{cccccc} \hline
$p_T$ & Yield & Stat. Error & \% & Sys. Error & \% \\ \hline
1.25 & 2.054E+00 & 1.461E-02 & 0.71 &  2.655E-01 & 12.93\\
1.75 & 4.137E-01 & 2.933E-03 & 0.71 &  4.616E-02 & 11.16\\
2.25 & 8.576E-02 & 7.654E-04 & 0.89 &  1.039E-02 & 12.11\\
2.75 & 2.028E-02 & 2.305E-04 & 1.14 &  2.612E-03 & 12.88\\
3.25 & 5.057E-03 & 7.980E-05 & 1.58 &  6.778E-04 & 13.40\\
3.75 & 1.665E-03 & 3.170E-05 & 1.90 &  1.995E-04 & 11.98\\
4.25 & 5.859E-04 & 1.511E-05 & 2.58 &  7.301E-05 & 12.46\\
4.75 & 2.253E-04 & 7.948E-06 & 3.53 &  3.003E-05 & 13.33\\
5.25 & 9.486E-05 & 4.369E-06 & 4.61 &  1.246E-05 & 13.14\\
5.75 & 4.651E-05 & 2.087E-06 & 4.49 &  6.696E-06 & 14.40\\
6.25 & 2.224E-05 & 1.249E-06 & 5.62 &  3.252E-06 & 14.62\\
6.75 & 1.109E-05 & 8.621E-07 & 7.78 &  1.899E-06 & 17.13\\
7.25 & 6.455E-06 & 5.485E-07 & 8.50 &  1.091E-06 & 16.90\\
7.75 & 3.568E-06 & 3.999E-07 & 11.21 &  7.173E-07 & 20.10\\
8.50 & 1.724E-06 & 1.718E-07 & 9.96 &  3.279E-07 & 19.01\\
9.50 & 6.318E-07 & 9.789E-08 & 15.49 &  1.144E-07 & 18.11\\
11.00 & 1.701E-07 & 3.347E-08 & 19.68 &  3.147E-08 & 18.51\\
13.00 & 5.093E-08 & 1.610E-08 & 31.62 &  9.747E-09 & 19.14 \\
\hline
\end{tabular}\end{ruledtabular}
\end{table}

\begin{table}
\caption{Final combined PbSc+PbGl $\pi^0$ invariant yields $vs.$ $p_T$ for centrality 20-30\%.
}
\begin{ruledtabular}\begin{tabular}{cccccc} \hline
$p_T$ & Yield & Stat. Error & \% & Sys. Error & \% \\ \hline
1.25 & 1.601E+00 & 9.668E-03 & 0.60 &  1.852E-01 & 11.57\\
1.75 & 2.879E-01 & 1.911E-03 & 0.66 &  3.260E-02 & 11.32\\
2.25 & 6.045E-02 & 5.117E-04 & 0.85 &  7.416E-03 & 12.27\\
2.75 & 1.429E-02 & 1.537E-04 & 1.08 &  1.761E-03 & 12.32\\
3.25 & 3.983E-03 & 5.534E-05 & 1.39 &  5.192E-04 & 13.04\\
3.75 & 1.233E-03 & 2.340E-05 & 1.90 &  1.546E-04 & 12.53\\
4.25 & 4.749E-04 & 1.158E-05 & 2.44 &  6.115E-05 & 12.88\\
4.75 & 1.732E-04 & 5.898E-06 & 3.41 &  2.258E-05 & 13.04\\
5.25 & 7.761E-05 & 3.503E-06 & 4.51 &  1.074E-05 & 13.84\\
5.75 & 3.573E-05 & 1.627E-06 & 4.55 &  4.870E-06 & 13.63\\
6.25 & 1.714E-05 & 9.568E-07 & 5.58 &  2.389E-06 & 13.94\\
6.75 & 9.015E-06 & 6.625E-07 & 7.35 &  1.384E-06 & 15.36\\
7.25 & 5.146E-06 & 4.423E-07 & 8.59 &  8.214E-07 & 15.96\\
7.75 & 2.878E-06 & 3.267E-07 & 11.35 &  5.465E-07 & 18.99\\
8.50 & 1.363E-06 & 1.452E-07 & 10.65 &  2.517E-07 & 18.46\\
9.50 & 6.216E-07 & 8.347E-08 & 13.43 &  1.088E-07 & 17.50\\
11.00 & 1.825E-07 & 2.972E-08 & 16.28 &  3.299E-08 & 18.08\\
13.00 & 3.552E-08 & 1.267E-08 & 35.68 &  6.852E-09 & 19.29\\
\hline
\end{tabular}\end{ruledtabular}
\end{table}

\begin{table}
\caption{Final combined PbSc+PbGl $\pi^0$ invariant yields $vs.$ $p_T$ for centrality 30-40\%.
}
\begin{ruledtabular}\begin{tabular}{cccccc} \hline
$p_T$ & Yield & Stat. Error & \% & Sys. Error & \% \\ \hline
1.25 & 1.040E+00 & 5.648E-03 & 0.54 &  1.244E-01 & 11.96\\
1.75 & 1.754E-01 & 1.100E-03 & 0.63 &  2.001E-02 & 11.41\\
2.25 & 3.833E-02 & 3.102E-04 & 0.81 &  4.567E-03 & 11.91\\
2.75 & 9.610E-03 & 9.930E-05 & 1.03 &  1.175E-03 & 12.23\\
3.25 & 2.670E-03 & 3.764E-05 & 1.41 &  3.512E-04 & 13.15\\
3.75 & 8.612E-04 & 1.667E-05 & 1.94 &  1.097E-04 & 12.74\\
4.25 & 3.270E-04 & 8.158E-06 & 2.49 &  4.185E-05 & 12.80\\
4.75 & 1.252E-04 & 4.421E-06 & 3.53 &  1.619E-05 & 12.94\\
5.25 & 5.266E-05 & 2.822E-06 & 5.36 &  7.394E-06 & 14.04\\
5.75 & 2.761E-05 & 1.348E-06 & 4.88 &  3.839E-06 & 13.90\\
6.25 & 1.189E-05 & 8.138E-07 & 6.85 &  1.949E-06 & 16.39\\
6.75 & 7.115E-06 & 5.804E-07 & 8.16 &  1.198E-06 & 16.84\\
7.25 & 3.705E-06 & 3.972E-07 & 10.72 &  6.264E-07 & 16.91\\
7.75 & 1.898E-06 & 2.549E-07 & 13.42 &  3.307E-07 & 17.42\\
8.50 & 1.168E-06 & 1.301E-07 & 11.13 &  1.967E-07 & 16.83\\
9.50 & 5.043E-07 & 8.312E-08 & 16.48 &  9.634E-08 & 19.10\\
11.00 & 1.541E-07 & 2.748E-08 & 17.83 &  2.910E-08 & 18.89\\
13.00 & 2.941E-08 & 1.278E-08 & 33.46 &  5.621E-09 & 19.11\\
\hline
\end{tabular}\end{ruledtabular}
\end{table}

\begin{table}
\caption{Final combined PbSc+PbGl $\pi^0$ invariant yields $vs.$ $p_T$ for centrality 40-50\%.
}
\begin{ruledtabular}\begin{tabular}{cccccc} 
$p_T$ & Yield & Stat. Error & \% & Sys. Error & \% \\ \hline
1.25 & 6.389E-01 & 3.367E-03 & 0.53 &  7.216E-02 & 11.29\\
1.75 & 1.156E-01 & 6.789E-04 & 0.59 &  1.315E-02 & 11.37\\
2.25 & 2.442E-02 & 1.926E-04 & 0.79 &  2.911E-03 & 11.92\\
2.75 & 6.172E-03 & 6.521E-05 & 1.06 &  7.890E-04 & 12.78\\
3.25 & 1.682E-03 & 2.455E-05 & 1.46 &  2.194E-04 & 13.04\\
3.75 & 5.822E-04 & 1.161E-05 & 1.99 &  7.179E-05 & 12.33\\
4.25 & 1.927E-04 & 6.113E-06 & 3.17 &  2.480E-05 & 12.87\\
4.75 & 8.818E-05 & 3.476E-06 & 3.94 &  1.169E-05 & 13.26\\
5.25 & 3.627E-05 & 2.166E-06 & 5.97 &  4.995E-06 & 13.77\\
5.75 & 1.611E-05 & 9.656E-07 & 5.99 &  2.261E-06 & 14.04\\
6.25 & 9.635E-06 & 6.880E-07 & 7.14 &  1.490E-06 & 15.47\\
6.75 & 4.467E-06 & 4.278E-07 & 9.58 &  7.232E-07 & 16.19\\
7.25 & 2.044E-06 & 2.585E-07 & 12.65 &  3.197E-07 & 15.64\\
7.75 & 1.363E-06 & 2.198E-07 & 16.13 &  2.882E-07 & 21.15\\
8.50 & 7.878E-07 & 1.056E-07 & 13.41 &  1.409E-07 & 17.88\\
9.50 & 2.197E-07 & 5.630E-08 & 25.62 &  4.969E-08 & 22.61\\
11.00 & 1.053E-07 & 2.280E-08 & 21.66 &  2.116E-08 & 20.10\\
13.00 & 2.792E-08 & 1.140E-08 & 40.82 &  6.121E-09 & 21.92\\
\hline
\end{tabular}\end{ruledtabular}
\end{table}

\begin{table}
\caption{Final combined PbSc+PbGl $\pi^0$ invariant yields $vs.$ $p_T$ for centrality 50-60\%.
For points with no errors given, data value represents
90\% confidence level upper limit. }
\begin{ruledtabular}\begin{tabular}{cccccc} \hline
$p_T$ & Yield & Stat. Error & \% & Sys. Error & \% \\ \hline
1.25 & 3.593E-01 & 1.941E-03 & 0.54 &  4.022E-02 & 11.19\\
1.75 & 6.197E-02 & 4.018E-04 & 0.65 &  7.069E-03 & 11.41\\
2.25 & 1.309E-02 & 1.175E-04 & 0.90 &  1.553E-03 & 11.87\\
2.75 & 3.479E-03 & 4.211E-05 & 1.21 &  4.205E-04 & 12.09\\
3.25 & 1.019E-03 & 1.695E-05 & 1.66 &  1.291E-04 & 12.67\\
3.75 & 3.480E-04 & 8.518E-06 & 2.45 &  4.380E-05 & 12.59\\
4.25 & 1.329E-04 & 4.558E-06 & 3.43 &  1.763E-05 & 13.26\\
4.75 & 4.959E-05 & 2.434E-06 & 4.91 &  6.310E-06 & 12.73\\
5.25 & 2.125E-05 & 1.585E-06 & 7.46 &  3.032E-06 & 14.27\\
5.75 & 9.917E-06 & 7.569E-07 & 7.63 &  1.540E-06 & 15.52\\
6.25 & 6.127E-06 & 5.471E-07 & 8.93 &  9.978E-07 & 16.29\\
6.75 & 3.246E-06 & 3.392E-07 & 10.45 &  4.965E-07 & 15.30\\
7.25 & 1.664E-06 & 2.449E-07 & 14.72 &  3.102E-07 & 18.65\\
7.75 & 1.129E-06 & 1.886E-07 & 16.70 &  2.114E-07 & 18.72\\
8.50 & 3.362E-07 & 7.419E-08 & 22.07 &  6.694E-08 & 19.91\\
9.50 & 1.817E-07 & 4.619E-08 & 25.42 &  3.329E-08 & 18.32\\
11.00 & 2.858E-08 & 1.112E-08 & 38.89 &  4.803E-09 & 16.81\\
13.00 & 2.311E-08 & {--} & {--} & {--} & {--} \\
\hline
\end{tabular}\end{ruledtabular}
\end{table}

\begin{table}
\caption{Final combined PbSc+PbGl $\pi^0$ invariant yields $vs.$ $p_T$ for centrality 60-70\%.
}
\begin{ruledtabular}\begin{tabular}{cccccc} \hline
$p_T$ & Yield & Stat. Error & \% & Sys. Error & \% \\ \hline
1.25 & 1.731E-01 & 1.121E-03 & 0.65 &  1.985E-02 & 11.47\\
1.75 & 3.022E-02 & 2.288E-04 & 0.76 &  3.425E-03 & 11.33\\
2.25 & 6.567E-03 & 7.011E-05 & 1.07 &  7.773E-04 & 11.84\\
2.75 & 1.644E-03 & 2.565E-05 & 1.56 &  2.057E-04 & 12.51\\
3.25 & 5.255E-04 & 1.158E-05 & 2.20 &  6.682E-05 & 12.72\\
3.75 & 1.801E-04 & 6.044E-06 & 3.36 &  2.259E-05 & 12.54\\
4.25 & 6.986E-05 & 3.184E-06 & 4.56 &  9.254E-06 & 13.25\\
4.75 & 2.312E-05 & 1.631E-06 & 7.06 &  3.101E-06 & 13.41\\
5.25 & 1.156E-05 & 1.145E-06 & 9.90 &  1.720E-06 & 14.87\\
5.75 & 4.884E-06 & 5.045E-07 & 10.33 &  7.560E-07 & 15.48\\
6.25 & 2.690E-06 & 3.650E-07 & 13.57 &  4.303E-07 & 16.00\\
6.75 & 1.822E-06 & 2.658E-07 & 14.58 &  3.369E-07 & 18.48\\
7.25 & 6.281E-07 & 1.480E-07 & 23.57 &  1.178E-07 & 18.76\\
7.75 & 2.446E-07 & 1.082E-07 & 44.22 &  4.632E-08 & 18.94\\
8.50 & 1.417E-07 & 4.482E-08 & 31.62 &  2.707E-08 & 19.10\\
9.50 & 1.094E-07 & 3.843E-08 & 35.14 &  2.106E-08 & 19.26\\
11.00 & 2.492E-08 & 1.114E-08 & 44.72 &  4.816E-09 & 19.33\\
13.00 & 4.728E-09 & 4.728E-09 & 100.00 & 9.226E-10 & 19.51\\
\hline
\end{tabular}\end{ruledtabular}
\end{table}

\begin{table}
\caption{Final combined PbSc+PbGl $\pi^0$ invariant yields $vs.$ $p_T$ for centrality 70-80\%.
}
\begin{ruledtabular}\begin{tabular}{cccccc} \hline
$p_T$ & Yield & Stat. Error & \% & Sys. Error & \% \\ \hline
1.25 & 7.416E-02 & 5.166E-04 & 0.70 &  8.842E-03 & 11.92\\
1.75 & 1.282E-02 & 1.189E-04 & 0.93 &  1.496E-03 & 11.67\\
2.25 & 2.721E-03 & 3.774E-05 & 1.39 &  3.245E-04 & 11.92\\
2.75 & 7.455E-04 & 1.514E-05 & 2.03 &  9.131E-05 & 12.25\\
3.25 & 2.461E-04 & 7.508E-06 & 3.05 &  3.248E-05 & 13.20\\
3.75 & 7.200E-05 & 3.689E-06 & 5.12 &  9.687E-06 & 13.46\\
4.25 & 2.609E-05 & 2.071E-06 & 7.94 &  4.034E-06 & 15.46\\
4.75 & 1.288E-05 & 1.308E-06 & 10.15 &  2.161E-06 & 16.78\\
5.25 & 4.650E-06 & 7.727E-07 & 16.62 &  9.050E-07 & 19.46\\
5.75 & 2.416E-06 & 3.897E-07 & 16.13 &  4.736E-07 & 19.60\\
6.25 & 1.763E-06 & 2.713E-07 & 15.39 &  2.795E-07 & 15.85\\
6.75 & 5.945E-07 & 1.651E-07 & 27.77 &  1.221E-07 & 20.53\\
7.25 & 4.817E-07 & 1.245E-07 & 25.84 &  8.088E-08 & 16.79\\
7.75 & 1.344E-07 & 6.718E-08 & 50.00 &  2.545E-08 & 18.94\\
8.50 & 1.135E-07 & 4.012E-08 & 35.36 &  2.167E-08 & 19.10\\
9.50 & 4.968E-08 & 2.484E-08 & 50.00 &  9.568E-09 & 19.26\\
11.00 & 5.060E-09 & 5.060E-09 & 100.00 &  9.778E-10 & 19.33\\
\hline
\end{tabular}\end{ruledtabular}
\end{table}

\begin{table}
\caption{Final combined PbSc+PbGl $\pi^0$ invariant yields $vs.$ $p_T$ for centrality 80-92\%.
}
\begin{ruledtabular}\begin{tabular}{cccccc} \hline
$p_T$ & Yield & Stat. Error & \% & Sys. Error & \% \\ \hline
1.25 & 3.494E-02 & 6.093E-04 & 1.74 &  4.504E-03 & 12.89\\
1.75 & 6.037E-03 & 1.291E-04 & 2.14 &  7.607E-04 & 12.60\\
2.25 & 1.319E-03 & 3.628E-05 & 2.75 &  1.701E-04 & 12.89\\
2.75 & 3.321E-04 & 1.243E-05 & 3.74 &  4.570E-05 & 13.76\\
3.25 & 1.059E-04 & 5.281E-06 & 4.99 &  1.483E-05 & 14.01\\
3.75 & 3.625E-05 & 2.408E-06 & 6.64 &  4.455E-06 & 12.29\\
4.25 & 1.233E-05 & 1.293E-06 & 10.48 &  1.730E-06 & 14.03\\
4.75 & 6.501E-06 & 7.988E-07 & 12.29 &  9.044E-07 & 13.91\\
5.25 & 3.018E-06 & 5.360E-07 & 17.76 &  4.224E-07 & 13.99\\
5.75 & 1.072E-06 & 2.315E-07 & 21.60 &  1.815E-07 & 16.94\\
6.25 & 3.265E-07 & 1.154E-07 & 35.36 &  5.945E-08 & 18.21\\
6.75 & 2.805E-07 & 9.918E-08 & 35.36 &  5.185E-08 & 18.48\\
7.25 & 2.231E-07 & 8.434E-08 & 37.80 &  4.187E-08 & 18.76\\
7.75 & 8.467E-08 & 4.888E-08 & 57.74 &  1.604E-08 & 18.94\\
8.50 & 3.602E-08 & 2.080E-08 & 57.74 &  6.880E-09 & 19.10\\
9.50 & 1.077E-08 & 1.077E-08 & 100.00 &  2.074E-09 & 19.26\\
11.00 & 4.375E-09 & 4.375E-09 & 100.00 &  8.455E-10 & 19.32\\
\hline
\end{tabular}\end{ruledtabular}
\end{table}

\begin{table}
\caption{Final combined PbSc+PbGl $\pi^0$ invariant yields $vs.$ $p_T$ for centrality 0-92\%.
}
\begin{ruledtabular}\begin{tabular}{cccccc} \hline
$p_T$ & Yield & Stat. Error & \% & Sys. Error & \% \\ \hline
1.25 & 1.078E+00 & 3.333E-03 & 0.31 &  1.205E-01 & 11.17\\
1.75 & 1.928E-01 & 6.847E-04 & 0.36 &  2.171E-02 & 11.26\\
2.25 & 4.038E-02 & 1.742E-04 & 0.43 &  4.822E-03 & 11.94\\
2.75 & 9.578E-03 & 5.293E-05 & 0.55 &  1.202E-03 & 12.55\\
3.25 & 2.564E-03 & 1.858E-05 & 0.72 &  3.375E-04 & 13.17\\
3.75 & 8.115E-04 & 7.353E-06 & 0.91 &  1.013E-04 & 12.48\\
4.25 & 2.906E-04 & 3.475E-06 & 1.20 &  3.729E-05 & 12.84\\
4.75 & 1.121E-04 & 1.806E-06 & 1.61 &  1.466E-05 & 13.08\\
5.25 & 4.924E-05 & 1.031E-06 & 2.09 &  6.494E-06 & 13.19\\
5.75 & 2.240E-05 & 4.723E-07 & 2.11 &  3.012E-06 & 13.45\\
6.25 & 1.190E-05 & 2.909E-07 & 2.44 &  1.647E-06 & 13.83\\
6.75 & 5.970E-06 & 1.943E-07 & 3.25 &  8.494E-07 & 14.23\\
7.25 & 3.246E-06 & 1.273E-07 & 3.92 &  4.758E-07 & 14.65\\
7.75 & 1.715E-06 & 9.049E-08 & 5.28 &  2.658E-07 & 15.49\\
8.50 & 8.583E-07 & 3.892E-08 & 4.53 &  1.285E-07 & 14.98\\
9.50 & 3.078E-07 & 2.351E-08 & 7.64 &  5.041E-08 & 16.38\\
11.00 & 9.178E-08 & 7.770E-09 & 8.47 &  1.417E-08 & 15.44\\
13.00 & 2.380E-08 & 3.856E-09 & 16.20 & 3.816E-09 & 16.03\\
\hline
\end{tabular}\end{ruledtabular}
\end{table}


\begin{table}[ht]
\caption{$\pi^0$ spectrum for combined centralities: 0-20\%
}
\begin{ruledtabular}\begin{tabular}{cccccc} \hline
$p_T$ & Yield & Stat. Error & \% & Sys. Error & \% \\ \hline
1.25 & 2.684E+00 & 1.455E-02 & 0.54 &  3.106E-01 & 11.58\\
1.75 & 5.059E-01 & 2.875E-03 & 0.57 &  5.131E-02 & 10.14\\
2.25 & 1.033E-01 & 7.343E-04 & 0.71 &  1.137E-02 & 11.01\\
2.75 & 2.373E-02 & 2.198E-04 & 0.93 &  2.837E-03 & 11.96\\
3.25 & 6.014E-03 & 7.501E-05 & 1.25 &  7.693E-04 & 12.79\\
3.75 & 1.912E-03 & 2.840E-05 & 1.49 &  2.137E-04 & 11.18\\
4.25 & 6.522E-04 & 1.307E-05 & 2.00 &  7.660E-05 & 11.74\\
4.75 & 2.484E-04 & 6.637E-06 & 2.67 &  3.055E-05 & 12.30\\
5.25 & 1.118E-04 & 3.683E-06 & 3.29 &  1.368E-05 & 12.23\\
5.75 & 5.034E-05 & 1.670E-06 & 3.32 &  6.775E-06 & 13.46\\
6.25 & 2.582E-05 & 9.994E-07 & 3.87 &  3.466E-06 & 13.43\\
6.75 & 1.194E-05 & 6.415E-07 & 5.37 &  1.825E-06 & 15.29\\
7.25 & 7.048E-06 & 4.237E-07 & 6.01 &  1.115E-06 & 15.82\\
7.75 & 4.099E-06 & 3.073E-07 & 7.50 &  7.159E-07 & 17.46\\
8.50 & 1.804E-06 & 1.247E-07 & 6.92 &  3.037E-07 & 16.84\\
9.50 & 8.445E-07 & 8.042E-08 & 9.52 &  1.491E-07 & 17.65\\
11.00 & 2.239E-07 & 2.714E-08 & 12.12 &  4.219E-08 & 18.84\\
13.00 & 5.517E-08 & 1.011E-08 & 18.32 &  1.044E-08 & 18.93\\
\hline
\end{tabular}\end{ruledtabular}
\end{table}

\begin{table}[ht]
\caption{$\pi^0$ spectrum for combined centralities: 20-60\%
}
\begin{ruledtabular}\begin{tabular}{cccccc} \hline
$p_T$ & Yield & Stat. Error & \% & Sys. Error & \% \\ \hline
1.25 & 9.097E-01 & 2.963E-03 & 0.33 &  9.041E-02 & 9.94\\
1.75 & 1.602E-01 & 5.854E-04 & 0.37 &  1.570E-02 & 9.80\\
2.25 & 3.407E-02 & 1.599E-04 & 0.47 &  3.579E-03 & 10.51\\
2.75 & 8.386E-03 & 4.969E-05 & 0.59 &  9.138E-04 & 10.90\\
3.25 & 2.339E-03 & 1.832E-05 & 0.78 &  2.711E-04 & 11.59\\
3.75 & 7.562E-04 & 8.034E-06 & 1.06 &  8.387E-05 & 11.09\\
4.25 & 2.819E-04 & 4.021E-06 & 1.43 &  3.259E-05 & 11.56\\
4.75 & 1.090E-04 & 2.126E-06 & 1.95 &  1.263E-05 & 11.58\\
5.25 & 4.695E-05 & 1.310E-06 & 2.79 &  5.959E-06 & 12.69\\
5.75 & 2.234E-05 & 6.107E-07 & 2.73 &  2.907E-06 & 13.01\\
6.25 & 1.120E-05 & 3.833E-07 & 3.42 &  1.605E-06 & 14.33\\
6.75 & 5.961E-06 & 2.591E-07 & 4.35 &  8.809E-07 & 14.78\\
7.25 & 3.140E-06 & 1.732E-07 & 5.52 &  4.937E-07 & 15.73\\
7.75 & 1.817E-06 & 1.264E-07 & 6.96 &  3.292E-07 & 18.12\\
8.50 & 9.139E-07 & 5.846E-08 & 6.40 &  1.581E-07 & 17.30\\
9.50 & 3.818E-07 & 3.462E-08 & 9.07 &  7.055E-08 & 18.48\\
11.00 & 1.176E-07 & 1.194E-08 & 10.15 &  2.061E-08 & 17.52\\
13.00 & 2.899E-08 & 6.031E-09 & 20.80 &  4.497E-09 & 15.51\\
\hline
\end{tabular}\end{ruledtabular}
\end{table}

\begin{table}[ht]
\caption{$\pi^0$ spectrum for combined centralities: 60-92\%
}
\begin{ruledtabular}\begin{tabular}{cccccc} \hline
$p_T$ & Yield & Stat. Error & \% & Sys. Error & \% \\ \hline
1.25 & 9.037E-02 & 4.484E-04 & 0.50 &  9.891E-03 & 10.94\\
1.75 & 1.571E-02 & 9.400E-05 & 0.60 &  1.680E-03 & 10.69\\
2.25 & 3.397E-03 & 2.836E-05 & 0.83 &  3.765E-04 & 11.08\\
2.75 & 8.712E-04 & 1.041E-05 & 1.19 &  1.025E-04 & 11.76\\
3.25 & 2.808E-04 & 4.746E-06 & 1.69 &  3.442E-05 & 12.26\\
3.75 & 9.236E-05 & 2.390E-06 & 2.59 &  1.073E-05 & 11.61\\
4.25 & 3.461E-05 & 1.282E-06 & 3.70 &  4.593E-06 & 13.27\\
4.75 & 1.369E-05 & 7.188E-07 & 5.25 &  1.857E-06 & 13.57\\
5.25 & 6.198E-06 & 4.761E-07 & 7.68 &  9.308E-07 & 15.02\\
5.75 & 2.683E-06 & 2.173E-07 & 8.10 &  4.415E-07 & 16.46\\
6.25 & 1.514E-06 & 1.486E-07 & 9.81 &  2.414E-07 & 15.94\\
6.75 & 8.605E-07 & 1.046E-07 & 12.16 &  1.589E-07 & 18.46\\
7.25 & 4.305E-07 & 6.822E-08 & 15.85 &  7.426E-08 & 17.25\\
7.75 & 1.502E-07 & 4.381E-08 & 29.17 &  2.719E-08 & 18.10\\
8.50 & 9.326E-08 & 2.035E-08 & 21.82 &  1.706E-08 & 18.29\\
9.50 & 5.374E-08 & 1.486E-08 & 27.65 &  9.997E-09 & 18.60\\
11.00 & 1.101E-08 & 4.162E-09 & 37.80 &  2.062E-09 & 18.73\\
13.00 & 1.478E-09 & 1.478E-09 & 100.00 &  2.883E-10 & 19.51\\
\hline
\end{tabular}\end{ruledtabular}
\end{table}

\endgroup




\clearpage



\end{document}